\documentclass[]{aa}
\usepackage{graphicx}
\usepackage{txfonts}
\usepackage{natbib}
\usepackage{bm}
\usepackage{url}
\def\farcs{\hbox{$.\!\!^{\prime\prime}$}}
\def\farcm{\hbox{$.\mkern-4mu^\prime$}}

\defcitealias{hbb05}{H05}

\begin{document}
   \title{Cosmic shear analysis of archival HST/ACS data\thanks{Based on observations made with the NASA/ESA Hubble Space Telescope, obtained from the data archives at the Space Telescope European Coordinating Facility and the Space Telescope Science Institute, which is operated by the Association of Universities for Research in Astronomy, Inc., under NASA contract NAS 5-26555.}}
   \subtitle{I. Comparison of early ACS pure parallel data to the HST/GEMS Survey
}
   \author{T. Schrabback
          \inst{1}
          \and
          T. Erben
          \inst{1}
          \and
          P. Simon
          \inst{1}
          \and
          J.-M. Miralles
          \inst{1,2,3}
          \and
          P. Schneider
          \inst{1}
          \and
          C. Heymans
          \inst{4}
          \and
          T. Eifler
          \inst{1}
          \and
          R.~A.~E. Fosbury
          \inst{5}
          \and
          W. Freudling
          \inst{5}
          \and
          M. Hetterscheidt
          \inst{1}
          \and
           H. Hildebrandt
          \inst{1}
           \and
           N. Pirzkal
          \inst{5,6}
          }

   \offprints{T. Schrabback}
    \institute{Argelander-Institut f\"ur Astronomie\thanks{Founded by merging of the Institut f\"ur Astrophysik
           und Extraterrestrische Forschung, the Sternwarte, and the
           Radioastronomisches Institut der Universit\"at Bonn.}, Universit\"at Bonn,
           Auf dem H\"ugel 71, D-53121 Bonn, Germany\\
           \email{schrabba@astro.uni-bonn.de}
              \and
              European Southern Observatory, Karl-Schwarzschild-Str. 2, 85741 Garching, Germany
              \and
              T\`ecniques d'Avantguarda, Avda. Carlemany 75, Les Escaldes, AD-700 Principat d'Andorra
               \and
              University of British Columbia, Department of Physics and Astronomy, 6224 Agricultural Road, Vancouver, Canada
              \and
              ST-ECF, European Southern Observatory, Karl-Schwarzschild Str. 2, 85741 Garching, Germany 
              \and
              Space Telescope Science Institute, 3700 San Martin Drive, Baltimore, MD 21218
            }

   \date{Received 24 June 2006 / Accepted 20 March 2007}
         
   \abstract{This is the first paper of a series describing our measurement of weak lensing by large-scale structure, also termed ``cosmic shear'', using archival observations from the Advanced Camera for Surveys (ACS) on board the Hubble Space Telescope (HST).}
{
In this work we present results from a pilot study testing the capabilities of the ACS for cosmic shear measurements with early parallel observations and presenting a re-analysis of HST/ACS data from the GEMS survey and the GOODS observations of the \textit{Chandra} Deep Field South (CDFS).}
{We describe the data reduction and, in particular, a new correction scheme for the time-dependent ACS point-spread-function (PSF) based on observations of stellar fields.
This is currently the only technique which takes the full time variation of the PSF between individual ACS exposures into account.
We estimate that our PSF correction scheme reduces the systematic contribution to the shear correlation functions due to PSF distortions to \mbox{$< 2 \times 10^{-6}$} for galaxy fields containing at least 10 stars,
which corresponds to $\lesssim 5\%$ of the cosmological signal expected on scales of a single ACS field.
}
{We perform a number of diagnostic tests indicating that the remaining level of systematics is consistent with zero for the GEMS and GOODS data confirming the success of our PSF correction scheme.
For the parallel data we detect a low level of remaining systematics which we interpret to be caused by a lack of sufficient dithering of the data.
Combining the shear estimate of the GEMS and GOODS observations using $96$ galaxies $\mathrm{arcmin}^{-2}$ with the photometric redshift catalogue of the GOODS-MUSIC sample, we determine a \textit{local single field estimate} for the mass power spectrum normalisation $\sigma_{8,\mathrm{CDFS}}=0.52^{+0.11}_{-0.15}\mathrm{(stat)}\pm0.07\mathrm{(sys)}$ (68\% confidence assuming Gaussian cosmic variance) at a fixed matter density $\Omega_\mathrm{m}=0.3$ for a $\Lambda$CDM cosmology marginalising over the uncertainty of the Hubble parameter and the redshift distribution.
We interpret this exceptionally low estimate to be due to a local under-density of the foreground structures in the CDFS.
}
{}
   \maketitle
\section{Introduction}
Cosmic shear, the weak gravitational lensing effect of the large-scale structure, provides a powerful tool to constrain the total matter power spectrum without any assumptions on the relation between luminous and dark matter.
Due to the weakness of the effect, it is challenging to measure, with the first detections only reported six years ago
\citep{bre00,kwl00,wme00,wtk00}.
Since then cosmic shear has developed into a flourishing field of cosmology yielding not only constraints on the matter content $\Omega_\mathrm{m}$ and the normalisation of the power spectrum $\sigma_8$ 
(\citealt{mwm01,wmr01,wmp02,wmh05,hyg02,rrg02,bmr03,btb03,jbf03,hms03,hbh04,hbb05,rrc04,mrb05,hss06}; \citealt{mhb07}),
but recently also on the equation of state parameter $w$ of dark energy 
(\citealt{jjb06,hmw06,smw06}; \citealt{kht07}), see \citet{bas01,mwb02,hyg02a,ref03} for reviews.

Due to the weakness of cosmological gravitational shear, proper correction for systematics, first of all the image point-spread-function (PSF) is indispensable.
Within the Shear TEsting Programme (STEP) a number of algorithms to measure the shear from faint and PSF-distorted galaxy images
are currently tested using a blind analysis of image simulations aiming to improve and quantify the accuracy of the different methods; see \citet{hwb06}; \citet{mhb07} for first results.

The majority of the previous cosmic shear measurements have been made with ground-based wide-field imaging data,
mainly probing the matter power spectrum on linear to moderately non-linear angular scales from several degrees down to several arcminutes.
In order to probe the highly non-linear power spectrum at arcminute and sub-arcminute scales, a high number density of usable background galaxies is required.
While deep ground-based surveys are typically limited to $\simeq 30$ galaxies/$\mathrm{arcmin}^2$ due to seeing,
significantly higher number densities of resolved galaxies can be obtained from space-based images.
Cosmic shear studies have already been carried out with the HST cameras WFPC2 \citep{rrg01,rrg02,crg03} and
STIS \citep{hms02,rrc04,meh05}.
With the installation of the \textit{Advanced Camera for Surveys} (ACS)
\textit{Wide-Field Channel} (WFC) detector, a camera combining improved sensitivity ($48\%$ total throughput at 660 nm) and a relatively large field-of-view \mbox{($\sim 3\farcm3 \times 3\farcm3$)} with good sampling (0\farcs05 per pixel) \citep{fch03,smc04}, the possibilities to measure weak lensing with HST have been further improved substantially.
ACS has already been used for weak lensing measurements of galaxy clusters
(\citealt{jwb05,jwf05,jwf06,lrb05,cbg06,bcg06}; \citealt{lgh07}) and galaxy-galaxy lensing \citep{hbr06,gtr07}.
The first cosmic shear analysis with ACS was presented by \citet{hbb05}, \citetalias{hbb05} henceforth, for the GEMS survey \citep{rbb04}, a \mbox{$\sim 28^\prime \times 28^\prime$} mosaic incorporating the HST/ACS GOODS observations of the \textit{Chandra} Deep Field South \citep[CDFS, ][]{gfk04}.
Recently \citet{lmk07} and \citet{mrl07} presented a cosmological weak lensing analysis for the ACS COSMOS\footnote{\url{http://www.astro.caltech.edu/~cosmos}} field.

In this work we present results from a pilot cosmic shear study using early data from the ACS Parallel Cosmic Shear Survey (proposals 9480, 9984; PI J. Rhodes).
Parallel observations provide many independent lines of sight reducing the impact of cosmic variance. 
With a separation of several arcminutes from the primary target (e.g. $\sim 6^\prime$ for WFPC2) they provide nearly random pointings for most classes of primary targets. 
This is important when measuring a statistical quantity like the shear. 
Still, for primary observations pointing at particularly over-dense regions, such as galaxy clusters, a significant selection bias might be introduced as they influence the shear field many arcminutes around them, which has to be checked carefully.

Analysing stellar fields we detect short-term variations of the ACS PSF, which are interpreted as focus changes due to thermal breathing of the telescope 
(see also \citealt{kri03,rma05,ank06}; \citealt{rma07}).
Whereas earlier studies with other HST cameras assumed temporal stability of the PSF, a fully time-dependent PSF correction is required for ACS due to these detected variations.
Given that only a low number of stars are present in high galactic latitude ACS fields \mbox{($\sim 10-30$)}, the correction cannot be determined from a simple interpolation across the field-of-view, but requires prior knowledge about possible PSF patterns.
In this work we apply such a correction scheme, which is based on PSF models derived from stellar fields. 
Our method takes the full PSF variation between individual exposures into account and can be applied for arbitrary dither patterns and rotations.
\citet{rma05,rma07} propose a different correction scheme, in which they fit co-added frames with theoretical single-focus PSF models created with a modified version of \texttt{TinyTim}\footnote{\url{http://www.stsci.edu/software/tinytim/tinytim.html}}.

\citetalias{hbb05} use a semi-time-dependent model to correct for the image PSF, based on the combined stars for each of the two GEMS observation epochs.
The tests for systematics presented by \citetalias{hbb05} indicate zero contamination with systematics for the GEMS only data, but remaining systematics at small scales if the GOODS data are included. 
In order to test our fully time-dependent PSF correction we present a re-analysis of the GEMS and GOODS data, yielding a level of systematics consistent with zero for the combined dataset.
We present a cosmological parameter estimation from the re-analysed GEMS and GOODS data in this work, whereas a 
parameter estimation from the parallel data will be provided in a future paper on the basis of the complete ACS Parallel Cosmic Shear Survey.

The paper is organised as follows: 
After summarising the weak lensing formalism applied in Sect.\thinspace\ref{se:method},
we describe the data and data reduction in Sect.\thinspace\ref{se:data}.
We present our analysis of the ACS PSF and the correction scheme
in Sect.\thinspace\ref{se:psf}.
Next we elaborate on the galaxy selection and determined redshift distribution (Sect.\thinspace\ref{se:galaxy_select_redshift}) and compute several estimators for the shear and systematics 
in Sect.\thinspace\ref{se:preliminary_cosmic_shear}.
After presenting the results of the cosmological re-analysis of the GEMS and GOODS data in Sect.\thinspace\ref{se:cosmo_para_estimate}, we conclude 
in Sect.\thinspace\ref{se:conclusion}.

\section{Method}
\label{se:method}

Cosmic shear provides a powerful tool to investigate the 3-dimensional power spectrum of matter fluctuations $P_\delta$ through the observable shear field \mbox{$\gamma=\gamma_1 + \mathrm{i} \gamma_2$} induced by the tidal gravitational field (see \citealt{bas01} for a broader introduction). 
They are related via the projected 2-dimensional shear (convergence) power spectrum
\begin{equation}
      P_\kappa(\ell) = \frac{9 H_0^4 \Omega_\mathrm{m}^2}{4 c^4} \int_o^{w_\mathrm{h}} \mathrm{d} w \frac{g^2(w)}{a^2(w)} P_\delta \left( \frac{\ell}{f_K(w)}, w \right) \, ,
\end{equation}
where $H_0$ is the Hubble parameter, $\Omega_\mathrm{m}$ the matter density parameter, $a$ the scale factor, $w$ the comoving radial distance, $w_\mathrm{h}$ the comoving distance to the horizon, 
$\ell$ the modulus of the wave vector,
and  $f_K(w)$ denotes the comoving angular diameter distance.
The source redshift distribution $p_w$ determines the weighted lens efficiency factor
\begin{equation}
    g(w) \equiv \int_w^{w_\mathrm{h}} \mathrm{d} w' \, p_w(w') \frac{f_K(w'-w)}{f_K(w')} ,
\end{equation}
see, e.g. \citet{kai98,swj98}.

\subsection{Cosmic shear estimators}
\label{se:theory:cosmic_shear_estimators}
In this work we measure the shear two-point correlation functions
\begin{equation}
     \langle \gamma_\mathrm{t} \gamma_\mathrm{t} \rangle (\theta) = \frac{\sum^{N}_{i,j}w_i w_j \gamma_{\mathrm{t},i}(\boldsymbol{x}_i)\cdot\gamma_{\mathrm{t},j}(\boldsymbol{x}_j)}{\sum^{N}_{i,j}w_i w_j} \, , 
\end{equation}
\begin{equation}
     \langle \gamma_\times \gamma_\times \rangle (\theta) = \frac{\sum^{N}_{i,j}w_i w_j \gamma_{\times,i}(\boldsymbol{x}_i)\cdot\gamma_{\times,j}(\boldsymbol{x}_j)}{\sum^{N}_{i,j}w_i w_j}
\end{equation}
from galaxy pairs separated by \mbox{$\theta=|\boldsymbol{x}_i-\boldsymbol{x}_j|$}, where the tangential component $\gamma_\mathrm{t}$ and the 45 degree rotated cross-component $\gamma_\times$ of the shear relative to the separation vector are estimated from galaxy ellipticities,
and $w_i$ denotes the weight of the $i$th galaxy.
It is useful to consider the combinations
\begin{equation}
    \xi_\pm (\theta) = \langle \gamma_t \gamma_t \rangle (\theta) \pm \langle \gamma_\times \gamma_\times \rangle (\theta) \, ,
\end{equation}
which are directly related to the convergence power spectrum
\begin{equation}
    \xi_+ (\theta) = \frac{1}{2 \pi} \int_0^\infty \mathrm{d} \ell \, \ell \, \mathrm{J}_0 (\ell \theta) P_\kappa (\ell) \, ,
\end{equation}
\begin{equation}
    \xi_- (\theta) = \frac{1}{2 \pi} \int_0^\infty \mathrm{d} \ell \, \ell \, \mathrm{J}_4 (\ell \theta) P_\kappa (\ell) \, ,
\end{equation}
where $\mathrm{J}_n$ denotes the $n^\mathrm{th}$-order Bessel function of the first kind.
\citet{cnp02} show that $\xi_\pm$ can be decomposed into a curl-free `E'-mode component $\xi^E(\theta)$ and a curl `B'-mode component $\xi^B(\theta)$ as
\begin{equation}
\label{eq:xi_eb}
  \xi^E(\theta)=\frac{\xi_+(\theta)+\xi^\prime(\theta)}{2}  \,, \quad  \xi^B(\theta)=\frac{\xi_+(\theta)-\xi^\prime(\theta)}{2}  \,,
\end{equation}
with
\begin{equation}
  \label{eq:xi_prime}
  \xi^\prime(\theta)=\xi_-(\theta)+4\int_\theta^\infty \frac{\mathrm{d}\vartheta}{\vartheta}\xi_-(\vartheta) -12 \theta^2 \int_\theta^\infty \frac{\mathrm{d}\vartheta}{\vartheta^3}\xi_-(\vartheta)  \,.
\end{equation}
As weak gravitational lensing only contributes to the curl-free E-modes, such a decomposition provides an important test for the remaining contamination of the data with systematics. 
Note, however, that the integral in (\ref{eq:xi_prime}) formally extends to infinity. 
Thus, due to finite field size, real data require the substitution of the measured $\xi_-(\theta)$ with theoretical predictions for large $\theta$.

This problem does not occur for the aperture mass statistics \citep{sch96m} 
 \begin{equation}
  M_{\mathrm{ap}/\bot}( \boldsymbol{\zeta} ) = \int \mathrm{d}^2 \theta^\prime \gamma_{\mathrm{t}/\times}(\boldsymbol{\theta}^\prime; \boldsymbol{\zeta}) Q(|\boldsymbol{\theta}^\prime-\boldsymbol{\zeta}|) \, ,
\end{equation}
which is defined using the tangential and cross-components of the shear relative to the aperture centre $\boldsymbol{\zeta}$
\begin{equation}
  \label{eq:gamma_t}
  \gamma_\mathrm{t}(\boldsymbol{\theta}^\prime; \boldsymbol{\zeta}) = - \Re \left[ \gamma (\boldsymbol{\theta}^\prime)\mathrm{e}^{-2 \mathrm{i} \phi} \right] \, ,
\end{equation}
\begin{equation}
  \label{eq:gamma_cross}
  \gamma_\times (\boldsymbol{\theta}^\prime; \boldsymbol{\zeta}) = - \Im \left[ \gamma (\boldsymbol{\theta}^\prime) \mathrm{e}^{-2 \mathrm{i} \phi} \right] \, 
\end{equation}
with $\boldsymbol{\theta}^\prime-\boldsymbol{\zeta}=|\boldsymbol{\theta}^\prime-\boldsymbol{\zeta}|(\cos{\phi}+\mathrm{i}\sin{\phi})$.
For the axially-symmetric weight function $Q(\vartheta)$ we use a form proposed in \citet{swj98} 
\begin{equation}
  \label{eq:q_generic}
  Q(\vartheta) = \frac{6}{\pi \theta^2} \left(\frac{\vartheta}{\theta}\right)^2 \left[ 1 - \left(\frac{\vartheta}{\theta}\right)^2\right]^2 \mathrm{H}(\theta - \vartheta)\, ,
\end{equation}
where $\mathrm{H}(x)$ denotes the Heaviside step function.

\citet{cnp02} show that $M_{\mathrm{ap}}$ purely measures the E-mode signal, whereas $M_{\bot}$ contributes to the B-mode only.

The dispersion of the aperture mass $\langle M_\mathrm{ap}^2 \rangle(\theta)$
is related to the convergence power spectrum by
\begin{equation}
  \label{eq:map2powergeneric}
  \langle M^2_\mathrm{ap} \rangle (\theta) = \frac{1}{2 \pi} \int_0^\infty \mathrm{d} \ell \,\ell\, P_\kappa(\ell) \,W_\mathrm{ap} (\theta \ell)  \, ,
\end{equation}
with
 $W_\mathrm{ap} (\eta) = 576\,\mathrm{J}_4^2 (\eta) \,\eta^{-4}$, 
and can in principle be computed by placing apertures on the shear field, yet in practice is preferentially calculated from the shear two-point correlation functions \citep{cnp02,swm02} to avoid problems with masked regions
\begin{equation}
  \label{eq:map2ofxi}
  \langle M^2_\mathrm{ap} \rangle (\theta) = \frac{1}{2}\int_0^{2\theta} \frac{\mathrm{d} \vartheta \, \vartheta}{\theta^2} \left [\xi_+ (\vartheta) T_+ \left( \frac{\vartheta}{\theta} \right)
+ \xi_- (\vartheta) T_- \left( \frac{\vartheta}{\theta} \right) \right]
\end{equation}
\begin{equation}
  \label{eq:mapbot2ofxi}
  \langle M^2_\bot \rangle (\theta) = \frac{1}{2}\int_0^{2\theta} \frac{\mathrm{d} \vartheta \, \vartheta}{\theta^2} \left [\xi_+ (\vartheta) T_+ \left( \frac{\vartheta}{\theta} \right)
- \xi_- (\vartheta) T_- \left( \frac{\vartheta}{\theta} \right) \right]\,,
\end{equation}
with 
$T_\pm$ given in \citet{swm02}.
%\begin{eqnarray}
% T_+ (x) & = &   \mathrm{H}(2-x) \left\{ \frac{6(2-15 x^2)}{5} \left[ 1 - \frac{2}{\pi} \arcsin{\left(\frac{x}{2}\right)} \right] \right. \nonumber \\
%& \, & \left. + \frac{x \sqrt{4-x^2}}{100 \pi} (120 + 2320x^2-754x^4+132x^6-9x^8) \right\} \, , \nonumber \\
%  T_- (x) & = & \frac{192}{35 \pi} x^3 \left(1-\frac{x^2}{4} \right)^{7/2} \mathrm{H}(2-x) \, .  \nonumber
%\end{eqnarray}

\subsection{The KSB formalism}
\label{se:su:ksb}
\citet*{ksb95}, \citet{luk97}, and \citet{hfk98} (KSB+) developed a formalism to estimate the reduced gravitational shear field 
\begin{equation}
  \label{eq:gofgamma}
g = g_1 + \mathrm{i} g_2 =\frac{\gamma}{1-\kappa} 
\end{equation}
from the observed images of background galaxies correcting for the smearing and distortion of the image PSF.
In this formalism the object ellipticity parameter
\begin{equation}
\label{eq:elli_e} 
   e = e_1 + \textrm{i} e_2 = \frac{Q_{11} - Q_{22} + 2 \textrm{i} Q_{12}}{Q_{11} + Q_{22}} 
\end{equation}
is defined in terms of second-order brightness moments
\begin{equation}
  \label{eq:qij_w}
  Q_{ij} = \int \mathrm{d}^2 \theta \, W_{r_\mathrm{g}}(|\boldsymbol{\theta}|) \, \theta_i \, \theta_j I(\boldsymbol{\theta}) \, , \quad i,j \in \{1,2\}  \, ,
\end{equation}
where $W_{r_\mathrm{g}}$ is a circular Gaussian weight function with filter scale $r_\mathrm{g}$.
The total response of a galaxy ellipticity to the reduced shear $g$ and PSF effects is given by
\begin{equation}
  \label{eq:elli_shear_psf}
  e_\alpha - e_\alpha^\mathrm{s} = P_{\alpha \beta}^g g_\beta + P_{\alpha \beta}^\mathrm{sm} q_\beta^*  \, ,
\end{equation}
with the intrinsic source ellipticity $e^\mathrm{s}$, the ``pre-seeing'' shear polarisability
\begin{equation}
  \label{eq:pg_def}
  P^g_{\alpha \beta} = P^{\mathrm{sh}}_{\alpha \beta} - P^{\mathrm{sm}}_{\alpha \gamma} \left[ \left( P^{\mathrm{sm}*} \right)^{-1}_{\gamma \delta} P^{\mathrm{sh}*}_{\delta \beta} \right]  \,,
\end{equation}
and the shear and smear polarisability tensors $P^\mathrm{sh}$ and $P^\mathrm{sm}$, which are calculated from higher-order brightness moments as detailed in \citet{hfk98}.
The anisotropy kernel $q^*(\boldsymbol{\theta})$ describes the anisotropic component of the PSF and has to be measured from stellar images (denoted with the asterisk throughout this paper), which are not affected by gravitational shear and have $e^{\mathrm{s}*}=0$:
\begin{equation}
  \label{eq:qalpha}
  q^*_\alpha = (P^{\mathrm{sm}*})^{-1}_{\alpha \beta} e_\beta^{*}  \,.
\end{equation}
We define the anisotropy corrected ellipticity 
\begin{equation}
  \label{eq:eani}
  e^\mathrm{ani}_\alpha = e_\alpha - P^{\mathrm{sm}}_{\alpha \beta} q^*_\beta  \,,
\end{equation}
and the fully corrected ellipticity as
\begin{equation}
  \label{eq:fully_corr}
  e^\mathrm{iso}_\alpha = (P^g)^{-1}_{\alpha \beta} e_\beta^\mathrm{ani}  \, ,
\end{equation}
which is an unbiased estimator for the reduced gravitational shear $\langle e^\mathrm{iso} \rangle = g$, assuming a random orientation of the intrinsic ellipticity $e^\mathrm{s}$.
For the weak distortions measured in cosmic shear $\kappa \ll 1$, and hence
\begin{equation}
  \label{eq:eiso_g_gamma}
  \langle e^\mathrm{iso} \rangle = g \simeq \gamma \, .
\end{equation}

The KSB+ formalism relies on the assumption that the image PSF can be described as a convolution of an isotropic part with a small anisotropy kernel.
Thus, it is ill-defined for several realistic PSF types \citep{kai00}, being of particular concern for diffraction limited space-based PSFs.
This shortcoming incited the development of alternative methods \citep{rrg00,kai00,bej02,reb03,mar05,kui06,nab07}. 
Nevertheless \citet{hfk98} demonstrated the applicability of the formalism for HST/WFPC2 images, if the filter scale $r_\mathrm{g}$ used to measure stellar shapes is matched to the filter scale used for galaxy images.

In this pilot study we restrain the analysis to the \citet{ewb01} implementation of the most commonly used KSB+ formalism.
There are currently several independent KSB implementations in use, which differ in the details of the computation, yielding slightly different results (see \citealt{hwb06} for a comparison of several implementations).
In particular, our implementation interpolates between pixel positions for the calculation of $Q_{ij}$, $P^{\mathrm{sm}}_{\alpha \beta}$, and $P^{\mathrm{sh}}_{\alpha \beta}$ and measures all stellar quantities needed for the correction of the galaxy ellipticities as a function of the filter scale $r_\mathrm{g}$ following \citet{hfk98}.
For the calculation of $P^g_{\alpha \beta}$ in (\ref{eq:pg_def}) and its inversion in (\ref{eq:fully_corr}) we use the approximations
\begin{equation}
  \label{eq:tensor_inversion}
 \left[ \left( P^{\mathrm{sm}*} \right)^{-1}_{\gamma \delta} P^{\mathrm{sh}*}_{\delta \beta} \right] \approx \frac{\mathrm{Tr}\left[P^{\mathrm{sh}*}\right]}{\mathrm{Tr}\left[P^{\mathrm{sm}*}\right]}\delta_{\gamma \beta} \, , \quad  (P^g)^{-1}_{\alpha \beta} \approx \frac{2}{\mathrm{Tr}\left[P^g\right]}\delta_{\alpha \beta} \, ,
\end{equation}
as the trace-free part of the tensor is much smaller than the trace
\citep{ewb01}.
To simplify the notation in the following sections we define
\begin{equation}
  \label{eq:t_pshpsm}
T^* \equiv \frac{\mathrm{Tr}\left[P^{\mathrm{sh}*}\right]}{\mathrm{Tr}\left[P^{\mathrm{sm}*}\right]} \, .
\end{equation}

We have tested this implementation with image simulations of the STEP project\footnote{\url{http://www.physics.ubc.ca/~heymans/step.html}}.
In the analysis of the first set of image simulations (STEP1) we identified significant biases \citep{hwb06}, which we eliminate with improved selection criteria (see Sect.\thinspace\ref{se:galaxy_selection}) and the introduction of a shear calibration factor 
\begin{equation}
  \label{eq:shear_calib}
 \langle \gamma_\alpha \rangle = c_\mathrm{cal} \langle e^\mathrm{iso}_\alpha \rangle \quad ,
\end{equation}
with $c_\mathrm{cal}=1/0.91$.
In a blind analysis of the second set of STEP image simulations (STEP2), which takes realistic ground-based PSFs and galaxy morphology into account, we find that the shear calibration of this improved method is on average accurate to the \mbox{$\sim 3$\%} level\footnote{Note that detected dependencies of the shear estimate on size and magnitude will lead to slightly different uncertainties for different surveys.}.
The method is capable to reduce the impact of the highly anisotropic ground-based PSFs which were analysed, to a level \mbox{$\lesssim7 \times 10^{-3}$} \citep{mhb07}.
We will also test this method on a third set of STEP simulations with realistic space-based PSFs (Rhodes et al. in prep.).
Depending on the results we will judge whether the systematic accuracy will be sufficient for the complete ACS Parallel Survey or if a different technique will be required for the final cosmic shear analysis.
For the GEMS Survey analysed in Sections \ref{se:preliminary_cosmic_shear} and \ref{se:cosmo_para_estimate}, a $\sim 3\%$ calibration error is well within the statistical noise.
In this work we use uniform weights \mbox{$w_i=1$} for all galaxies in order to keep the analysis as similar to our original STEP2 analysis as possible.
\section{Data}
\label{se:data}

The ACS Wide-Field-Channel detector (WFC) consists of two \mbox{$2\mathrm{k} \times 4\mathrm{k}$} CCD chips with a pixel scale of 0\farcs05 yielding a field-of-view (FOV) of \mbox{$\sim 3\farcm3 \times 3\farcm3$} \citep{fch03}.

In this pilot project we use pure parallel ACS/WFC F775W observations from HST proposal 9480 (PI J. Rhodes), denoted as the ``parallel data'' for the rest of this paper.

For comparison we also apply our data reduction and analysis to the combined F606W ACS/WFC observations of the GEMS field \citep{rbb04} and the GOODS observations of the \textit{Chandra} Deep Field South \citep[CDFS,][]{gfk04}.
A cosmic shear analysis of this \mbox{$\sim 28^\prime \times 28^\prime$} mosaic has already been presented by \citetalias{hbb05}, allowing us to compare the different correction schemes applied.

Both datasets were taken within the first operational year of ACS (August 2002 to March 2003 for the parallel data; July 2002 to February 2003 for the GEMS and GOODS observations).
Therefore these data enable us to test the feasibility of cosmic shear measurements with ACS at an early stage, when the charge-transfer-efficiency (CTE) has degraded only slightly \citep{rim04,rie04,mus05}.

\subsection{The ACS parallel data}
\label{se:acs_parallel}
The data analysed consist of 860 
WFC exposures, which we associate to fields by joining exposures dithered by less than 
a quarter of the field-of-view observed with the guiding mode \texttt{FINE\_LOCK}.
In order to permit cosmic ray rejection we only process associations containing at least two exposures.
Furthermore, in this pilot study we only combine exposures observed within one visit and with the same role-angle in order to achieve maximal stability of the observing conditions.
With these limitations, which are similar to those used by \citet{pce01} for the STIS Parallel Survey, we identify  208
associated fields (including re-observations of the same field at different epochs), combining 835
exposures.

For a weak lensing analysis, accurate guiding stability is desired in order to minimise variations of the PSF.
In case of parallel observations, differential velocity aberration between the primary and secondary instrument can lead to additional drifts during observations with the secondary instrument \citep{cox97}.
In order to verify the guiding stability for each exposure we determine the size of the telemetry jitter-ball, which describes the deviation of the pointing from the nominal position.
While the jitter-balls typically have shapes of moderately elliptical \mbox{($\langle b/a \rangle = 0.68$)} distributions with \mbox{$\langle \mathrm{FHWM} \rangle = 9.8 \,\mathrm{mas}$} (0.196 WFC pixel), we have verified that \mbox{$\mathrm{FHWM} < 20 \,\mathrm{mas}$} (0.4 WFC pixel) and \mbox{$b/a > 0.4$} for all selected exposures. 
Therefore the tracking accuracy is sufficiently good and expected to affect the image PSF only slightly. 
Any residual impact on the PSF will be compensated by our PSF correction scheme, which explicitly allows for an additional ellipticity contribution due to jitter (see Sect.\thinspace\ref{se:psf_correction_templates}).

\subsubsection{Data reduction}
The data retrieved was bias and flat-field corrected.
We use \texttt{MultiDrizzle}\footnote{MultiDrizzle 2.7.0, \url{http://stsdas.stsci.edu/multidrizzle/}} \citep{kfh02} for the rejection of cosmic rays, the correction for geometric camera distortions and differential velocity aberration \citep{cog02}, and the co-addition of the exposures of one association.

We refine relative shifts and rotations of the exposures by applying the IRAF task \texttt{geomap} to matched windowed \texttt{SExtractor} \citep{bea96} positions of compact sources detected in separately drizzled frames.
For the star and galaxy fields selected for the analysis (see Sect. \ref{se:field_select}) the median shift refinement relative to the first exposure of an association is $0.17$ WFC pixels, with $7.3\%$ of the exposures requiring shifts larger than $0.5$ WFC pixels.
Refinements of rotations were in most cases negligible with a median of $1.6\times 10^{-4}$ degrees corresponding to a displacement of $\simeq 0.008$ WFC pixels near the edges of the FOV.
Only in $1.5\%$ of the exposures rotation refinements exceeded $3\times 10^{-3}$ degrees, corresponding to displacements of $\simeq 0.15$ WFC pixels.

Deviating from the default parameters, we 
use the \texttt{minmed} algorithm \citep{pkm06} during the creation of the median image as it is more efficient to reject cosmic rays for a low number of co-added exposures.
For the cosmic ray masks we let rejected regions grow by one pixel in each direction (\texttt{driz\_cr\_grow}=3) in order to improve the rejection of neighbouring pixels affected due to charge diffusion and pixels with cosmic ray co-incidences in different exposures.

Furthermore we use a finer pixel scale of 0\farcs03 per pixel in combination with the \texttt{SQUARE} kernel for the final drizzle procedure
in order to increase the resolution in the co-added image and reduce 
the impact of aliasing.
For the default pixel scale (0\farcs05 per pixel) resampling adds a strong artifical 
noise component to the shapes of un- and poorly resolved objects (aliasing), which depends on the position of the object centre relative to the pixel grid and most strongly affects the $e_1$-ellipticity component.
According to our testing with stellar field images, the \texttt{GAUSSIAN} kernel leads to even lower shape noise caused by aliasing.
However, as it leads to stronger noise correlations between neighbouring pixels, we decided to use the \texttt{SQUARE} kernel for the analysis.

Aliasing most strongly affects unresolved stars, which is critical if one aims to derive PSF models from a low number of stars in drizzled frames.
However, since we determine our PSF model from undrizzled images (see Sect.\thinspace\ref{se:psf_correction_templates}), this does not affect our analysis.
Consistent with the results from \citet{rma05,rma07} we find that a further reduction of the pixel scale does not further reduce the impact of aliasing significantly, while unnecessarily increasing the image file size.

In this paper the term \textit{pixel} refers to the scale of the drizzled images (0\farcs03 per pixel) unless we explicitly allude to \textit{WFC pixels}.

\subsubsection{Field selection}
\label{se:field_select}
\begin{figure*}
   \centering
   \includegraphics[height=4.4cm]{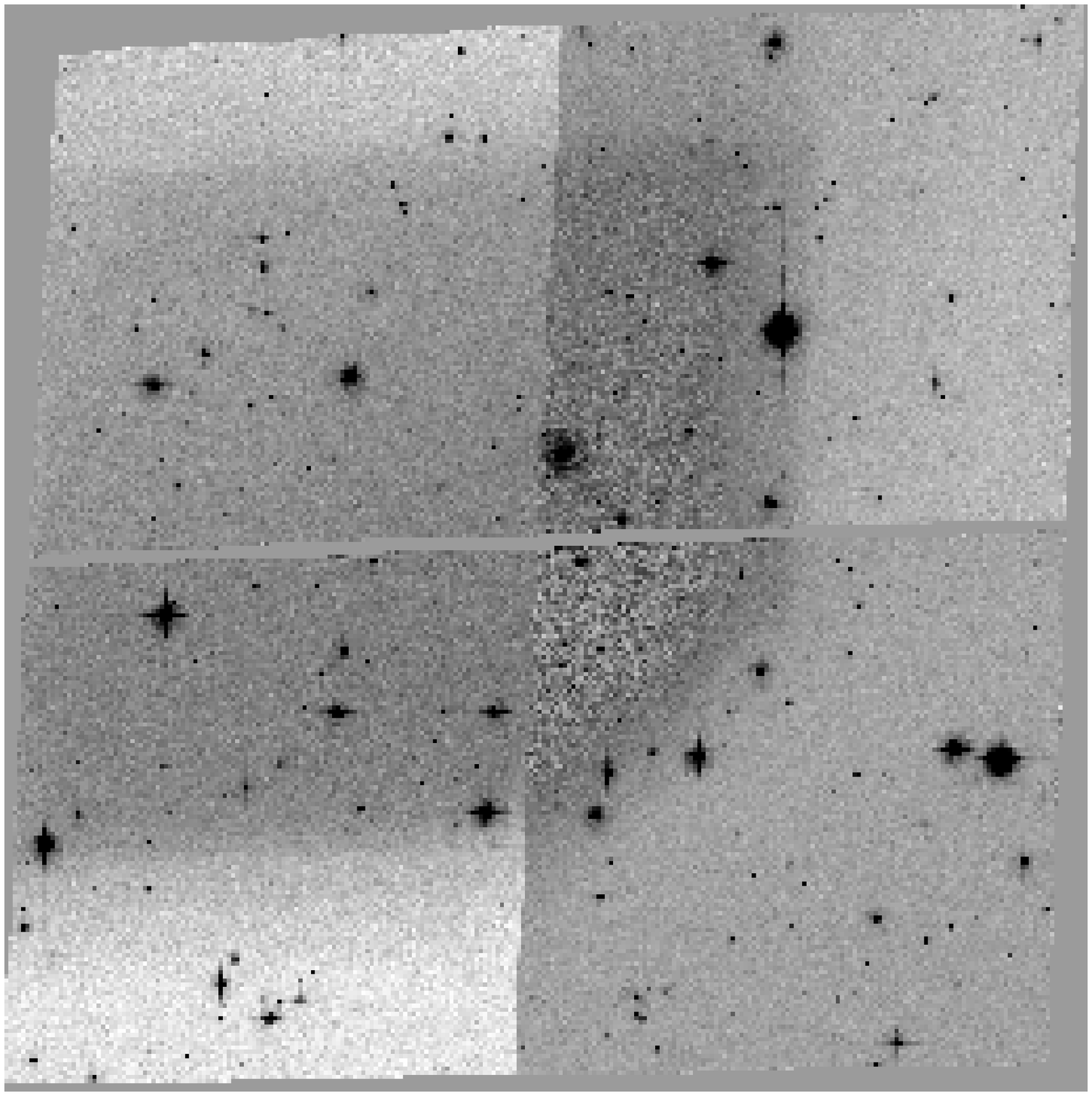}
   \includegraphics[height=4.4cm]{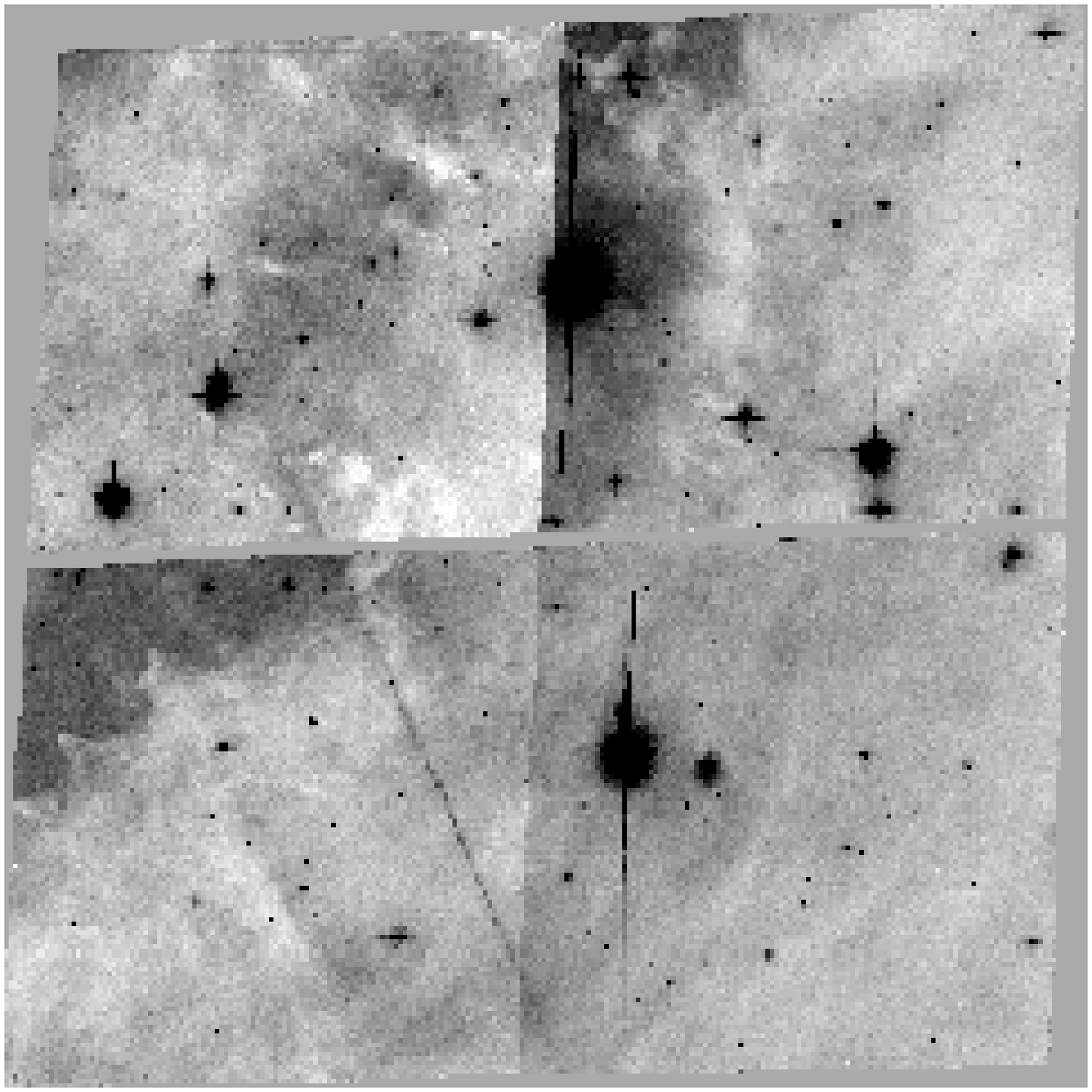}
   \includegraphics[height=4.4cm]{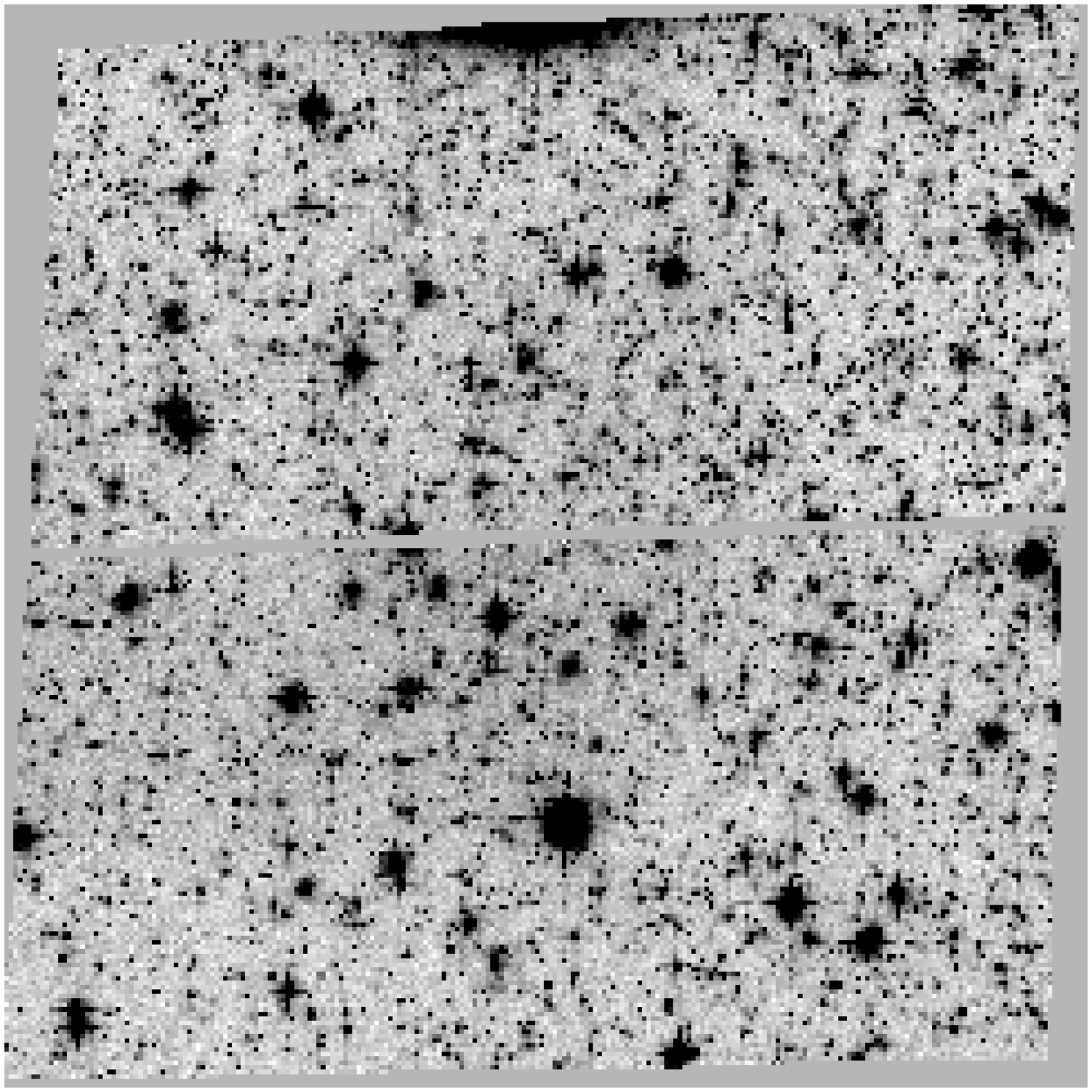}
   \includegraphics[height=4.4cm]{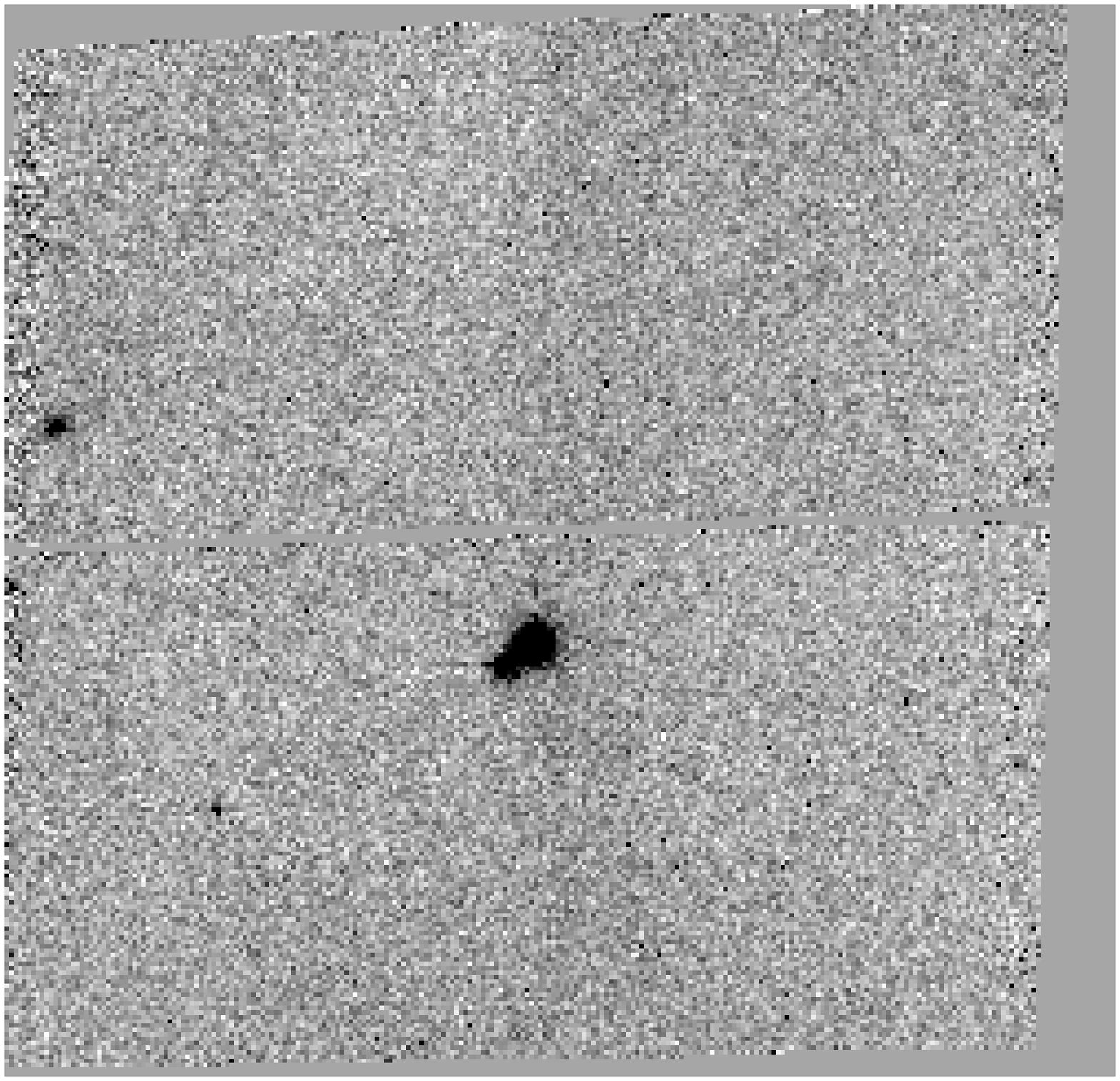}
   \caption{Examples for fields which were rejected by visual inspection for the following reasons (panels left to right): erroneous calibration, galactic nebula, many saturated stars, almost empty field.}
   \label{fi:reject_visually}
    \end{figure*}

The 
208 associations were all visually inspected.
We discard in total 
31
 fields for the following reasons:
\begin{itemize}
  \item Fields which show a strong variation of the background in the pre-processed exposures 
(4 fields).
  \item Fields containing galactic nebulae (10 fields).
   \item Fields of significantly poorer image quality (6 fields).
   \item Fields which contain a high number of saturated stars with extended diffraction spikes (5 fields).
   \item Fields in M31 and M33 with a very high number density of stars, resulting in a strong crowding of the field, which makes them even unsuitable for star fields (4 fields).
   \item Almost empty galactic fields affected by strong extinction (2 fields).
\end{itemize}
Examples of the discarded fields are shown in Fig.\thinspace\ref{fi:reject_visually}.

After this pre-selection, fields fulfilling the following criteria are selected as galaxy fields for the cosmic shear analysis: 
\begin{itemize}
  \item Fields have to be located at galactic latitudes $|b| \ge 25^\circ$ in order to be affected only weakly by galactic extinction. 
  \item Only fields co-added from at least three individual exposures are used, facilitating sufficiently good cosmic ray rejection. 
   \item Fields are required not to be dominated by a single object or stars resolved in a local group galaxy.
   \item In the case of re-observations of the same field at different visits, the observation with the longest exposure time is used.
\end{itemize}
55 independent fields fulfil these selection criteria. Additionally four fields with \mbox{$20^\circ < |b| < 25^\circ$} are included, which contain a high number density of galaxies indicating rather low extinction, making a total of 59 galaxy fields.
This corresponds to 28.4\% of the fields and 36.2\% of the co-added exposures.

All fields passing the preselection and containing at least 300 stars are used as star fields for the PSF analysis (see Sect.\thinspace\ref{se:psf}).
These 61 fields consisting of 205 exposures amount to 29.3\% of the fields and 24.5\% of the co-added exposures.

\subsection{The GEMS+GOODS data}

\begin{figure*}[htb]
   \centering
    \sidecaption
   \includegraphics[width=5.95cm]{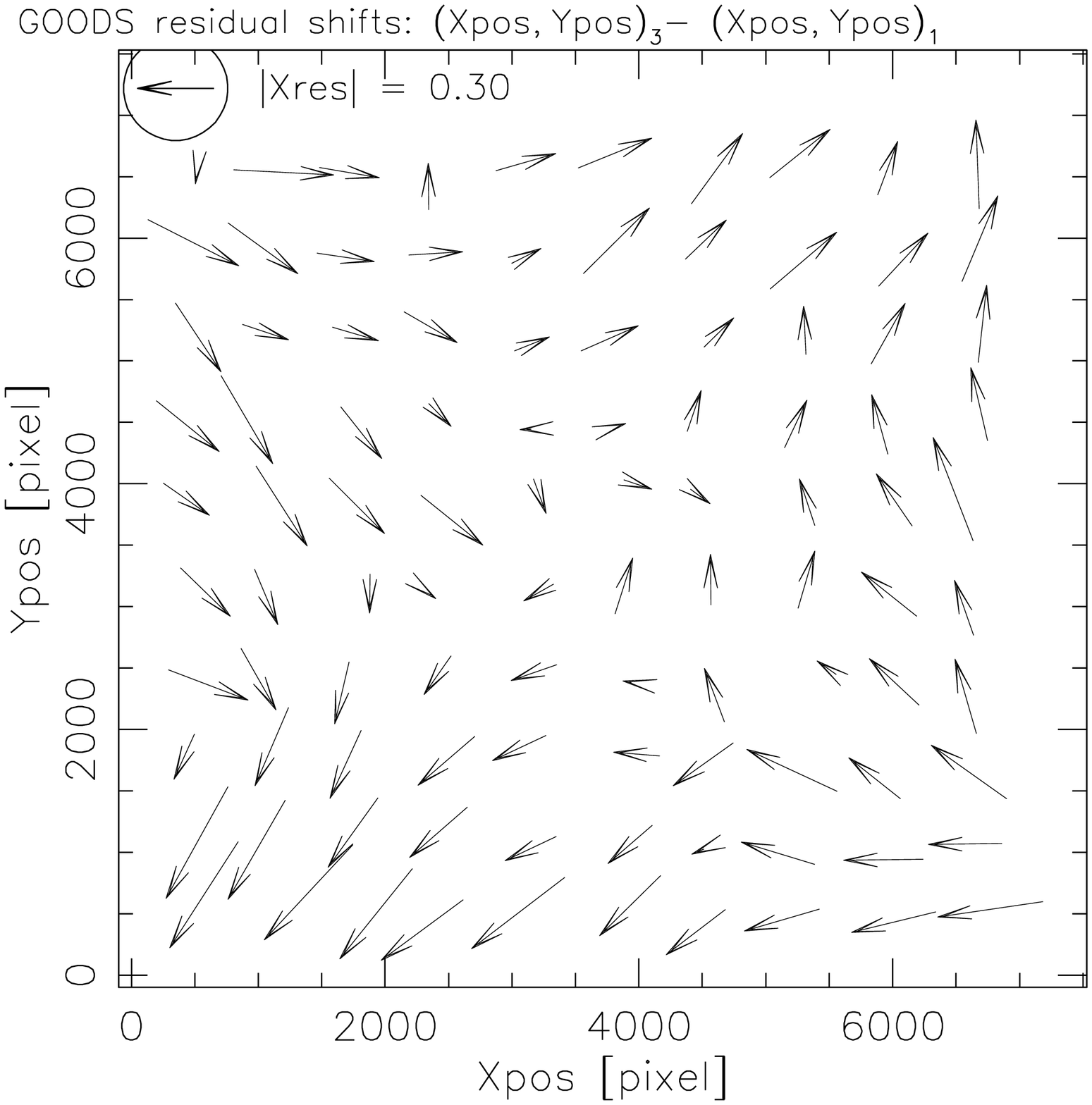}
   \includegraphics[width=5.95cm]{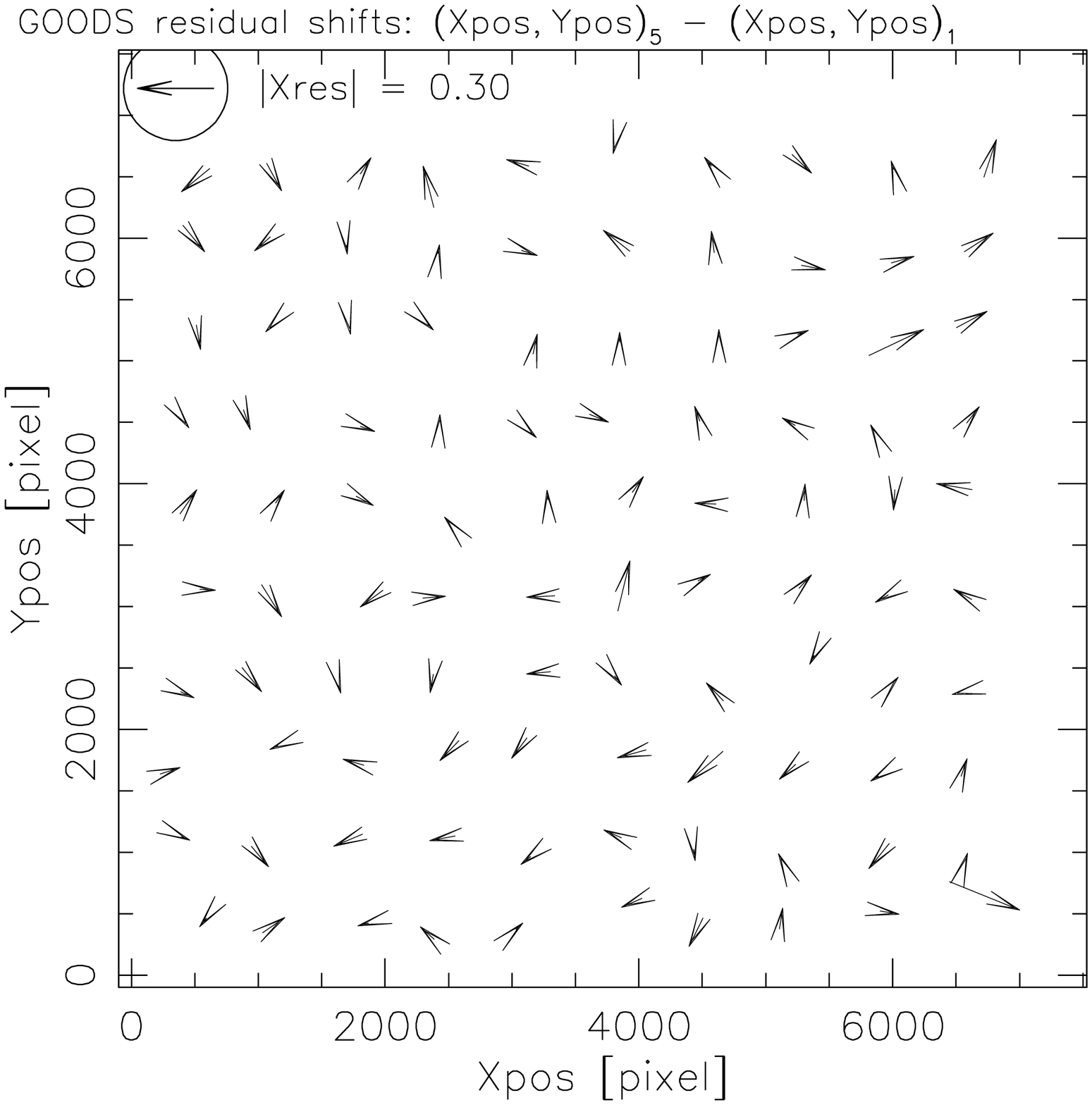}
   \caption{Residual shifts [pixel] computed from windowed SExtractor positions of compact sources between epochs 3 and 1 (left) and between epoch 5 and 1 (right) of the F606W ACS GOODS/CDFS observations. For these plots compact objects from all 15 tiles are used, and residual shifts are averaged in bins of $700^2$ pixels. For each tile the exposures of each epoch were drizzled onto one output pixel grid, with a common WCS per tile defined by epoch 1.
Possible interpretations for the residual shifts in the left panel are slight temporal changes in the ACS geometric distortion or a slightly imperfect treatment of the distortion correction in the \texttt{MultiDrizzle} version used.
} 
   \label{fi:residual_shifts_GOODS}
    \end{figure*}

The GEMS F606W data consist of 63 ACS/WFC tiles imaged with three exposures of 720 to 762 seconds each.
They are arranged around the ACS GOODS/CDFS observations, which have been imaged in five epochs with different position angles (see Tab.\thinspace\ref{ta:goodsobs}) consisting of two exposures per tile and epoch with 480 to 520 seconds per exposure.
\begin{table}[tb]
\begin{center}
\caption{Observation dates and position angles (\texttt{ORIENTAT}) of the ACS/WFC F606W GOODS/CDFS observations.
}
\begin{tabular} {|c|c|c|}
\hline
Epoch & Observation dates & Position angle\\% & Comment \\
\hline
1 & 2002-07-31--2002-08-04 & $-112^\circ$\\
2 & 2002-09-19--2002-09-22 & $-67^\circ$\\
3 & 2002-10-31--2002-11-03 & $-22^\circ$\\
4 & 2002-12-19--2002-12-22 & $23^\circ$\\
5 & 2003-02-01--2003-02-05 & $68^\circ$\\
\hline
\end{tabular}
\label{ta:goodsobs}
\end{center}
\end{table}
In total the ACS GOODS/CDFS field is covered with 15 tiles during epochs 1, 3, and 5, whereas 16 tiles were used for epochs 2 and 4.
\citetalias{hbb05} limit their analysis to the epoch 1 data.
In order to reach a similar depth for the used GOODS and GEMS data we decided to combine the data of epoch 1 with either epoch 3 or 5 as they have an optimal overlap.
The combination of epoch 1 and epoch 5 exposures is unproblematic.
In contrast we find significant, FOV dependent residual shifts between matched object positions in exposures from epochs 1 and 3 after applying refined image shifts and rotations (Fig.\thinspace\ref{fi:residual_shifts_GOODS}).
Possible interpretations for these remaining shifts are slight medium-term temporal changes in the ACS geometric distortion or a slightly imperfect treatment of the distortion correction in the \texttt{MultiDrizzle} version used.
\citet{psb05} report similar effects for two epochs of \textit{Hubble} Ultra Deep Field (UDF) data.
As remaining shifts also occur for UDF images observed with position angles that are $\sim 90^\circ$ apart, the \texttt{MultiDrizzle} interpretation might be more plausible.
The largest residual shifts have a comparable magnitude of $\sim 0.5$ pixels for both the GOODS and UDF data.
A combination of exposures with remaining shifts would in any case degrade the PSF of the combined image.
Additionally central pixels of some stellar images could falsely be flagged as cosmic rays by \texttt{MultiDrizzle}. 
Therefore we only use the combined epoch 1 and 5 exposures for the cosmic shear analysis that follows.

In order to investigate the ACS F606W PSF, we additionally analysed 184 archival F606W exposures of dense stellar fields containing at least 300 stars, which were observed between July 2002 and July 2003. 

\subsection{Catalogue creation}
\label{se:catalog_creation}
We use \texttt{SExtractor} \citep{bea96} for the detection of objects and the \citet{ewb01} implementation of the KSB formalism for shape measurements.
We analyse the images of galaxies in the combined drizzled images.
However, for the time-dependent PSF correction described in Sect.\thinspace\ref{se:psf_correction_templates}, we additionally perform stellar shape measurements in the undrizzled but cosmic ray-cleaned \textit{COR}-images, which are also created by \texttt{MultiDrizzle}, and the drizzled uncombined frames (\textit{DRZ}-images).

The \texttt{SExtractor} object detection and deblending parameters are summarised in Table\thinspace\ref{te:sexpara}.
We use a rather low detection threshold for the galaxies in order to minimise the impact of PSF-based selection bias \citep{kai00,bej02}.
Spurious detections are later rejected with cuts in the signal-to-noise ratio.
We find that the deblending parameters applied perform well except for 
the case of spiral galaxies extended by several arcseconds, for which substructure components are in some cases detected as separate objects.
Thus, we mask these galaxies manually.
If more than one object is detected within 1\farcs2,
only the brighter component is kept.
We furthermore reject galaxies containing pixels with low values in the \texttt{MultiDrizzle} weight image (\mbox{$w_\mathrm{min}=100 \,\mathrm{s}$})\footnote{The weight image pixel value corresponds to the effective exposure time contributing for the pixel, scaled with the relative area of output and input pixels.} within their \texttt{SExtractor} isophotal area and also semi-automatically create masks to reject bright stars with diffraction spikes and extended image artifacts like ghost-images.

We use different detection parameters for the star fields (see Table\thinspace\ref{te:sexpara}).
Due to the increased detection threshold \texttt{DETECT\_THRESH}, the object detection becomes less sensitive to the faint and extended stellar diffraction spikes, reducing the time needed for masking.

\begin{table}[tb]
\begin{center}
\caption{Relevant parameters for the object detection with \texttt{SExtractor} for the galaxy fields and the star fields. 
Note that the number of pixels for a detection \texttt{DETECT\_MINAREA} corresponds to the subpixel of the drizzled images except for the values in brackets, which are used for object detection in the undrizzled \textit{COR}-images.
}
\begin{tabular} {|l|c|c|}
\hline
Parameter & Galaxy fields & Star fields \\% & Comment \\
\hline
\texttt{BACK\_TYPE} & AUTO & MANUAL \\% & type of background determination \\
\texttt{BACK\_SIZE} & 100 & -- \\%& Size of background mesh \\
\texttt{BACK\_FILTERSIZE} & 5 & -- \\%& Size of background filter \\
\texttt{BACK\_VALUE} & -- & 0.0  \\%& Fixed background value \\
\texttt{DETECT\_MINAREA} & 16 & 16 (5) \\%& min. pixel number above threshold \\   
\texttt{DETECT\_THRESH} & 1.5 & 3 (4)\\%& sigmas over background \\    
\texttt{DEBLEND\_NTHRESH} & 16 & 32  \\%& Deblending sub--thresholds \\
\texttt{DEBLEND\_MINCONT} & 0.05 & 0.1  \\%& Minimum contrast for deblending \\
\hline
\multicolumn{1}{|l|}{\texttt{FILTER\_NAME}} & \multicolumn{2}{|c|}{gauss\_2.5\_5x5 (gauss\_2.0\_3x3)}  \\
\hline
\end{tabular}
\label{te:sexpara}
\end{center}
\end{table}

We use the \texttt{SExtractor} \texttt{FLUX\_RADIUS} parameter as Gaussian filter scale $r_\mathrm{g}$ for the shape measurements of the galaxies.
Here the integration is carried out to a radius of $3 r_\mathrm{g}$ from the 
centroid.
This truncation was introduced to speed up the algorithm and is justified due to the strong down-weighting of the outer regions in KSB. We also verified from the data that it does not bias the shape measurement.
For the stellar \textit{DRZ}-images we repeat the shape measurements for 18 different filter scales ranging from 2.0 to 15 pixels, which are later matched to the filter scales of the galaxies. 
For larger filter scales we find that it is essential to continue the integration out to sufficiently large radii due to the wide diffraction wings of the PSF.
Therefore, we employ a stellar integration limit of $4.5\times \texttt{FLUX\_RADIUS}^* \simeq 9$ pixels.
For the stellar shape measurements in the \textit{COR}-images we use a fixed Gaussian filter scale $r_\mathrm{g}=1.5$ WFC pixels, which according to our testing roughly maximises the signal-to-noise of the stellar ellipticity measurement for most of the occurring PSF anisotropy patterns (see Sect.\thinspace\ref{se:psf}).

For object selection we use the signal-to-noise definition from \citet{ewb01}
\begin{equation}
\label{eq:snratio}
\mathrm{S}/\mathrm{N}=\frac{ \int \mathrm{d}^2 \theta \, W_{r_\mathrm{g}}(|\boldsymbol{\theta}|) \, I(\boldsymbol{\theta})}{ \sigma_\mathrm{sky}\sqrt{\int \mathrm{d}^2 \theta \, W_{r_\mathrm{g}}^2(|\boldsymbol{\theta}|)}} \, ,
\end{equation}
which is based on the same filter function as the one used for shape measurements.
In the computation of S/N we do not take the correlation of noise in adjacent pixels into account, which is created by drizzling.
However, a correction for the noise in a large area (e.g. the extent of a galaxy) can be assessed from Eq.\thinspace A19 in \citet{cmd00}, which for our drizzle parameters yields
\begin{equation}
\frac{\sigma_1}{\sigma_2}=m \left[ \frac{25}{12} \left( 1 - \frac{5}{9m} \right) \right] \, ,
\end{equation}
where $\sigma_2=\sigma_\mathrm{sky}$ is the single pixel background dispersion, while $\sigma_1$ denotes the dispersion computed from areas of size $m^2$ (drizzled) pixels.
The expression in squared brackets gives the correction factor to the area scaling expected for uncorrelated noise. 
Using the effective area of the Gaussian weight function $A=2 \pi r_g^2$ and $m=\sqrt{A}$ we estimate a noise correction factor which increases from 1.86 for unresolved sources ($r_g\simeq 2.1$ pixels) to 2.05 for the largest galaxies considered.
Hence, the true S/N will be lower than the directly computed value by this factor.
The cuts applied to the data refer to the directly computed value.

\section{PSF analysis and correction}
\label{se:psf}
Due to the low number of stars \mbox{($\sim 10-30$)} present in galaxy fields at high galactic latitudes we examined the ACS PSF from stellar fields (see Section \ref{se:field_select}) containing \mbox{$\sim300-20000$} stars.
We do this analysis on the basis of single exposures instead of combined images, in order to optimally investigate possible temporal PSF variations. 
We investigate the PSF both in the undrizzled, but cosmic ray cleansed \textit{COR}-images created by \texttt{MultiDrizzle}, and also the drizzled and cosmic ray cleansed single exposures (\textit{DRZ}).
Here we limit the discussion to the F775W data.
Our analysis of the F606W PSF was performed in an identical fashion with only minor differences in the resulting PSF models.
A detailed KSB+ analysis of the F606W PSF can be found in \citetalias{hbb05}.

\subsection{Star selection}
In the \textit{DRZ}-images (\textit{COR}-images) we select stars with 0.6 pixel (0.45 WFC pixel) wide cuts in half-light radius $r_\mathrm{h}$ \citep{ewb01} and cuts in the signal-to-noise ratio \mbox{$\mathrm{S}/\mathrm{N} > 40$} (\mbox{$\mathrm{S}/\mathrm{N} > 30$}).
We furthermore reject stars with saturated pixels using magnitude cuts, and, 
in the case of crowed fields, stars with a neighbour closer than 20 (10) pixels, 
which would otherwise affect the shape measurements for large $r_\mathrm{g}$.

\subsection{PSF anisotropy variation}
\label{se:psf_anis}
   \begin{figure*}
   \centering
   \includegraphics[width=5.8cm]{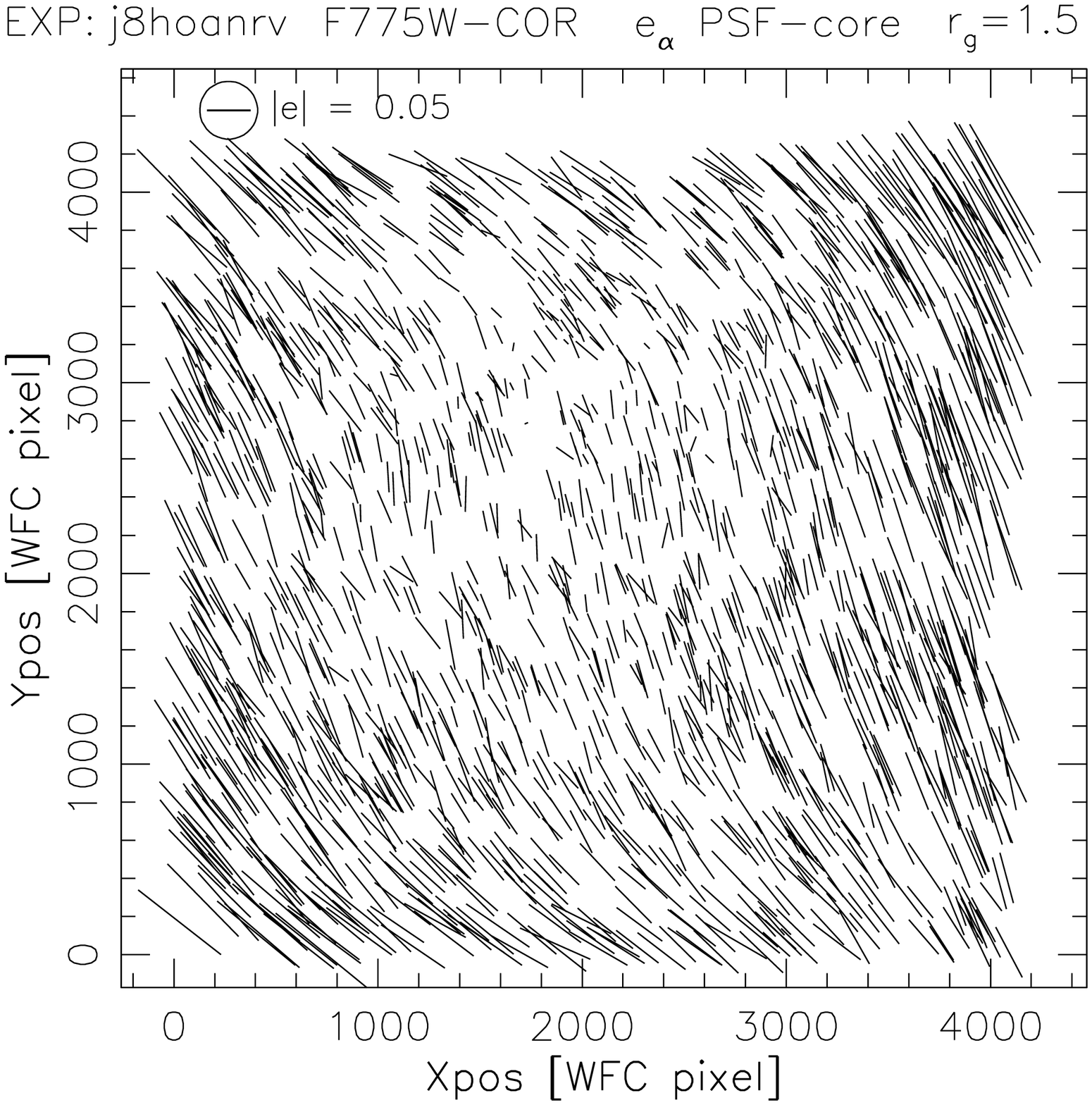}
   \includegraphics[width=5.8cm]{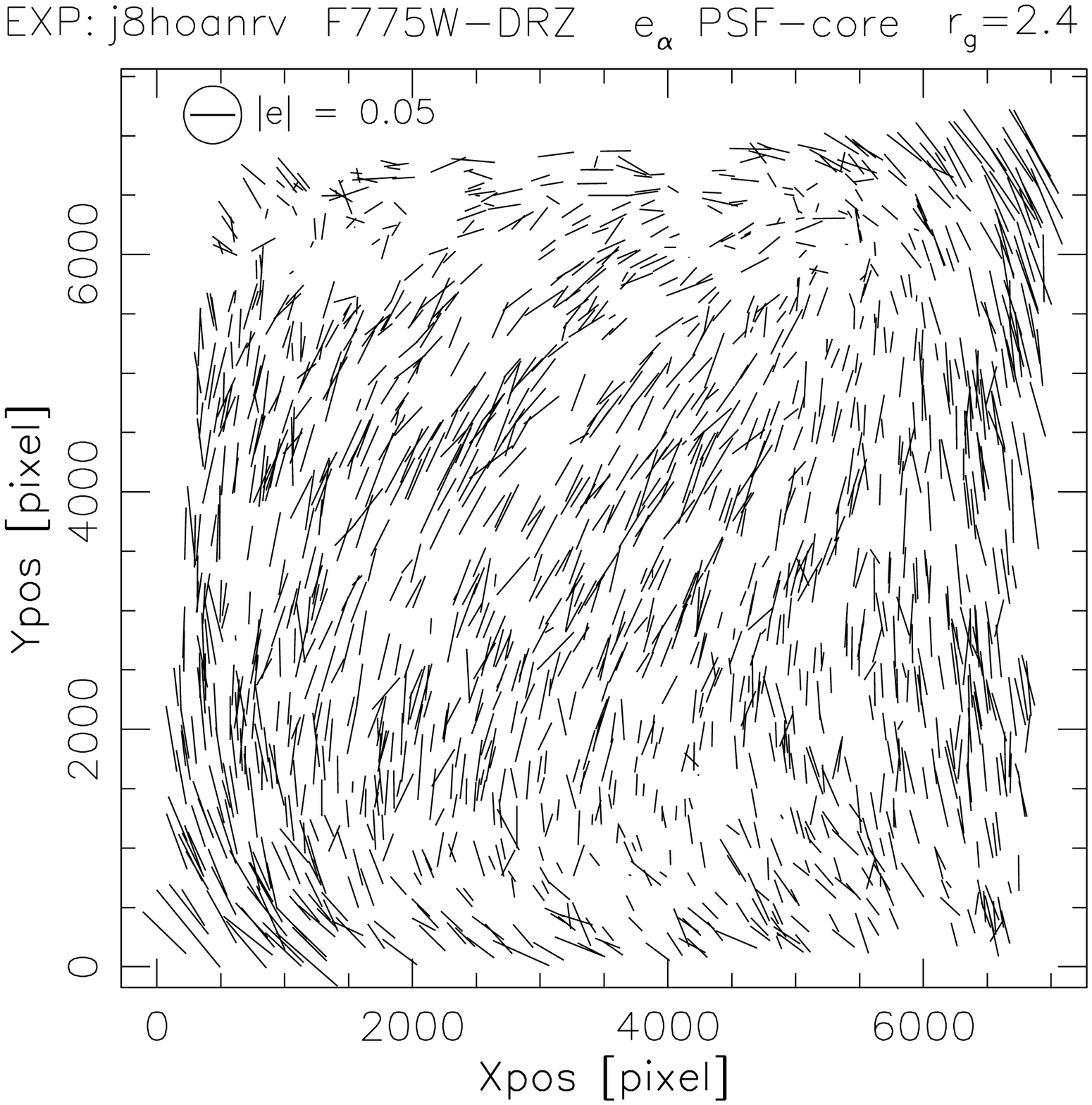}
   \includegraphics[width=5.8cm]{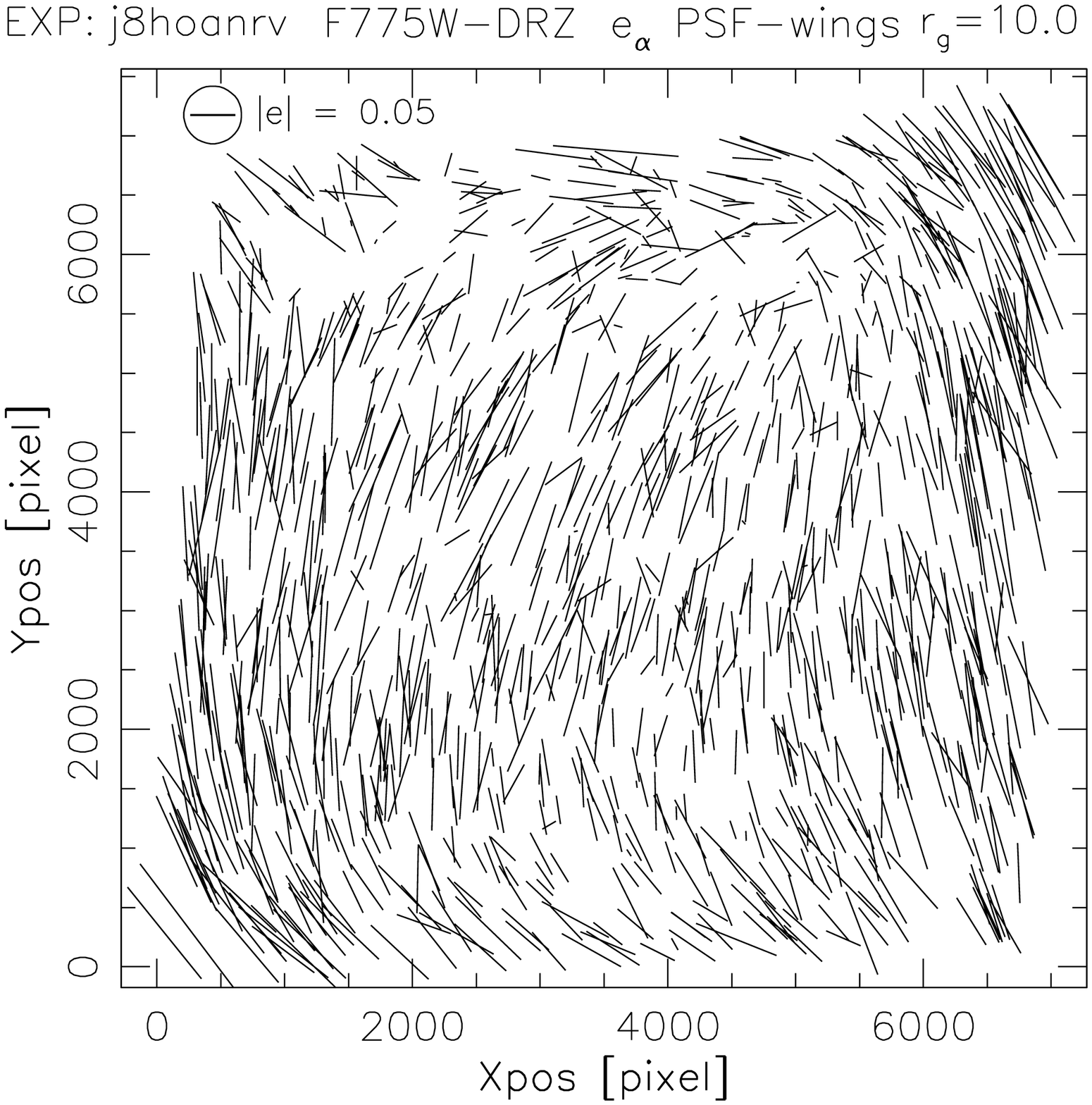}
 
   \caption{Stellar ``whisker plots'' for an example F775W stellar field exposure. Each whisker represents a stellar ellipticity. The left panel shows $e^*_\alpha$ in the undrizzled \textit{COR}-image measured with \mbox{$r_\mathrm{g}=1.5$} WFC pixels (PSF core).
The middle and right panels correspond to the drizzled \textit{DRZ}-image showing the PSF core (\mbox{$r_\mathrm{g}=2.4$} pixels, middle) and the PSF wings (\mbox{$r_\mathrm{g}=10.0$} pixels, right).
The fit to the ellipticities in the middle panel is shown in the lower right panel of Fig.\thinspace\ref{fi:psf:short_term_variation}.
}
   \label{fi:psf:whisker}
    \end{figure*}
   \begin{figure*}
   \centering
    \sidecaption
   \includegraphics[width=6.83cm]{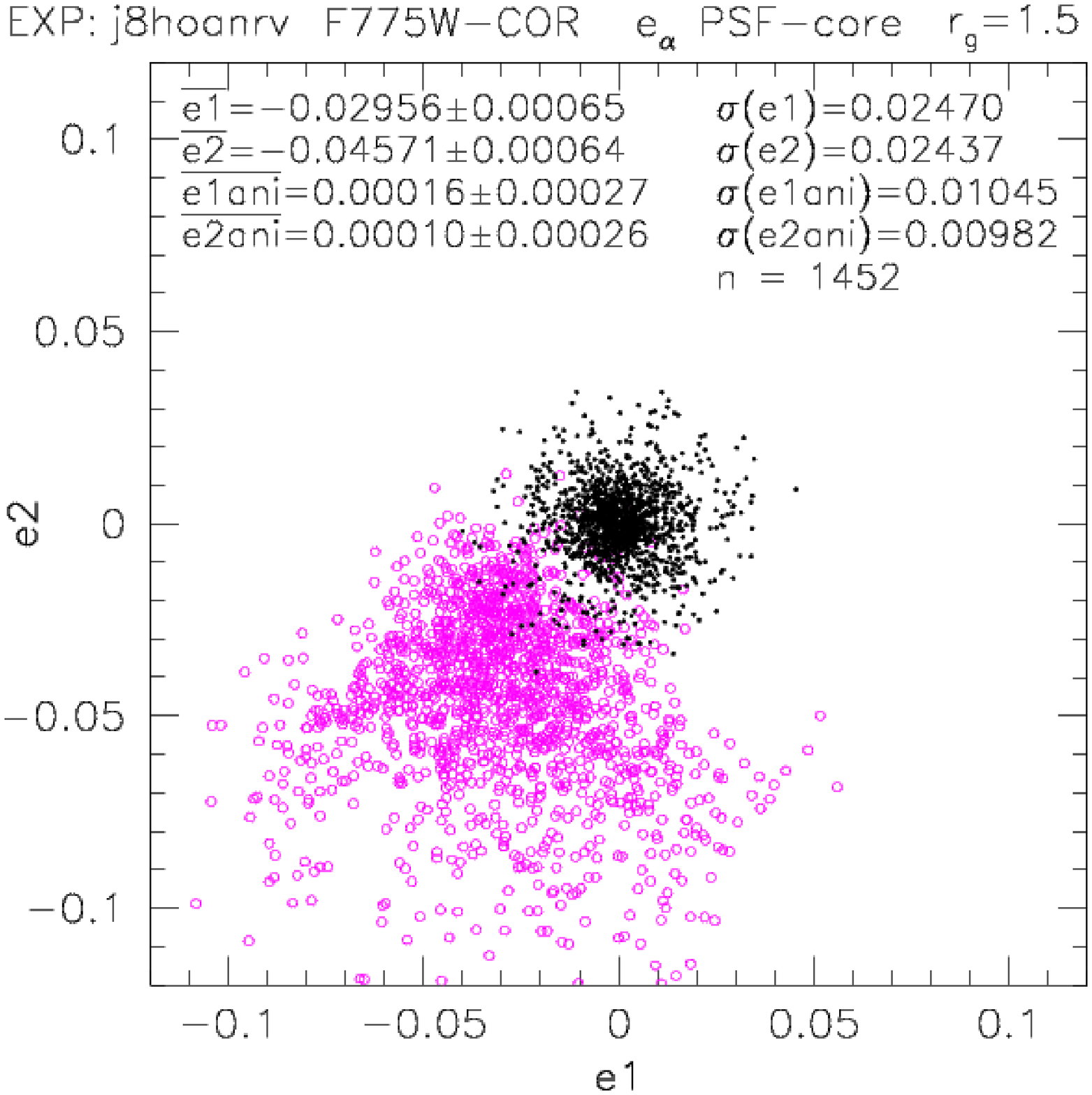}
   \includegraphics[width=6.83cm]{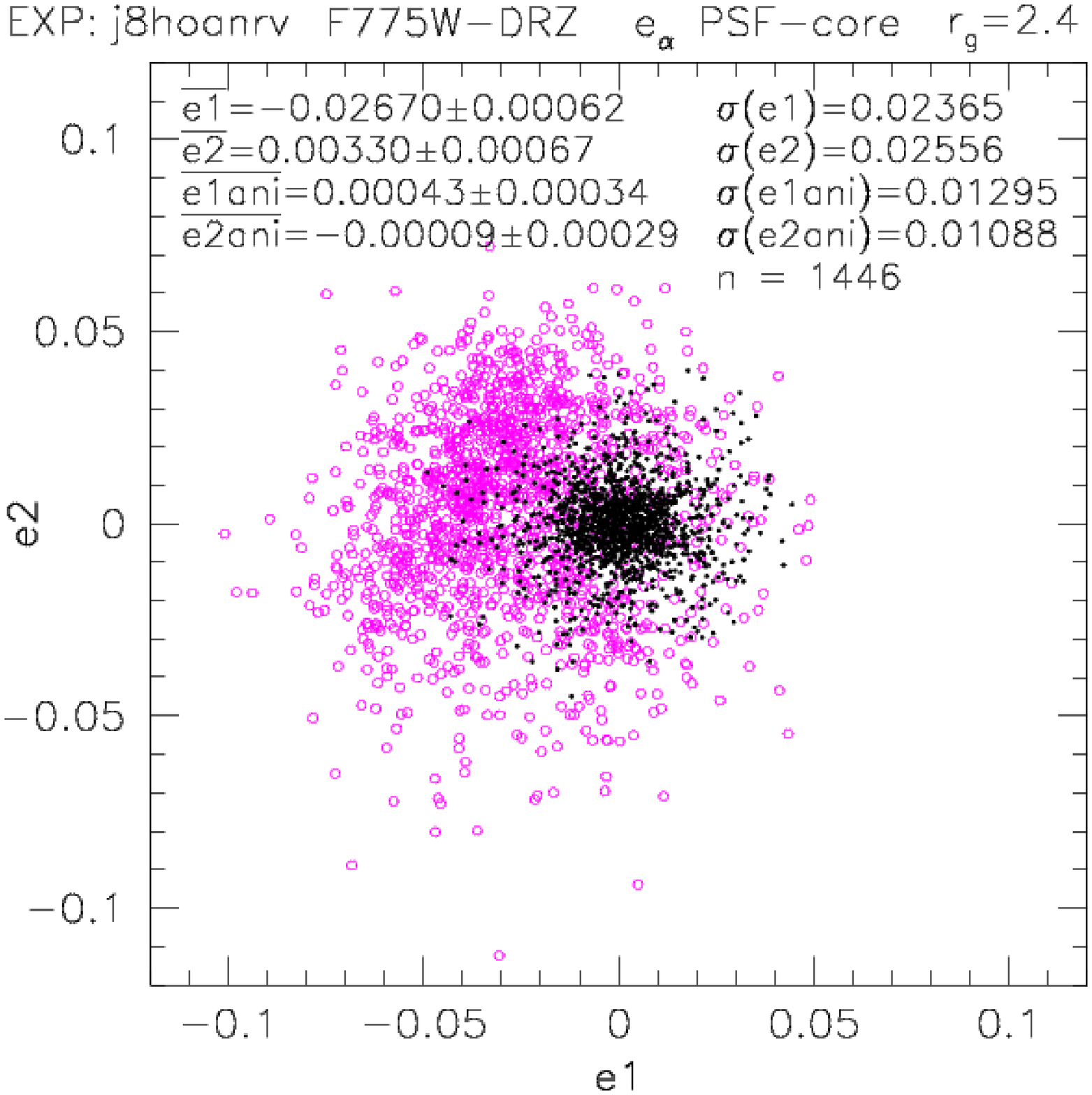}
   \caption{Stellar ellipticity distribution (PSF core) for an example F775W stellar field exposure %\textit{j8hoanrv} 
measured in the undrizzled \textit{COR}-image (left) and the drizzled \textit{DRZ}-image (right).
The open circles represent the uncorrected ellipticities $e^*_\alpha$, whereas the black points show the ellipticities $e^{\mathrm{ani}*}_\alpha$ corrected with a third-order polynomial for each chip. 
In the right panel $\sigma(e^{\mathrm{ani}*}_1)$ is significantly increased, which is a result of the
re-sampling in the drizzle algorithm.
In these plots outliers have been rejected at the $3\sigma$ level.
}
   \label{fi:psf:e1_e2}
    \end{figure*}

Investigating stellar fields we find that the stellar ellipticity $e^*_\alpha$ and anisotropy kernel $q^*_\alpha$ vary smoothly across each WFC chip and can well be fit with third-order polynomials.
Fig.\thinspace\ref{fi:psf:whisker} shows the FOV variation of $e^*_\alpha$ for a 400 second stellar field exposure both for the undrizzled \textit{COR}-image (left panel) and the drizzled and thus distortion corrected \textit{DRZ}-image, where the middle panel corresponds to the PSF core measured with \mbox{$r_\mathrm{g}=2.4$} pixels, whereas the right panel shows the PSF wings (\mbox{$r_\mathrm{g}=10.0$} pixels).
The observed differences between the PSF core and wings, which mainly constitute in a stronger ellipticity for larger $r_\mathrm{g}$, underline the importance to measure stellar quantities as a function of filter scale $r_\mathrm{g}$ (see also \citealt{hfk98}; \citetalias{hbb05}).

In Fig.\thinspace\ref{fi:psf:e1_e2} we compare the stellar ellipticity distribution in the \textit{COR}-image and the \textit{DRZ}-image for similar Gaussian filter scales of $r_\mathrm{g}=1.5$ WFC pixels and $r_\mathrm{g}=2.4$ pixels, both uncorrected and after the subtraction of a third-order polynomial model for each chip.
Here drizzling with the \texttt{SQUARE} kernel increases the corrected ellipticity dispersion $\sigma(e^{\mathrm{ani}*}_1)$ by $\simeq 24\%$ and thus decreases the accuracy of the ellipticity estimate.
For the galaxy fields we therefore determine the PSF model from the undrizzled \textit{COR}-images (see Sect.\thinspace\ref{se:psf_correction_templates}).
Note that the stellar ellipticities in the \textit{COR}-images (left panels in Fig.\thinspace\ref{fi:psf:whisker} and \ref{fi:psf:e1_e2}) are created by the combined image PSF and geometric camera distortion, whereas the \textit{DRZ}-image ellipticities correspond to pure image PSF. 
However, since the resulting pattern can in both cases well be fit with third-order polynomials, the corrected ellipticity dispersions are directly comparable. 

Note that we always plot the FOV variation in terms of $e^*_\alpha$ in order to simplify the comparison to other publications.
However, for the actual correction scheme we employ fits of $q^*_\alpha$ defined in (\ref{eq:qalpha}) due to a slight 
PSF width variation leading to a
 variation of $P^{\mathrm{sm}*}_{\alpha \beta}$ (see Sect.\thinspace\ref{se:psf_sizevariation}).

\begin{figure*}
   \centering
   \includegraphics[width=5.8cm]{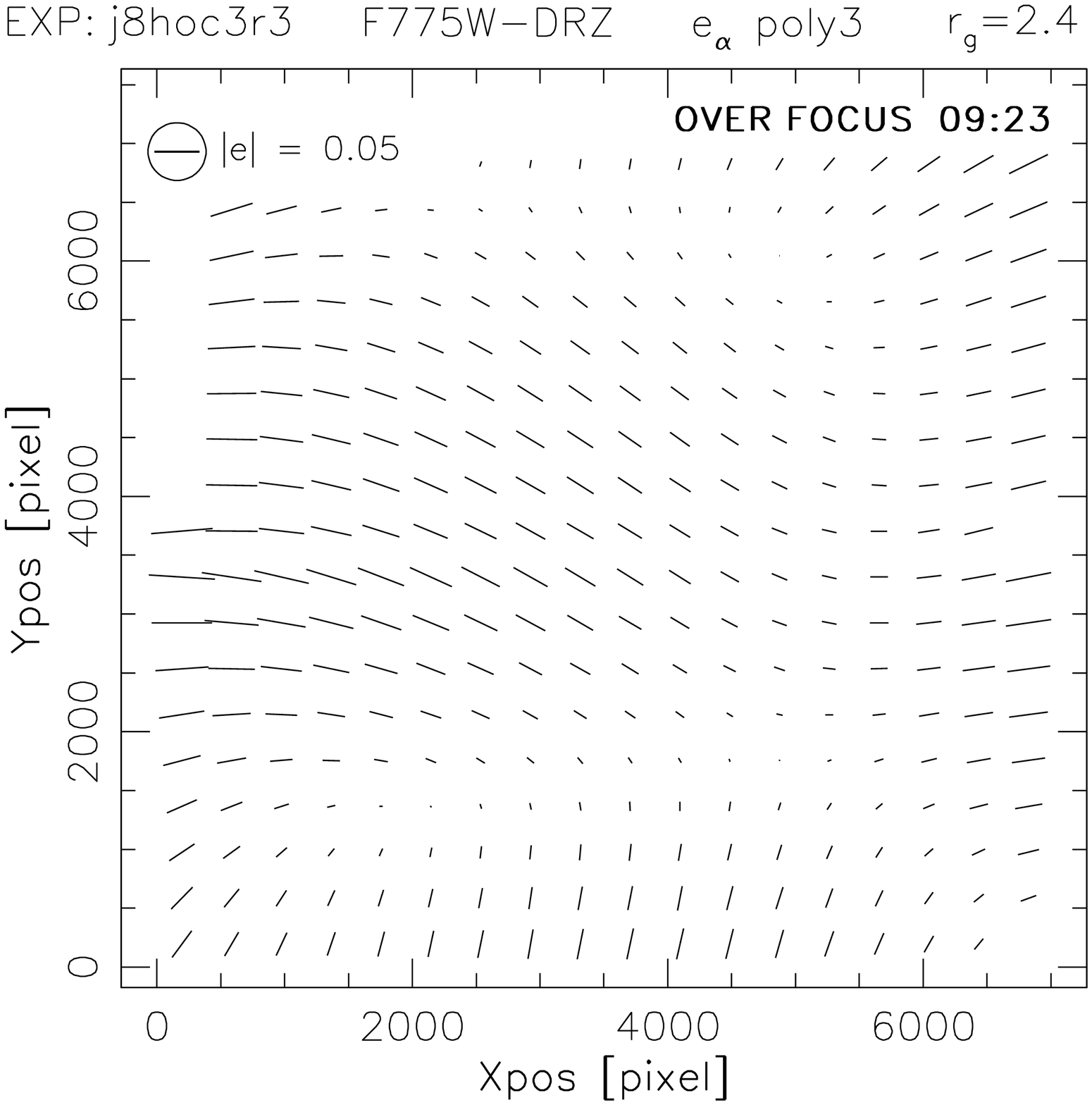}
   \includegraphics[width=5.8cm]{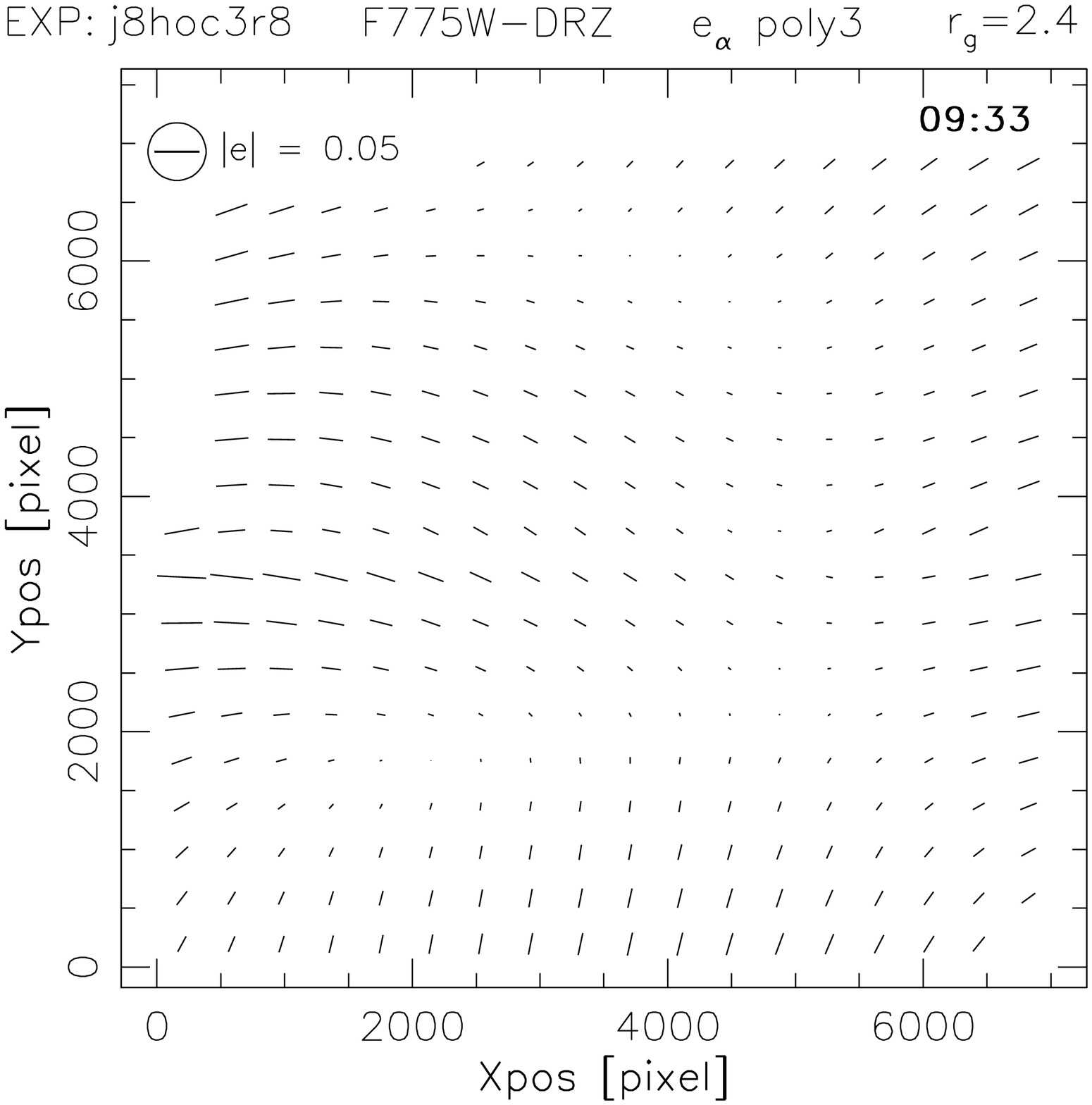}
   \includegraphics[width=5.8cm]{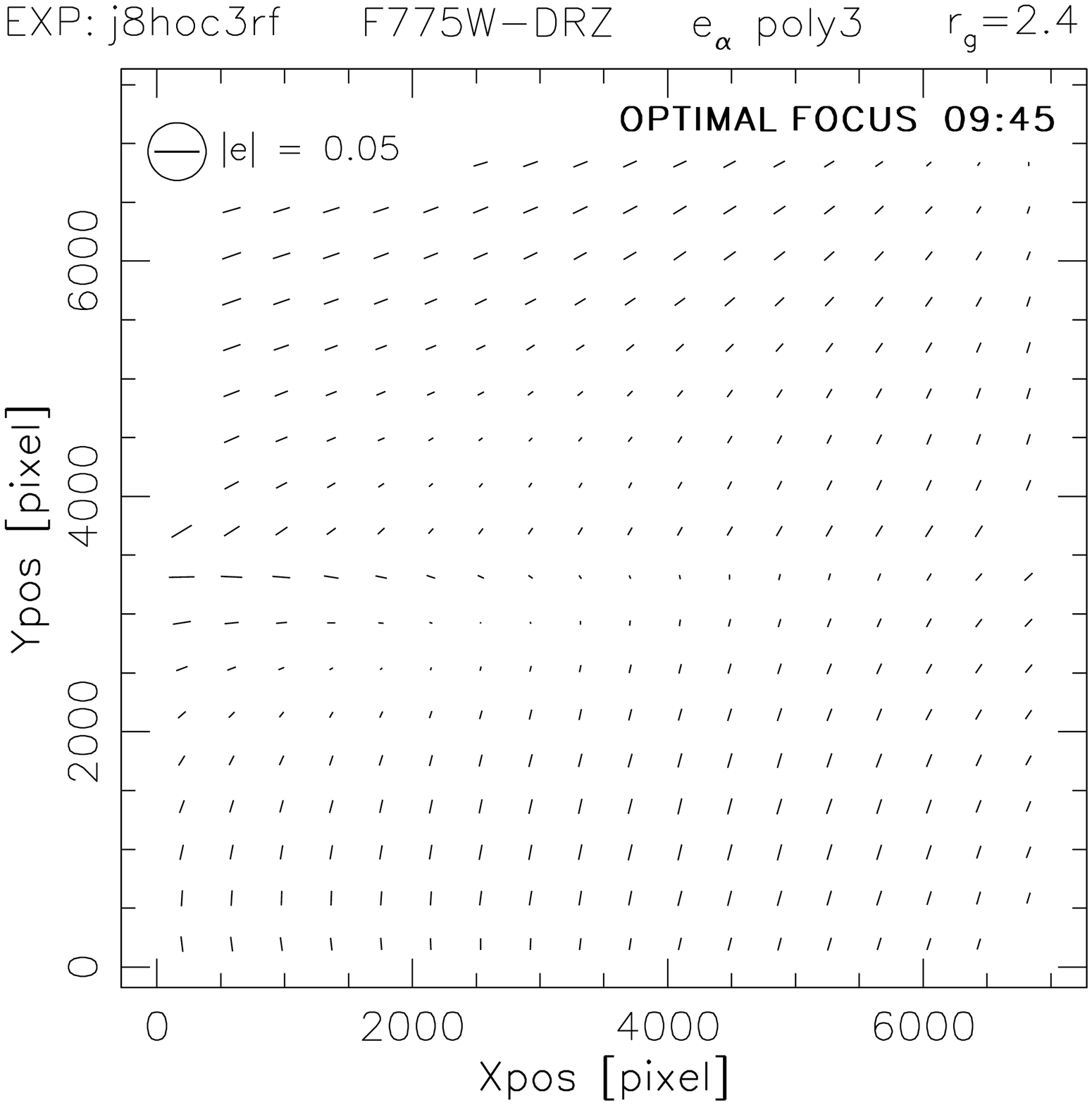}
   \includegraphics[width=5.8cm]{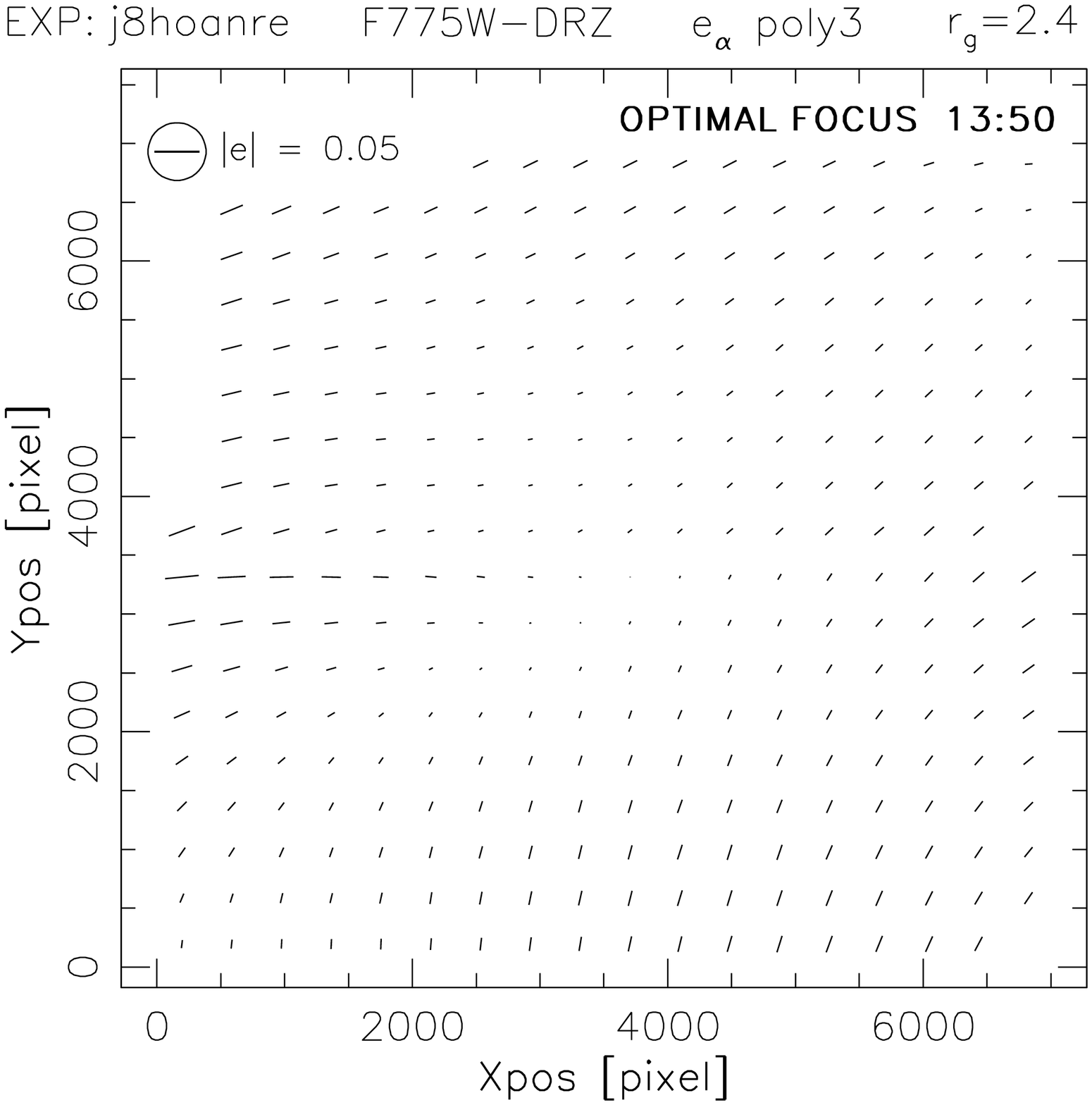}
   \includegraphics[width=5.8cm]{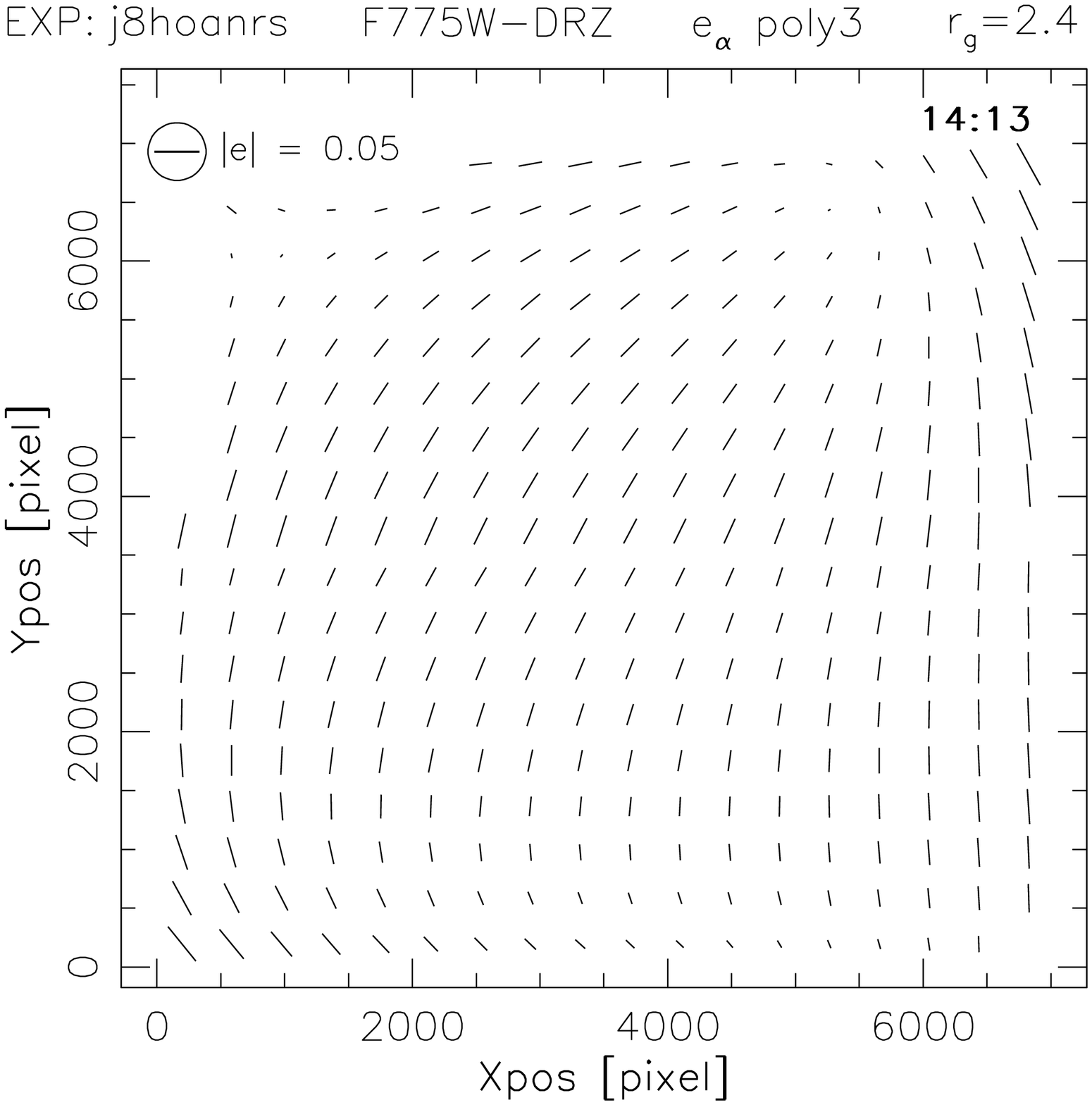}
   \includegraphics[width=5.8cm]{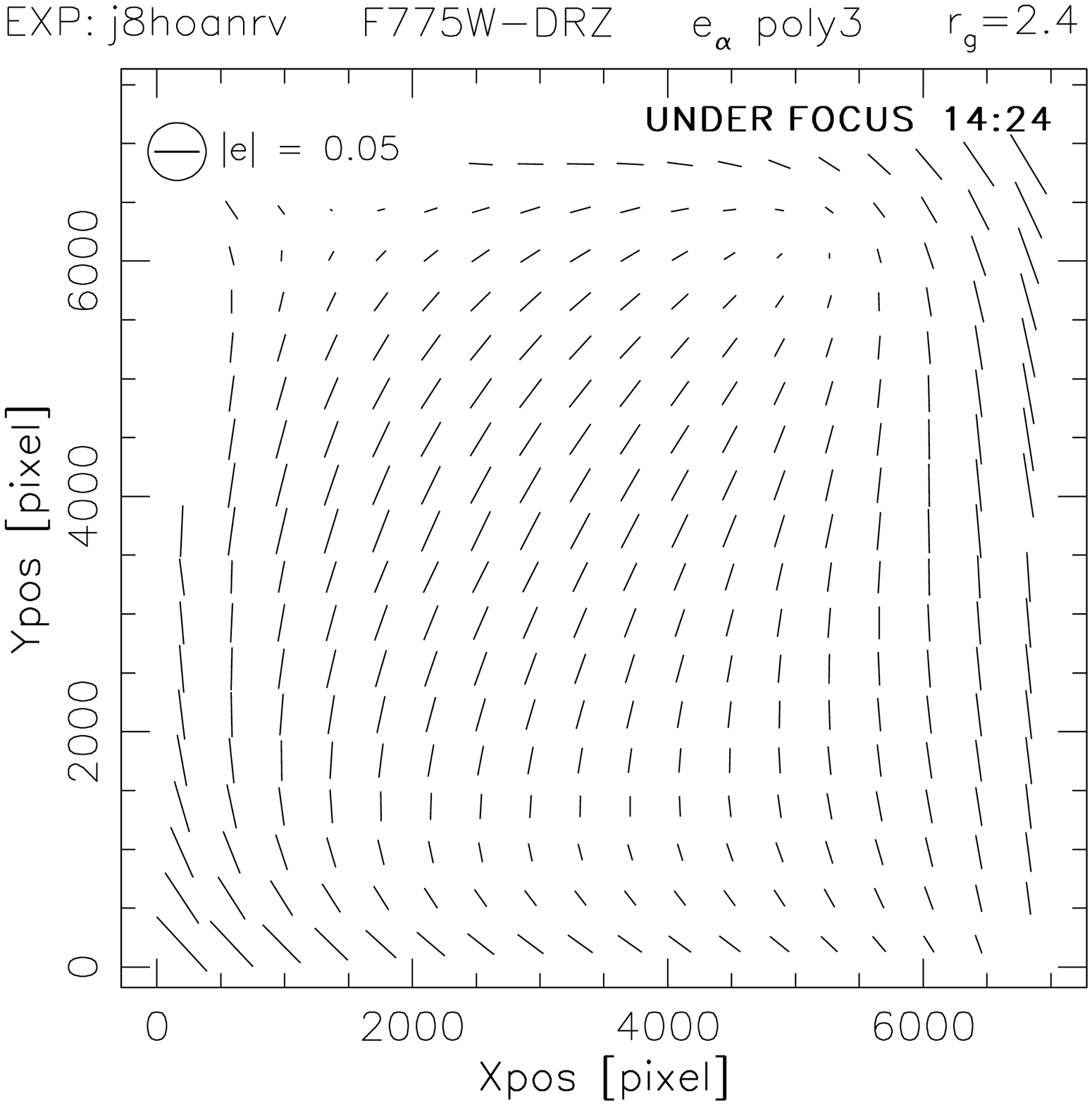}

   \caption{Third-order polynomial fits to stellar ellipticities in the \textit{DRZ}-images of two series of subsequent exposures.
The 400 second exposures were taken on 2002-08-28 (upper panels) and 2002-08-17 (lower panels), where the time indicated corresponds to the middle of the exposure (UT). The variations are interpreted as thermal breathing of the telescope.
The upper right and lower left plots are near the optimal focus position, whereas the other exposures represent positive focus offsets (upper left panel) or negative focus offsets (lower right panel).
}
   \label{fi:psf:short_term_variation}
    \end{figure*}

Comparing stellar field exposures observed at different epochs, we detect 
significant temporal variations of the PSF anisotropy already within one orbit.
Time variations of the ACS PSF were also reported by \citet{kri03,jwb05,jwf05,jwf06}; \citetalias{hbb05}; \citet{rma05,rma07} and \citet{ank06}, and are expected to be caused by focus changes due to thermal breathing of the telescope.
\citet{kri03} illustrates the variation of PSF ellipticity induced by astigmatism, which increases for larger focus offsets and changes orientation by 90$^\circ$ when passing from negative to positive offsets.
This behaviour is approximately reproduced in Fig.\thinspace\ref{fi:psf:short_term_variation} showing polynomial fits to stellar ellipticities in two series of subsequent exposures.

   \begin{figure*}
   \centering
   \includegraphics[width=5.8cm]{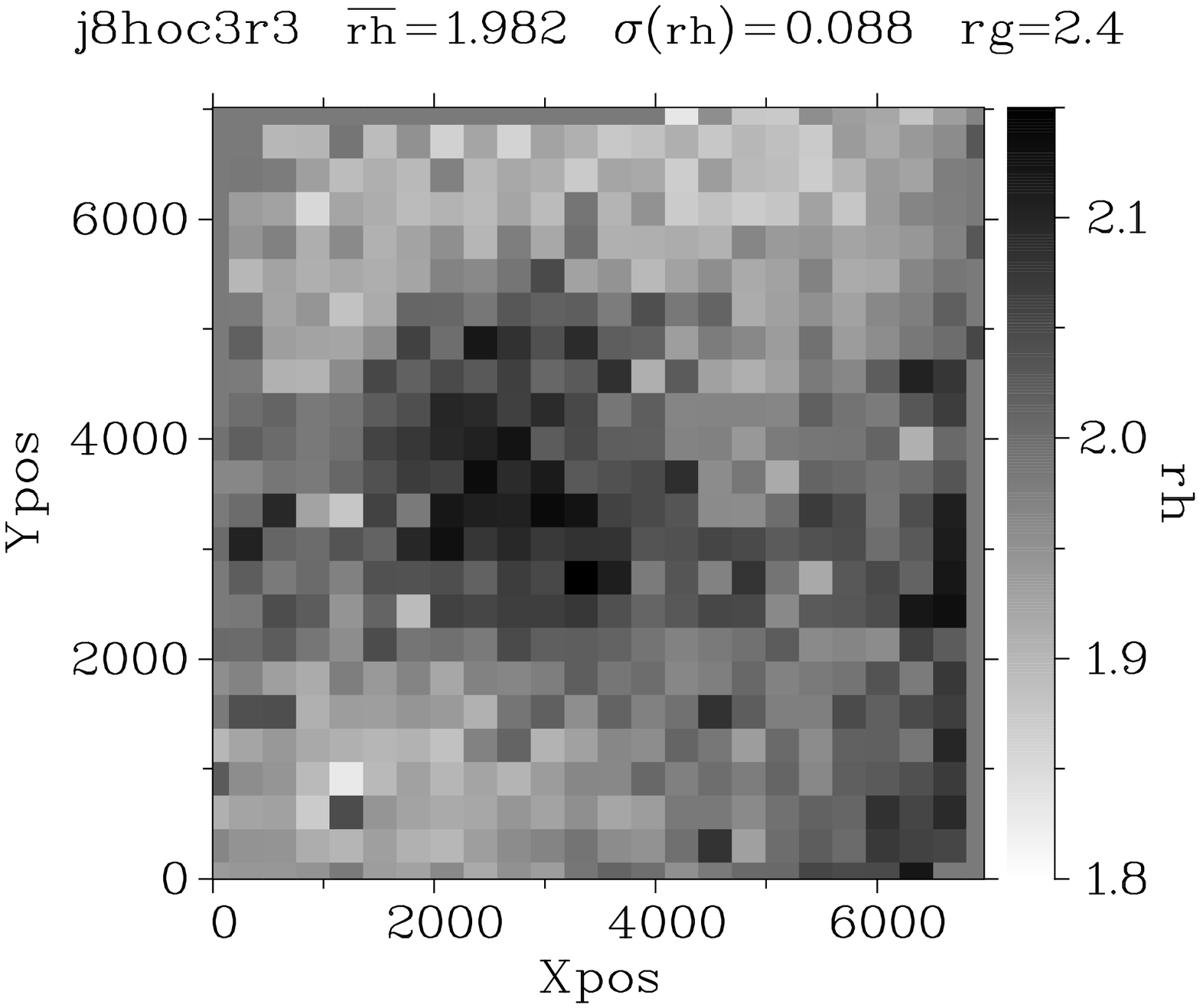}
   \includegraphics[width=5.8cm]{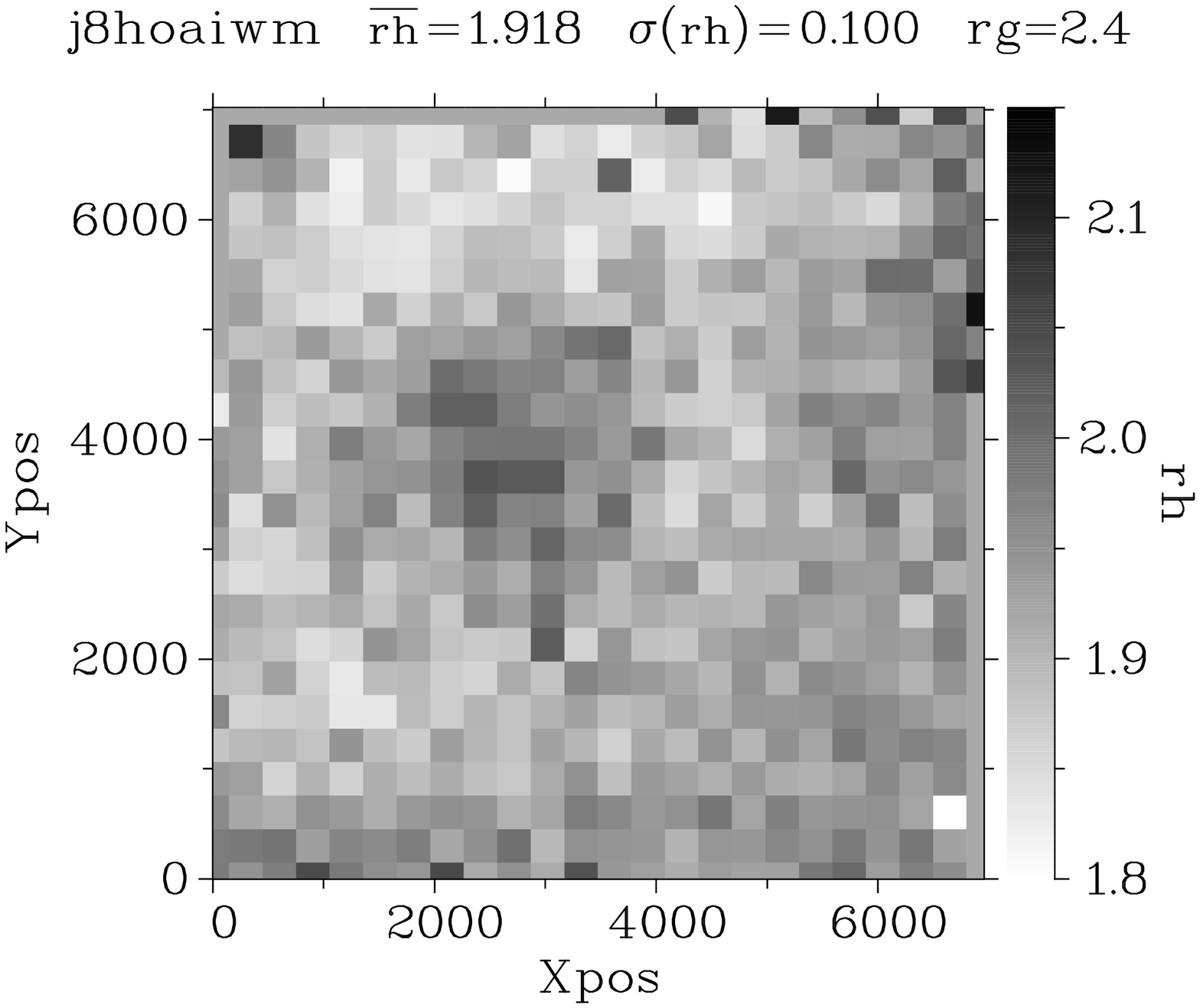}
   \includegraphics[width=5.8cm]{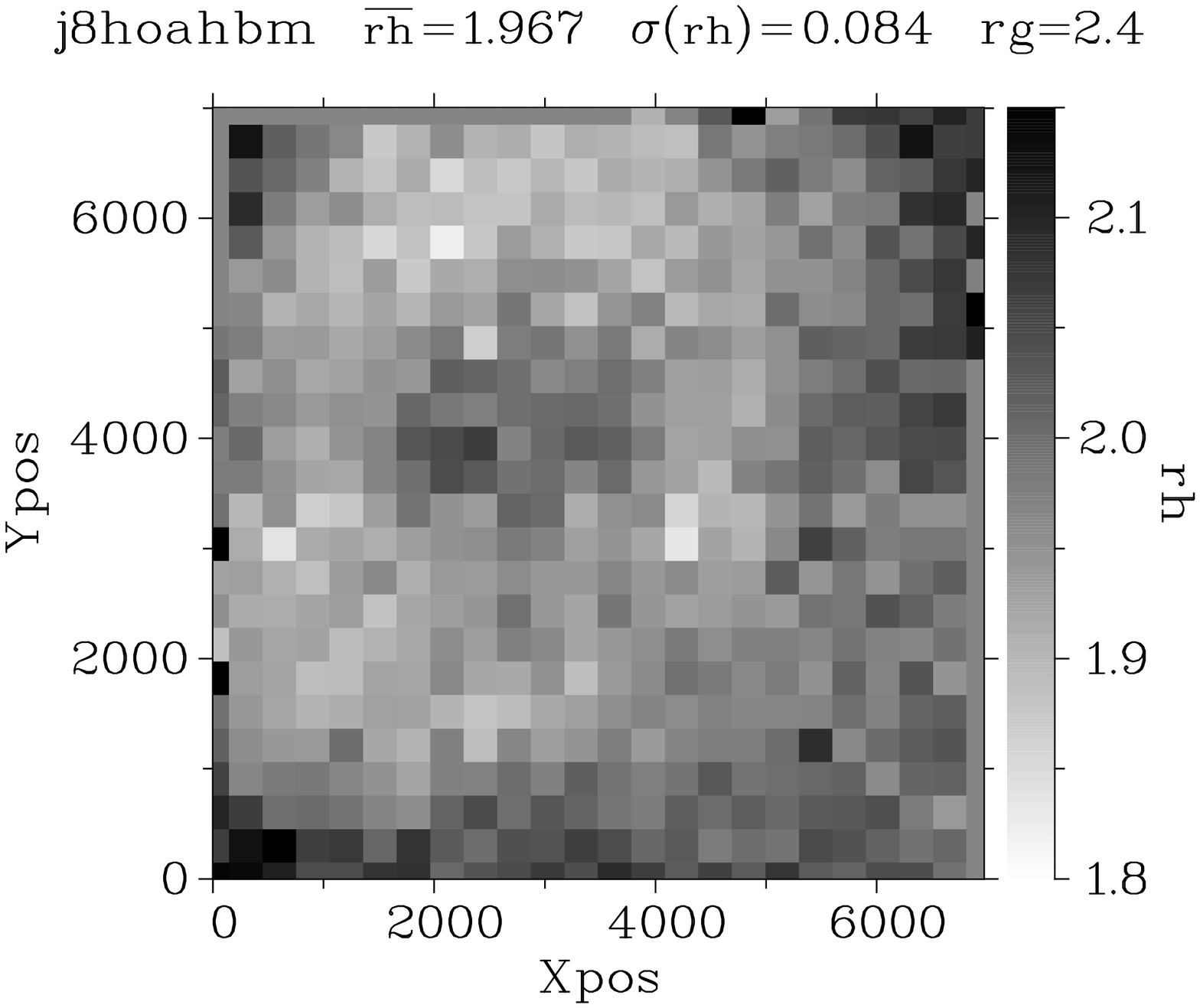}

   \caption{Field-of-view variation of the stellar half-light radius $r_\mathrm{h}$ in the \textit{DRZ}-images of three F775W stellar field exposures with positive focus offset (left), near optimal focus (middle) and negative focus offset (right).
For these plots $r_\mathrm{h}$ was averaged within bins of $300\times300$ pixels. Bins without any stars show the average value.}
\label{fi:psf:rh_variation}
\end{figure*}

\subsection{PSF width variation}
\label{se:psf_sizevariation}
Additional to the PSF ellipticity variation we also detect time and FOV variations of the PSF width.
Fig.\thinspace\ref{fi:psf:rh_variation} shows the FOV dependence of the stellar half-light radius $r_\mathrm{h}$ for three different exposures.
Among all F775W stellar field exposures the average half-light radius varied between $1.89\le \overline{r_\mathrm{h}} \le 2.07$, with an average FOV variation $\overline{\sigma(r_\mathrm{h})}=0.085$.
We find that the variation of the stellar quantity $T^*$ needed for 
the PSF correction of the galaxy ellipticities (\ref{eq:t_pshpsm}) 
closely follows the variation of $r_\mathrm{h}$ and can well be fitted with fifth-order polynomials in each chip.
For a further discussion of the PSF width variation see \citet{kri03}.

\subsection{PSF correction scheme}
\label{se:psf_correction_templates}

   \begin{figure}
   \centering
   \includegraphics[width=6.cm]{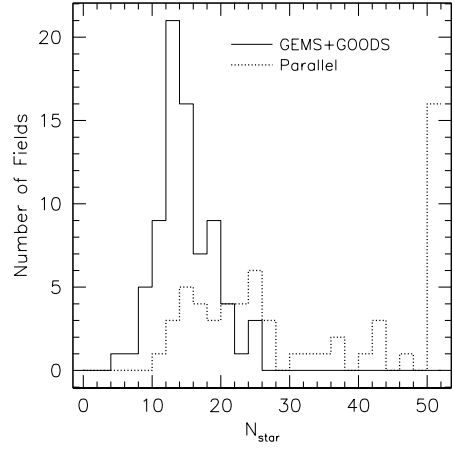}
 
   \caption{Histogram of the number of galaxy fields with $N_\mathrm{star}$ selected stars in the co-added images for the Parallel Survey (dashed line) and the GEMS+GOODS data (solid line).
}
   \label{fi:nstarsgalfields}
    \end{figure}

In order to correct for the detected temporal PSF variations using the low number of stars present in most galaxy fields (see Fig.\thinspace\ref{fi:nstarsgalfields}),
we apply a new correction scheme,
in which we determine the best-fitting stellar field PSF model for each galaxy field exposure separately.

\subsubsection{Description of the algorithm}
\label{se:psf_correction_templates:algorithm}
Due to the low number of stars present in galaxy fields, we require a PSF fitting method with as few free parameters as possible, excluding the possibility to use for example a direct polynomial interpolation.
As the main PSF determining factor is the focus position, we expect a nearly 1-parameter family of PSF patterns.
With the high number of stellar field exposures analysed $N_\mathrm{sf}=205$ for F775W and $N_\mathrm{sf}=184$ for F606W, we have a nearly continuous database of the varying PSF patterns.
This database consists of well constrained third- or fifth-order polynomial fits to $q_{\alpha}(x,y,r_\mathrm{g})$ and $T(x,y,r_\mathrm{g})$ for numerous values of $r_\mathrm{g}$, both for the \textit{COR}- and \textit{DRZ}-images.
In this section we omit the asterisk when we refer to these polynomial fits derived from the stellar fields in order to allow for a clear distinction to $q_{\alpha}^*$ measured from the stars in the galaxy fields.

Given the noisier $e_{\alpha}^*$ and $q_{\alpha}^*$ measurement in drizzled images (see Sect.\thinspace\ref{se:psf_anis}), we estimate the PSF correction for a galaxy field from the stellar images in each \textit{COR}-exposure of the galaxy field.
However, we apply the corresponding \textit{DRZ}-image PSF model as galaxy shapes are also measured on the drizzled co-added image.

In order to determine the correction for a co-added galaxy field, we fit the measured \mbox{$q_{\alpha}^{*,\mathrm{COR}}(r_\mathrm{g}=1.5)$} of the $N_{\mathrm{stars},k}$ stars present in galaxy field exposure $k$ with the stellar field PSF models $q_{\alpha,j}^{\mathrm{COR}}(x, y,r_\mathrm{g})$, with \mbox{$j \in {1,...,N_\mathrm{sf}}$} and identify the best fitting stellar field exposure $j_k$ with minimal 
\begin{equation}
\chi^2_{k,j}=\sum_{i=1}^{N_{\mathrm{stars},k}}\left[q_{\alpha,i}^{*,\mathrm{COR}}(r_\mathrm{g}=1.5)-q_{\alpha,j}^{\mathrm{COR}}(x_i, y_i,1.5)\right]^2\, .
\end{equation}
Here we choose the Gaussian window scale $r_\mathrm{g}=1.5$ WFC pixels to maximise the signal-to-noise in the shape measurement (see Sect.\thinspace\ref{se:catalog_creation}).
In this fit we reject outliers at the $2.5 \sigma$ level to ensure that stars in the galaxy field with noisy ellipticity estimates do not dominate the fit.

Having found the ``most similar'' (best-fitting) \textit{COR}-PSF model $j_k$ for each galaxy field exposure,
we next have to match the coordinate systems of the corresponding \textit{DRZ}-image and the co-added galaxy field.
This is necessary, as the single \textit{DRZ}-images used to create the PSF models are always drizzled without extra shifts in the default orientation of the camera, whereas the galaxy field exposures are aligned by \texttt{MultiDrizzle} according to their dither position.
For this we trace the position of each object in the co-added galaxy field back to the position it would have in the single drizzled exposure $k$ without shift and rotation.
Let $\phi_k$ and $(x_0,y_0)_k$  denote the rotation and shift applied by \texttt{MultiDrizzle} for exposure $k$.
For a galaxy with coordinates $(x,y)$ in the co-added image we then compute the ``single \textit{DRZ}''-coordinates
\begin{eqnarray}
\left( \begin{array}{c}
\tilde{x}\\ \tilde{y}\end{array} \right)_k
=
\left( \begin{array}{cc}
\cos{\phi_k} & \sin{\phi_k}\\ 
-\sin{\phi_k} & \cos{\phi_k} 
\end{array} \right)
\left( \begin{array}{c}
x - x_{0,k}\\ y - y_{0,k}\end{array} \right)
\end{eqnarray}
and the PSF model
\begin{eqnarray}
q_{k}^{\mathrm{DRZ}}(x, y, r_\mathrm{g}) &=&q_{j_k}^{\mathrm{DRZ}}(\tilde{x}, \tilde{y}, r_\mathrm{g}) \, \mathrm{e}^{2\mathrm{i}\phi_k} \\
T^{\mathrm{DRZ}}_{k}(x, y, r_\mathrm{g})&=&T^{\mathrm{DRZ}}_{j_k}(\tilde{x}, \tilde{y}, r_\mathrm{g}) \,,
\end{eqnarray}
where we denote the components of $q_{k}^{\mathrm{DRZ}}$ as $q_{\alpha,k}^{\mathrm{DRZ}}$.

In order to estimate the combined PSF model for the co-added galaxy image, we then compute the exposure time $t_k$-weighted average 
\begin{eqnarray}
q_{\alpha,\mathrm{comb}}^{\mathrm{DRZ}}(x, y, r_\mathrm{g})&=&\left(\sum_k t_k q_{\alpha,k}^{\mathrm{DRZ}}(x, y, r_\mathrm{g}) \Delta_k\right)/\sum_k t_k \Delta_k \\
T^{\mathrm{DRZ}}_\mathrm{comb}(x, y, r_\mathrm{g}) &=&\left(\sum_k t_k T^{\mathrm{DRZ}}_{k}(x, y, r_\mathrm{g}) \Delta_k\right)/\sum_k t_k \Delta_k  
\end{eqnarray}
of all shifted and rotated single exposure models, with $\Delta_k=1$ if the galaxy is located within the chip boundaries for exposure $k$ and $\Delta_k=0$ otherwise.

  \begin{figure*}
   \centering
   \includegraphics[width=5.8cm]{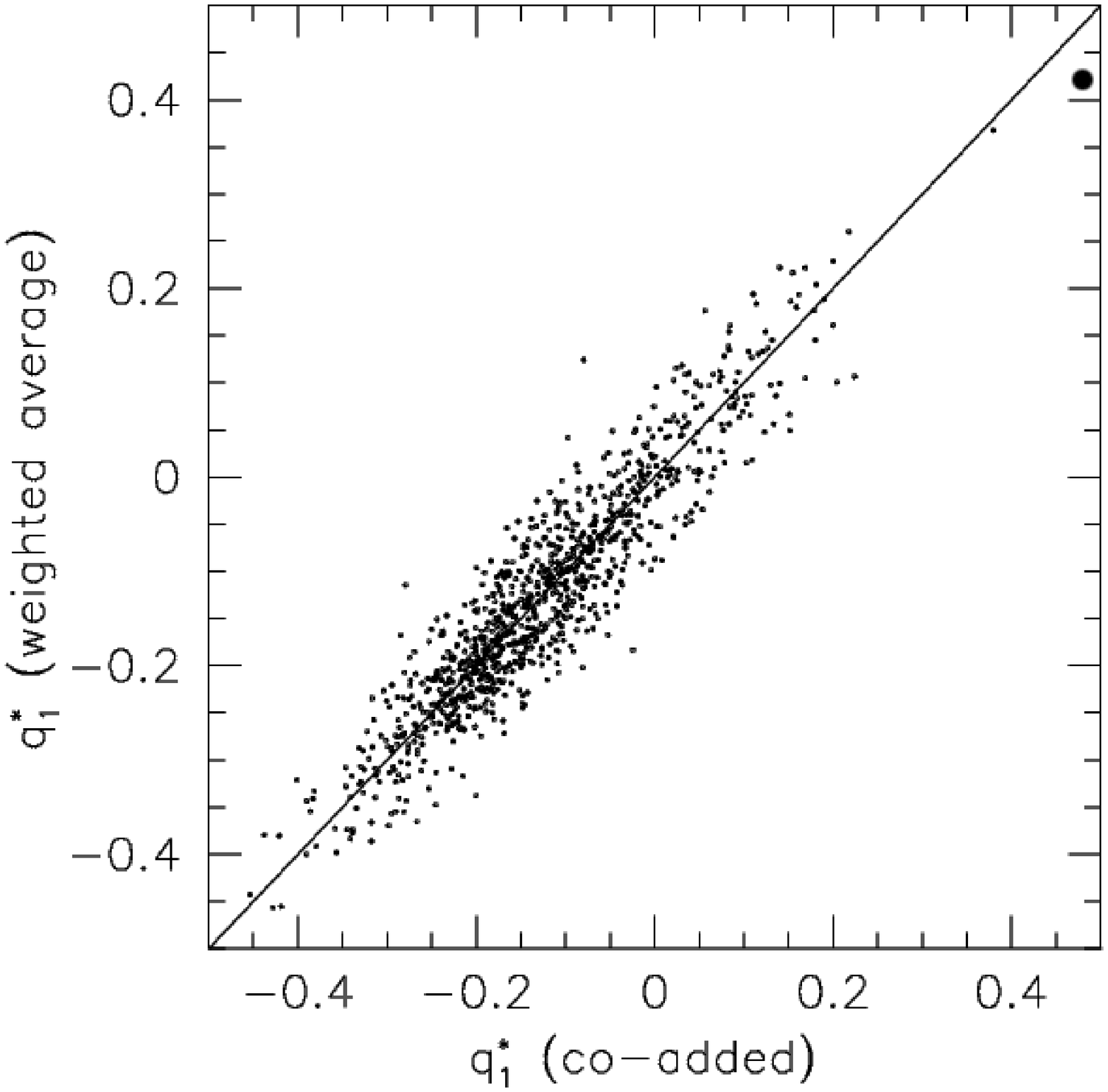}
   \includegraphics[width=5.8cm]{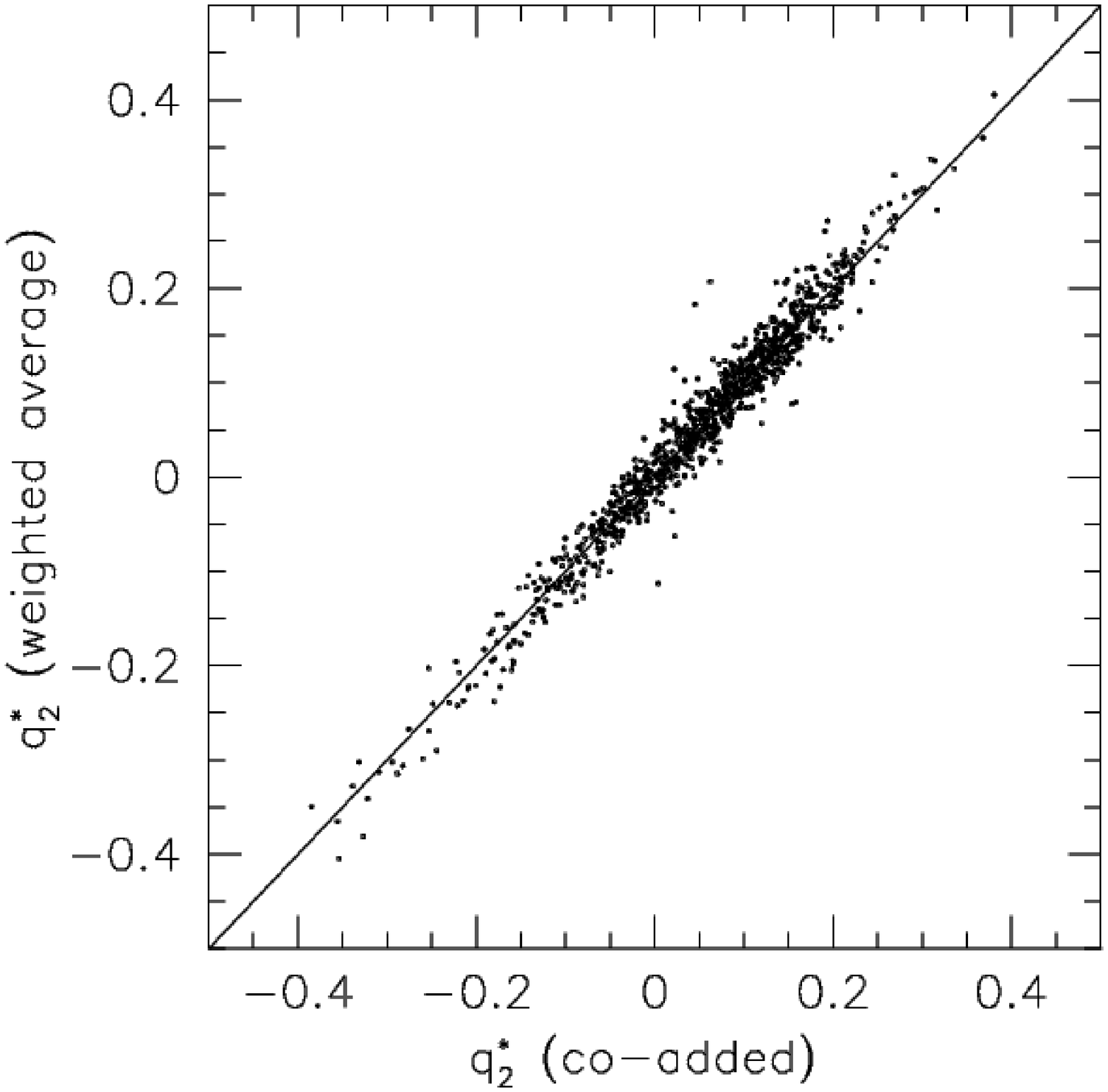}
   \includegraphics[width=5.8cm]{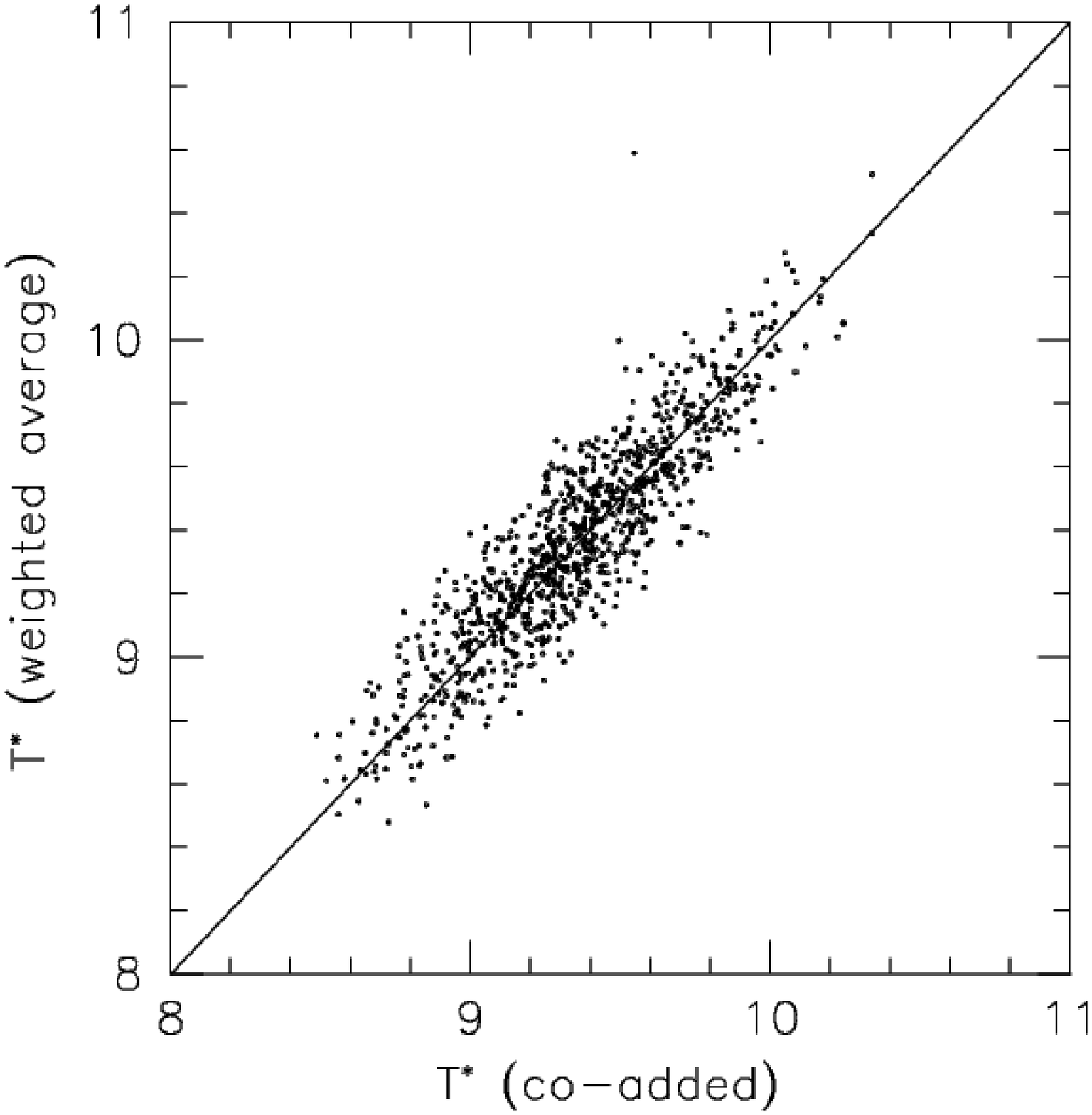}
 
   \caption{Comparison of the stellar quantities $q^*_\alpha$ and $T^*$ measured on individual stars with $r_\mathrm{g}=2.4$ pixels from a co-added stellar field to the same quantities computed as an exposure time-weighted average of the estimates in the single \textit{DRZ}-images. The good and unbiased agreement justifies the direct use of these quantities in the PSF correction scheme without the need to work on individual moments. The plotted points correspond to the three stellar field exposures shown in the bottom of Fig.\thinspace\ref{fi:psf:short_term_variation}. Note the larger scatter for $q^*_1$ compared to $q^*_2$ which is mainly due to the noise created by re-sampling.
}
   \label{fi:psf:demonstrate_average}
    \end{figure*}

Another factor which is expected to influence the image PSF besides focus changes are jitter variations created by tracking inaccuracies (Sect.\thinspace\ref{se:data}).
To take those into account we fit an additional, position-independent jitter term $q^{0}_{\alpha}(r_\mathrm{g})$.
We already take this constant into account while fitting the galaxy field stars with the stellar field models to ensure that a large jitter term does not bias the identification of the best-fitting star field.
Yet, as the number of stars with sufficient signal-to-noise is higher in the co-added image
and since only the combined jitter effect averaged over all exposures is relevant for the analysis, we re-determine the jitter term in the co-added drizzled image after subtraction of the combined PSF model $q_{\alpha,\mathrm{comb}}^{\mathrm{DRZ}}(x, y, r_\mathrm{g})$.
The final PSF model used for the correction of the galaxies is then given by 
\begin{equation}
q_{\alpha,\mathrm{total}}^{\mathrm{DRZ}}(x, y, r_\mathrm{g})=q_{\alpha,\mathrm{comb}}^{\mathrm{DRZ}}(x, y, r_\mathrm{g})+q^{0}_{\alpha}(r_\mathrm{g})
\end{equation}
and \mbox{$T^{\mathrm{DRZ}}_\mathrm{comb}(x, y, r_\mathrm{g})$}.

Note that this correction scheme assumes that the PSF model quantities $q_{\alpha,k}^{\mathrm{DRZ}}(x, y, r_\mathrm{g})$ and $T^{\mathrm{DRZ}}_{k}(x, y, r_\mathrm{g})$ determined for each galaxy field exposure can directly be averaged to determine the correction for the co-added image.
While only brightness moments add exactly linearly, this computation simplifying approach is still justified, as both the PSF size variation and the absolute value of the stellar ellipticities are small (see Sections \ref{se:psf_anis} and \ref{se:psf_sizevariation}).
Computing the flux-normalised trace of the stellar second brightness moments
\begin{equation}
  \hat{Q}\equiv \frac{Q_{11}+Q_{22}}{\mathrm{FLUX}^*}
\end{equation}
for all stars in the F775W stellar field exposures with fixed \mbox{$r_\mathrm{g}=2.4$} pixels,
we find that $\hat{Q}$ has a relative variation of 3\% only ($1\sigma$).
Therefore we can well neglect non-linear terms induced by the denominator in (\ref{eq:elli_e}).
The same holds for non-linear contributions of $P^{\mathrm{sm}*}$ and $T^*$,
which show $1\sigma$-variations only by 6\% ($\mathrm{Tr}\left[P^{\mathrm{sm}*}\right]$) and 5\% ($T^*$).
As a (very good) first-order approximation we can therefore simply average $q_{\alpha,k}^{\mathrm{DRZ}}(x, y, r_\mathrm{g})$ and $T^{\mathrm{DRZ}}_{k}(x, y, r_\mathrm{g})$ linearly, as also demonstrated in Fig.\thinspace\ref{fi:psf:demonstrate_average}, where we compare the exposure time-weighted average of the quantities measured from stars in individual drizzled frames to the value measured in the co-added image.

\subsubsection{Test with star fields}
\label{se:Startest}

\begin{figure}
   \centering
   \includegraphics[width=8.cm]{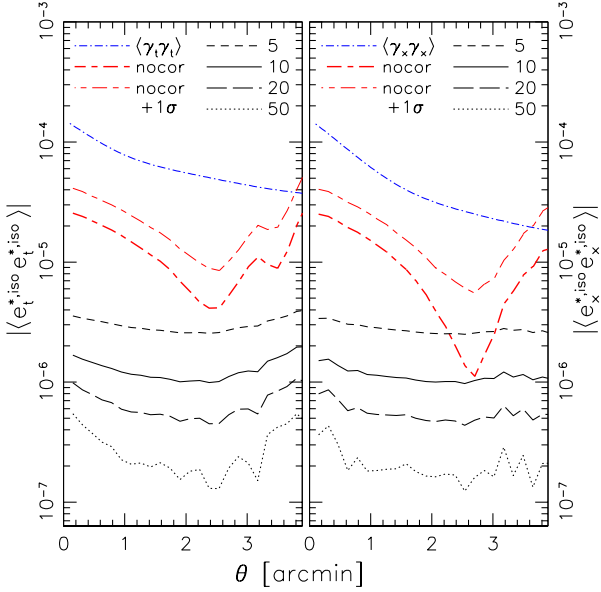} 
   \caption{Estimate for the PSF fitting accuracy: In order to simulate the low number of stars in galaxy fields, the PSF correction technique was applied to the 61 parallel data star fields, from which only small random subsets of $N_\mathrm{star}$ stars were used to determine the fit. 
%For the fitting of a particular stellar field exposure, its own entry in the PSF model database was ignored. 
We plot the correlation functions $\langle e^{*,\mathrm{iso}}_{\mathrm{t}} e^{*,\mathrm{iso}}_{\mathrm{t}}\rangle$ (left) and $\langle e^{*,\mathrm{iso}}_{\times} e^{*,\mathrm{iso}}_{\times}\rangle$ (right) of the ``transformed'' and corrected stellar ellipticity $e^{*,\mathrm{iso}}_{\alpha}$ (\ref{eq:e_star_galaxy_elli}), which accounts for the different susceptibility of stars and galaxies to PSF effects. 
The numbers \mbox{$N_\mathrm{star}=(5, 10, 20, 50)$} indicate the number of random stars used in each subset. 
Note that the uncorrected PSF signal computed from the transformed but \textit{not} anisotropy corrected ellipticity (nocor) and its $1\sigma$ upper limit (nocor$+1\sigma$) are only slightly lower than $\Lambda$CDM predictions 
for the cosmological lensing signal shown by the dashed-dotted curves for \mbox{$\sigma_8=0.7$, $z_\mathrm{m}=1.34$}.
}
   \label{fi:psf:starfield_test}
    \end{figure}

In order to estimate the accuracy of our fitting scheme we test it on all co-added stellar field images.
For each stellar field we randomly select subsets of $N_\mathrm{star}$ stars from the \textit{COR}-images and the co-added image simulating the low number of stars present in galaxy fields.
This subset of stars is used to derive the PSF model as described in Sect.\thinspace\ref{se:psf_correction_templates:algorithm} which we then apply to the entirety of stars in the co-added image.
For the fitting of a particular stellar field exposure, we ignore its own entry in the PSF model database and only consider the remaining models.
The strength of any coherent pattern left in the stellar ellipticities after model subtraction provides an estimate of the method's accuracy.
In order to determine the actual impact of the remaining PSF anisotropy on the cosmic shear estimate, one has to
consider that although galaxy ellipticities are on the one hand less affected by PSF anisotropy than stars, they are additionally scaled with the $P^g$ correction (\ref{eq:eani},\ref{eq:fully_corr}).
We thus ``transform'' the remaining stellar ellipticity into a corrected galaxy ellipticity \citep[see e.g. ][]{hoe04}
\begin{eqnarray}
\label{eq:e_star_galaxy_elli}
e^{*,\mathrm{iso}}_{\alpha} =  \frac{2 c_\mathrm{cal}}{\mathrm{Tr}P^g_{\mathrm{gal}}} \left[ \frac{\mathrm{Tr}P^{\mathrm{sm}}_\mathrm{gal}}{\mathrm{Tr}P^{\mathrm{sm}*}(r_\mathrm{g})} e^{*}_\alpha(r_\mathrm{g}) - P^{\mathrm{sm}}_{\alpha\beta,\mathrm{gal}} \, q_{\beta,\mathrm{total}}^{\mathrm{DRZ}}(x, y, r_\mathrm{g}) \right]\, ,
\end{eqnarray}
where we randomly assign to each star the value of $P^g_{\mathrm{gal}}$, $P^{\mathrm{sm}}_\mathrm{gal}$, and $r_\mathrm{g}$ from one of the parallel data galaxies used for the cosmic shear analysis (see Sect.\thinspace\ref{se:preliminary_cosmic_shear}).
Fig.\thinspace\ref{fi:psf:starfield_test} shows the estimate of the two-point correlation functions of $e^{*,\mathrm{iso}}_{\alpha}$ averaged over all star fields and 30 randomisations for different numbers of random stars $N_\mathrm{star}$.
This plot indicates that already for \mbox{$N_\mathrm{star}=10$} stars present in a galaxy field the contribution of remaining PSF anisotropy is expected to be reduced to a level \mbox{$\langle e^{*,\mathrm{iso}}_{t/\times} e^{*,\mathrm{iso}}_{t/\times} \rangle < 2 \times 10^{-6}$} corresponding to $\simeq 1-5\%$ of the cosmological shear correlation function expected on scales probed by a single ACS field.
Since all of the examined parallel fields and the large majority of the GEMS+GOODS fields contain more than 10 stars (see Fig.\thinspace\ref{fi:nstarsgalfields}), we are confident that the systematic accuracy of this fitting technique will be sufficient also for the complete ACS parallel data.

The further reduction of the remaining systematic signal for larger $N_\mathrm{star}$ shows that the accuracy is mainly limited by the number of available stars and not by a too narrow coverage of our PSF database or the linear averaging of $q_{\alpha,k}^{\mathrm{DRZ}}(x, y, r_\mathrm{g})$.

For comparison we also plot in Fig.\thinspace\ref{fi:psf:starfield_test} the correlation functions calculated from the $P^g$-scaled, but \textit{not} anisotropy corrected stellar ellipticity, which for larger scales is of the same order of magnitude as the expected shear signal.
Note that the plotted values depend on the selection criteria for the galaxies (see Sect.\thinspace\ref{se:galaxy_selection}).
Particularly, the inclusion of smaller, less resolved galaxies would increase both the corrected and uncorrected signal.
Additionally, it is assumed that the distribution of PSFs occurring is the same for the star and galaxy fields.
For more homogeneous surveys (e.g. the GEMS+GOODS data) one might expect that more similar PSFs occur more frequently than for the quasi random parallel star fields, for which the stellar correlation function is expected to partially cancel out.
Thus, we also plot the one sigma upper limit of the uncorrected correlation function in Fig.\thinspace\ref{fi:psf:starfield_test}, which might be a more realistic estimate for the uncorrected systematic signal for such surveys.

\subsubsection{Discussion of the algorithm}
\label{se:discussion_psfalgorithm}
The applicability of the proposed algorithm relies on the assumption that the stellar fields densely cover the parameter space of PSF patterns occurring in the galaxy fields.
This is likely to be the case if 
\begin{enumerate}
  \item both datasets roughly cover the same time span,
  \item the number of star field exposures is sufficiently large,
  \item and no significant additional random component occurs besides the constant jitter offset that we have considered.
\end{enumerate}
For both the F606W and F775W data (1.) is fulfilled and from the ensemble of observed stellar field PSFs we are confident that (2.) and (3.) are also well satisfied.
This is also confirmed by the test presented in Sect.\thinspace\ref{se:Startest}.
Yet, the reader should note that datasets might exist for which conditions (1.) to (3.) are not well fulfilled,
e.g. due to observations in a rarely used filter with only a low number of observed stellar fields.
In such a case the described algorithm might be adjusted using a principal component analysis \citep{jaj04} or theoretical PSF models \citep{rma05,rma07}. 
Note that the differences in the observed PSFs are interpreted to be mainly driven by different focus offsets.
However, the suggested algorithm will work just as well if further factors play a role, as long as sufficient stellar field exposures are available.

\subsubsection{Advantages of our PSF correction scheme}
\label{se:advantages_compared_to_other}
Finally we want to summarise the advantages our method provides 
for the high demands of a cosmological weak lensing analysis 
on accurate PSF correction:
\begin{enumerate}
\item Our technique deals very well with the low number of stars present in typical high galactic latitude fields, which inhibits direct interpolation across the field-of-view.
\item The ACS PSF shows substantial variation already between consecutive exposures (see Fig.\thinspace\ref{fi:psf:short_term_variation}), which is adequately taken into account in our technique. When exposures from different focus positions are combined, a single-focus PSF model, as e.g. used by \citet{rma07}, is no longer guaranteed to be a good description for the co-added image.
\item Our PSF models are based on actually observed stellar fields and are thus not affected by possible limitations of a theoretical PSF model.
\item We determine the PSF fits in the undrizzled \textit{COR}-images, which excludes any impact from additional shape noise introduced by re-sampling.
\item The algorithm is applicable for arbitrary dither patterns and rotations, and can easily be adapted for other weak lensing techniques (e.g. Nakajima et al. in prep.).
\end{enumerate}

\section{Galaxy catalogue and redshift distribution}
\label{se:galaxy_select_redshift}
\subsection{Galaxy selection}
\label{se:galaxy_selection}
We select galaxies with cuts in the signal-to-noise ratio \mbox{$\mathrm{S}/\mathrm{N}>4$},
half-light radius \mbox{$2.8<r_\mathrm{h}<15$} pixels, corrected galaxy ellipticity \mbox{$|e^\mathrm{iso}|<2.0$}, and \mbox{$\mathrm{Tr}P^g/2 > 0.1$}.
From the analysis of the STEP1 image simulations \citep{hwb06} we find no indications for a significant bias in the shear estimate introduced by these conservative cuts for $|e^\mathrm{iso}|$ and $\mathrm{Tr}P^g$.
However, due to a detected correlation of the shear estimate both with $r_\mathrm{h}$ and the \texttt{SExtractor} \texttt{FLUX\_RADIUS}, 
cuts in these quantities introduce a significant selection bias.
For the analysis of the STEP2 image simulations we therefore chose $r_\mathrm{h}$-cuts closely above the stellar sequence \citep{mhb07}.
   \begin{figure}
   \centering
   \includegraphics[width=8.cm]{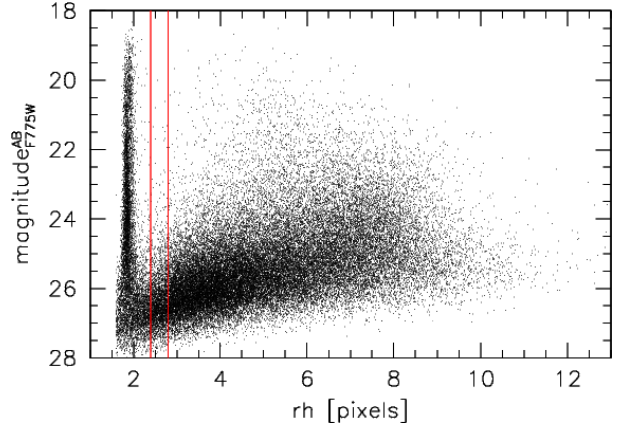}
 
   \caption{$r_\mathrm{h}$--magnitude distribution of the Parallel data F775W objects after applying a cut $\mathrm{S}/\mathrm{N}>4$. The vertical lines indicate two different cuts for the galaxy selection: Although a cut $r_\mathrm{h}>2.4$ pixels is sufficient to reliably exclude stars, we additionally reject very small galaxies (\mbox{$2.4\,\mathrm{pixels}<r_\mathrm{h}<2.8\,\mathrm{pixels}$}), which are most strongly affected by the PSF.
}
   \label{fi:galaxy_rh_mag}
    \end{figure}
Yet, as the magnitude-size relation is very different for ground- and space-based images, we expect that a cut at larger $r_\mathrm{h}$ will introduce a smaller shear selection bias for space-based images.
In Fig.\thinspace\ref{fi:galaxy_rh_mag} we plot the $r_\mathrm{h}$--magnitude distribution of the objects in the F775W galaxy fields after 
a cut $\mathrm{S}/\mathrm{N}>4$ was applied.
Considering the PSF size variation (Sect.\thinspace\ref{se:psf_sizevariation}) and increased noise in the $r_\mathrm{h}$ measurement for faint objects,
stars can reliably be rejected with a cut $r_\mathrm{h}\gtrsim2.4$ pixels.
With the cuts in $|e^\mathrm{iso}|$, and $\mathrm{Tr}P^g$ applied, 
increasing the size cut to $r_\mathrm{h}>2.8$ pixels rejects only 6.1\% of the remaining galaxies.
As these galaxies are most affected by the PSF, and considering the possible limitations for the application of the KSB formalism for a diffraction limited PSF (Sect.\thinspace\ref{se:method}), we decided to use the more resolved galaxies with $r_\mathrm{h}>2.8$ pixels.
We plan to investigate whether this introduces a significant shear selection bias with the space-based STEP3 simulations 
(Rhodes et al. in prep.).

With these cuts we select in total 39898 (77749) galaxies corresponding to an average galaxy number density of $63 \,\mathrm{arcmin}^{-2}$ ($96 \,\mathrm{arcmin}^{-2}$)
for the parallel F775W fields (GEMS+GOODS F606W tiles) with a 
corrected ellipticity dispersion 
\mbox{$\sigma(c_\mathrm{cal}e^\mathrm{iso}_\alpha)=0.32$ ($0.33$)} for each component.

\citetalias{hbb05} found that the faintest galaxies in their catalogue were very noisy diluting the shear signal.
Therefore they use a conservative rejection of faint galaxies (\mbox{$m_{606}<27.0$}, \mbox{$\mathrm{SNR} > 15$})
leading to a lower galaxy number density of \mbox{$\sim~60\,\mathrm{arcmin}^{-2}$} for the GEMS and GOODS F606W data.
For our primary analysis we use a rather low cut \mbox{$\mathrm{S}/\mathrm{N}>4$} (see above) to be consistent with our analysis of the STEP simulations.
In order to assess the impact of the faintest galaxies and ease the comparison to \citetalias{hbb05},
we repeat the cosmological parameter estimation in Sect.\thinspace\ref{se:cosmo_para_estimate} with more conservative cuts 
\mbox{$\mathrm{S}/\mathrm{N}>5, m_{606}<27.0$} leading to 
a number density of $72 \,\mathrm{arcmin}^{-2}$, which is roughly comparable to the value found by \citetalias{hbb05} given the deeper combined GOODS images in our analysis.

\begin{figure}
   \centering
   \includegraphics[width=6.cm]{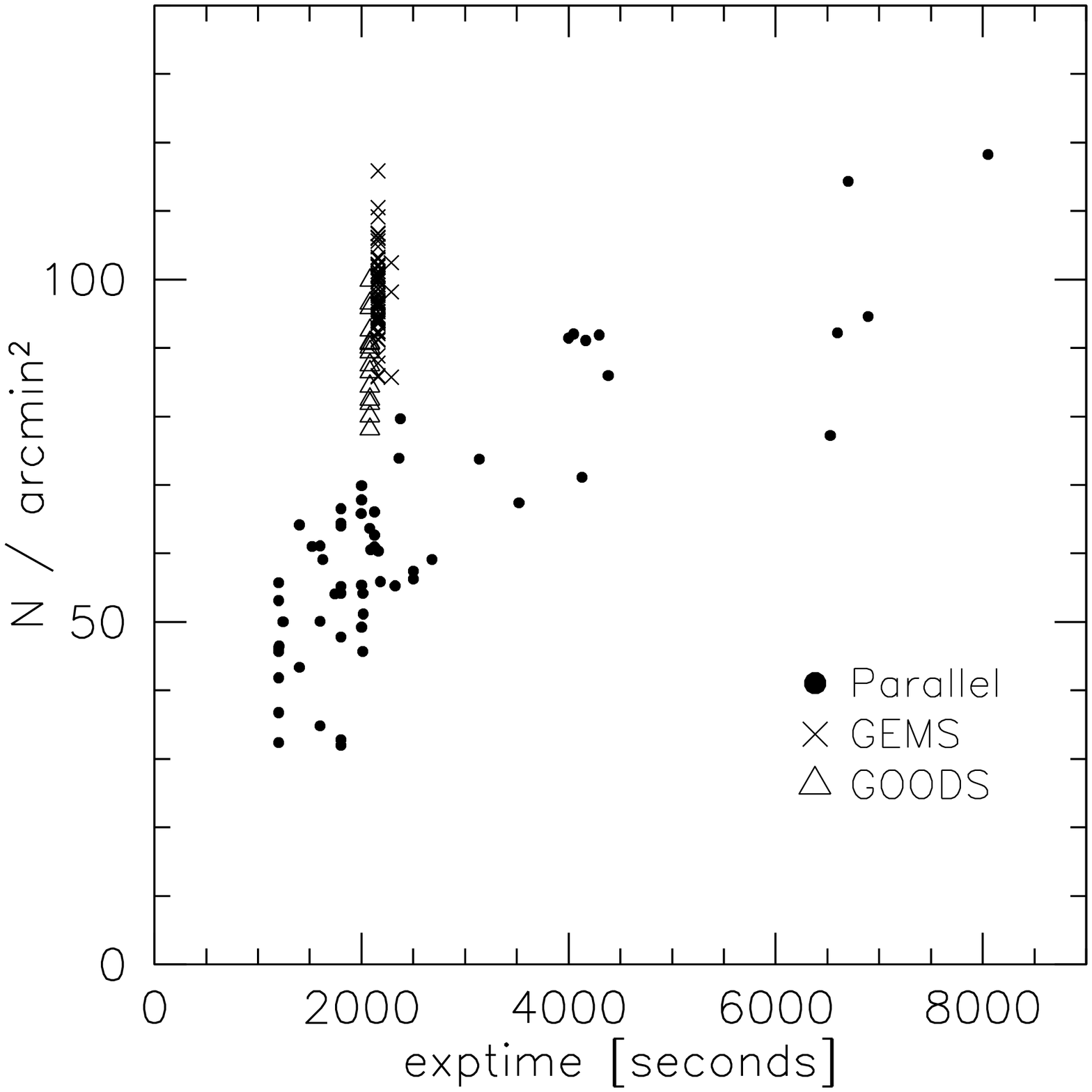}
 
   \caption{Number density of selected galaxies (\mbox{$\mathrm{S}/\mathrm{N}>4$}) for the parallel data F775W fields and the GEMS/GOODS F606W tiles as a function of exposure time.
}
   \label{fi:galaxy_N_exptime}
    \end{figure}

We plot the average galaxy number density as a function of exposure time for the different datasets in Fig.\thinspace\ref{fi:galaxy_N_exptime},
indicating that F606W is more efficient than F775W in terms of the average galaxy number density.
However one should keep in mind that the parallel fields are subject to varying extinction and more inhomogeneous data quality.

For the GEMS+GOODS tiles we rotate the galaxy ellipticities to a common coordinate system and reject double detections in overlapping regions which leaves 71682 galaxies for \mbox{$\mathrm{S}/\mathrm{N}>4$} and 53447 galaxies for \mbox{$\mathrm{S}/\mathrm{N}>5, m_{606}<27.0$}.

\subsection{Comparison of shear catalogues}
\label{su:shear_comparison}

\begin{figure*}
   \centering
    \sidecaption
   \includegraphics[width=6.5cm]{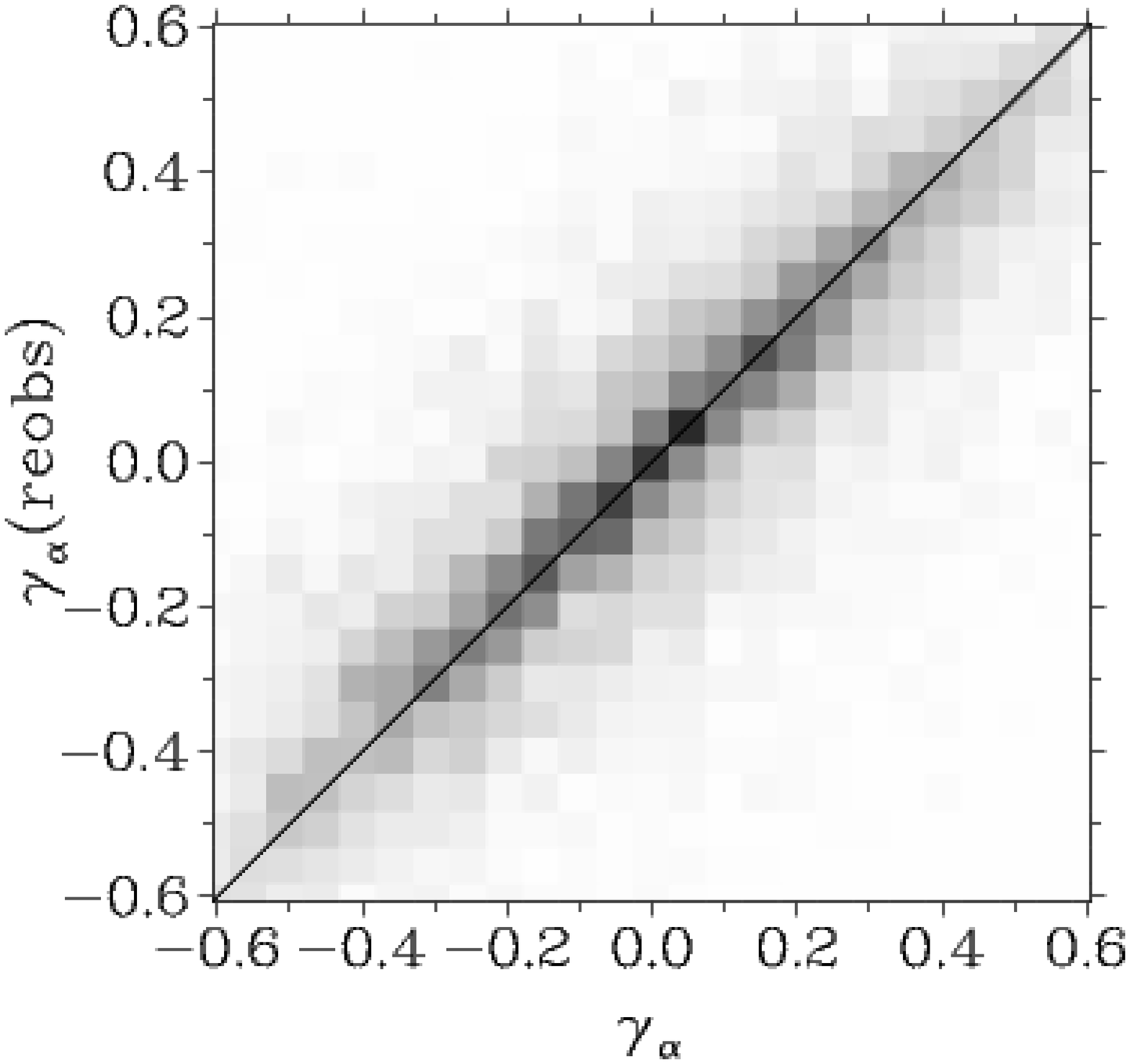}
   \includegraphics[width=6.5cm]{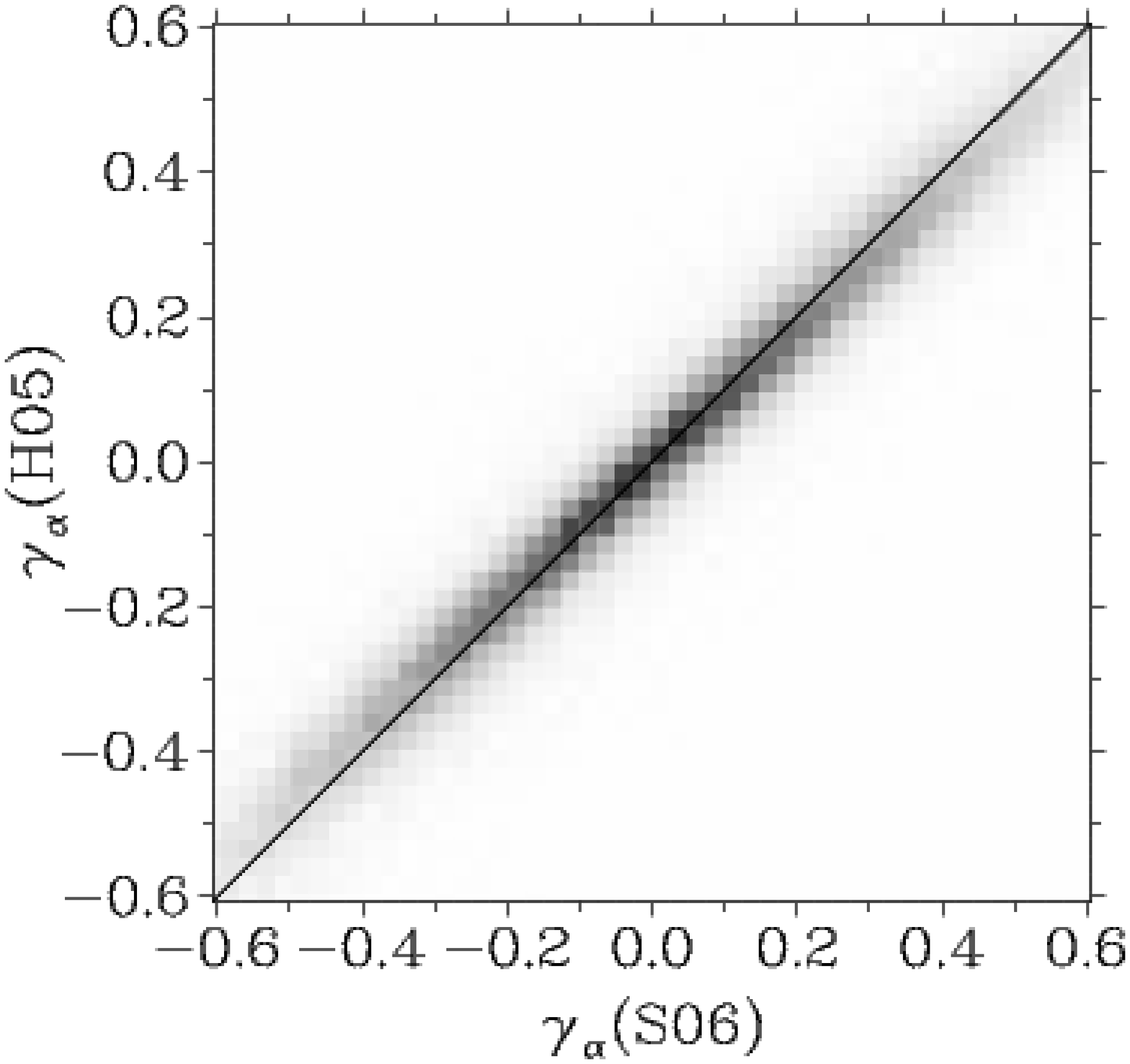}
   \caption{Comparison of the shear estimates between overlapping ACS tiles (left) and between the \citetalias{hbb05} and our catalogue (right). The grey-scale indicates the number of galaxies. Note the slight difference in the shear calibration between the two pipelines ($\sim 3.3\%$). 
In the left panel galaxies from different noise realisations are compared, leading to the larger scatter.
The solid line shows a 1:1 relation.
}
   \label{fi:elli_compare}
    \end{figure*}

In regions where different GEMS and GOODS tiles overlap, we have two independent shear estimates from the same galaxies with different noise realisations corrected for different PSF patterns.
This provides us with a good consistency check for our shear pipeline.
We compare the two shear estimates in the left panel of Fig.\thinspace\ref{fi:elli_compare}.
Although there is a large scatter created by the faint galaxies, which are strongly affected by noise, the shear estimates agree very well on average confirming the reliability of the pipeline.

Additionally, we match our shear catalogue to the \citetalias{hbb05} catalogue, which stems from an independent data reduction and weak lensing pipeline, and compare the two shear estimates in the right panel of Fig.\thinspace\ref{fi:elli_compare}.
Overall there is good agreement between the two pipelines with a slight difference in the shear calibration, where our shear estimate is in average larger by 3.3\%.
This is also consistent with results of the STEP project given a 3\% under-estimation of the shear for the Heymans pipeline in STEP1 \citep{hwb06} and an error of the average shear calibration consistent with zero for the Schrabback pipeline in STEP2 \citep{mhb07}.
The slightly different results for the two KSB+ pipelines are likely to be caused by the shear calibration factor used in our pipeline and the different treatment of measuring shapes from pixelised data, where we interpolate across pixels while \citetalias{hbb05}  evaluate the integrals at the pixel centres.
See also \citet{mhb07} for a discussion of the impact of pixelisation based on the STEP2 results.
For the GEMS and GOODS data a shear calibration error of $\sim 3\%$ is well within the statistical noise.

\subsection{Redshift distribution}
\label{se:redshift_distribution}

\begin{figure}
   \centering
   \includegraphics[width=6.cm]{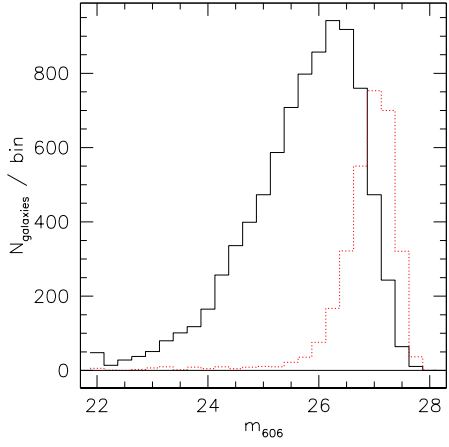}
 
   \caption{Number of selected GOODS-CDFS galaxies as a function of $m_{606}$. The solid line corresponds to galaxies for which spectroscopic or photometric redshift are available from the GOODS-MUSIC sample \citep{gfs06}, whereas the dotted line shows galaxies in the shear catalogue without redshift estimate located in the same area.
}
   \label{fi:n_mag_music}
    \end{figure}

In order to estimate cosmological parameters from cosmic shear data, accurate knowledge of the source redshift distribution is required.
This is of particular concern if the redshift distribution is constrained from external fields \citep[see e.g.][]{wwh06,htb06}.
However, as the \textit{Chandra} Deep Field South has been observed with several instruments including infrared observations, accurate photometric redshifts can directly be obtained for a significant fraction of the galaxies without the need for external calibration.
In this work we use the photometric redshift catalogue of the GOODS-MUSIC sample presented by \citet{gfs06}.
This catalogue combines the F435W, F606W, F775W, and F850LP ACS GOODS/CDFS images \citep{gfk04}, the $JHK$s VLT data (Vandame et al. in prep.), the Spitzer data provided by the IRAC instrument at 3.6, 4.5, 5.8, and 8.0 $\mu$m (Dickinson et al. in prep.), and $U$--band data from the MPG/ESO 2.2m and VLT-VIMOS.
Additionally the GOODS-MUSIC catalogue contains spectroscopic data from several surveys \citep{caa00,cwg01,wmr01a,bse03,dsg04,fvp04,sbh04,sbm04,srd04,wfd04,mcz05,vcd05}, which are also compiled in a Master\footnote{\url{http://www.eso.org/science/goods/spectroscopy/CDFS_Mastercat/}} catalogue by the ESO-GOODS team.
We match the GOODS-MUSIC catalogue to our filtered galaxy shear catalogue, yielding in total 8469 galaxies with a photometric redshift estimate, including 408 galaxies with additional spectroscopic redshifts and a redshift quality flag $qz\le 2$.
In the area covered by the GOODS-MUSIC catalogue 95.0\% of the galaxies in our shear catalogue with $m_{606}<26.25$ have a redshift estimate, and only for fainter magnitudes substantial redshift incompleteness occurs (Fig.\thinspace\ref{fi:n_mag_music}).
\citet{gfs06} estimate the photometric redshifts errors from the absolute scatter between photometric and spectroscopic redshifts to be $\langle | \Delta z / (1+z)|\rangle=0.045$.

%galaxies brighter than 26.0: with redshift: 4596 (ratio of all 0.964128) no redshifts: 171
%galaxies brighter than 26.25: with redshift: 5532 (ratio of all 0.950189) no redshifts: 290

In cosmic shear studies the redshift distribution is often parametrised as
\begin{equation}
\label{eq:redshift_distribution}
p(z)=\frac{\beta}{z_0 \Gamma\left(\frac{1+\alpha}{\beta}\right)} \left(\frac{z}{z_0}\right)^\alpha \exp{\left[-\left(\frac{z}{z_0}\right)^\beta\right]} 
\end{equation}
\citep[e.g. ][]{bbs96,smw06,hmw06}.
In order to extrapolate the redshift distribution for the faint and redshift incomplete magnitudes we consider $p(z)=p(z,m_{606})$ and assume a linear relation between the magnitude $m_{606}$ and the median redshift $z_\mathrm{m}$ of an ensemble of galaxies with this magnitude
\begin{equation}
\label{eq:redshift_zm_parametrisation}
z_\mathrm{m}=r z_0=a (m_{606}-22) + b \, ,
\end{equation}
where $r(\alpha,\beta)$ is calculated from numerical integration of (\ref{eq:redshift_distribution}).
For a single galaxy of magnitude $m_{606}$, (\ref{eq:redshift_distribution}) corresponds to the redshift probability distribution given the parameter set $(\alpha,\beta,a,b)$.
Thus, we can constrain these parameters via a maximum likelihood analysis, for which we marginalise over the photometric redshift errors $\Delta z$.
The total redshift distribution of the survey with $N$ galaxies is then constructed as 
\begin{equation}
\label{eq:redshift_total}
\phi(z)=\frac{\sum_{i=1}^{i=N}p(z,m_{606}(i))}{N} \, .
\end{equation}
Note that this approach is similar to the one used by \citetalias{hbb05}, but does not require magnitude or redshift binning.

For the maximum likelihood analysis we apply the CERN Program Library \texttt{MINUIT}\footnote{\url{http://wwwasdoc.web.cern.ch/wwwasdoc/minuit/}} and use all galaxies with redshift estimates in the magnitude range \mbox{$21.75<m_{606}<26.25$}.
Varying all four parameters $(\alpha,\beta,a,b)$ we find the best fitting parameter combination \mbox{$(\alpha,\beta,a,b)=(0.563, 1.716, 0.299, 0.310)$}, for which \mbox{$z_\mathrm{m}=0.7477 z_0$}.
In order to estimate the fit accuracy, we fix $\alpha$ and $\beta$ to the best fitting values and identify the 68\% (95\%) confidence intervals for $a$ and $b$: $a=0.299^{+0.006(0.013)}_{-0.007(0.014)}, \, b=0.310^{+0.018(0.037)}_{-0.017(0.033)}$.

\begin{figure}
   \centering
   \includegraphics[width=7.cm]{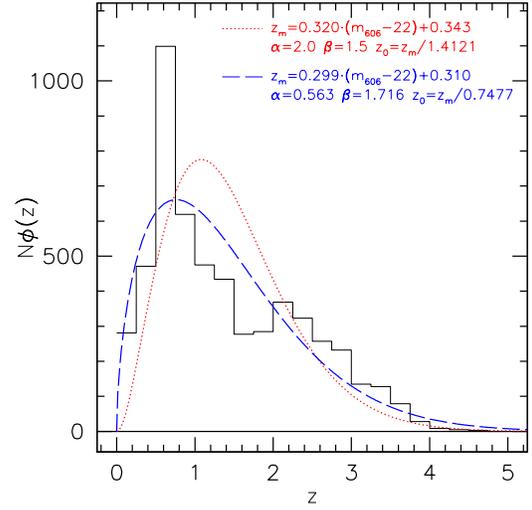} 
   \caption{Redshift distribution for the matched shear catalogue galaxies with redshift estimate from the GOODS-MUSIC sample in the magnitude range \mbox{$21.75<m_{606}<26.25$} (solid line histogram). 
The dashed curve shows the reconstructed redshift distribution $N\phi(z)$ for these galaxies using the best fitting values for \mbox{$(\alpha,\beta,a,b)=(0.563, 1.716, 0.299, 0.310)$}.
The dotted curve was computed for fixed \mbox{$(\alpha, \beta) = (2,1.5)$}.
}
   \label{fi:phi_for_matching}
\end{figure}

Using these parameter estimates, we reconstruct the redshift distribution of the galaxies used for the fitting (Fig.\thinspace\ref{fi:phi_for_matching}).
The reconstruction fits the actual redshift distribution very well except for a prominent galaxy over-density at $z\simeq 0.7$ and an under-density at $z\gtrsim 1.5$, which are known large-scale structure features of the field \citep{gcd03,gdz05,sbh04,fvp04,ami05,vcd06,gfs06}.
Yet, given that the reconstruction and the photometric redshift distribution have almost identical average redshifts \mbox{$\langle z_\mathrm{recon}(\mathrm{fit\,sample})\rangle=1.39$}, \mbox{$\langle z_\mathrm{photo}(\mathrm{fit\, sample})\rangle=1.41$}, we estimate that the impact of the large-scale structure on the cosmic shear estimate via the source redshift distribution will be small.
However, the large-scale structure significantly influences the estimate of the median redshift \mbox{$z_{\mathrm{m},\mathrm{recon}}(\mathrm{fit\, sample})=1.23$}, \mbox{$z_{\mathrm{m},\mathrm{photo}}(\mathrm{fit\, sample})=1.10$}.
Thus, a redshift distribution determined from the computed median redshift of the galaxies would most likely 
be biased to too low redshifts.
Note that in Fig.\thinspace\ref{fi:phi_for_matching} the reconstruction falls off slower for high $z$ than the actual distribution of the data. 
To exclude a possible bias we thus always truncate the high redshift tail for $z>4.5$.

For comparison we also determine a reconstruction from the best fitting values for $(a,b)$ with fixed values $(\alpha,\beta)=(2,1.5)$, which are sometimes used in the literature (e.g. \citealp{bae94}; \citetalias{hbb05}).
While they seem to provide a good parametrisation for shallower surveys \citep[see e.g.][]{btb03},
they lead to a 
distribution that is too narrowly peaked with a maximum at too high redshifts for the deep GEMS and GOODS data (Fig.\thinspace\ref{fi:phi_for_matching}).

\begin{figure}
   \centering
   \includegraphics[width=8.cm]{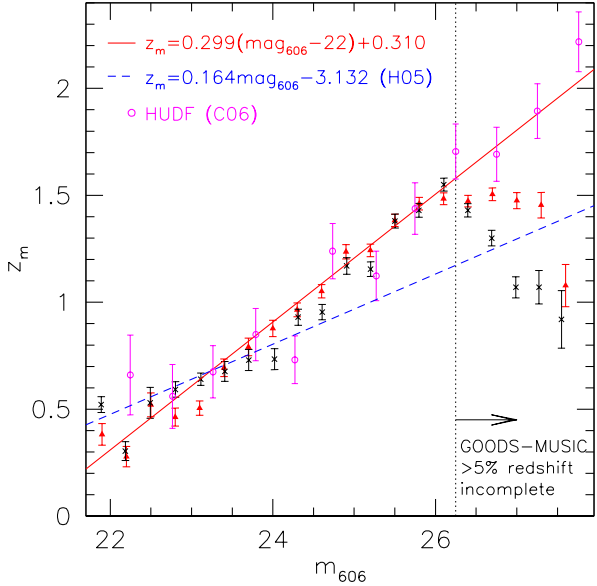} 
   \caption{Median redshift of the matched galaxies in $m_{606}$ bins computed directly from the data (thin crosses) and determined from a maximum likelihood fit for $z_\mathrm{m}$ with fixed \mbox{$(\alpha,\beta)=(0.563, 1.716)$} (triangles), with errors-bars indicating the error of the mean or the $1\sigma$ confidence region, respectively.
The solid line corresponds to the best fitting parameters of the joint likelihood fit, whereas the dashed line shows the fit determined by \citetalias{hbb05} for the magnitude range \mbox{$21.8<m_{606}<24.4$}.
Note that a large-scale structure peak at \mbox{$z\simeq 0.7$} induces both the flatter slope for the directly computed $z_\mathrm{m}$ for \mbox{$m_{606}\lesssim24.7$} and the increased spread for the fitted points for \mbox{$m_{606}\lesssim23.3$}. For \mbox{$m_{606}\gtrsim 26.25$} substantial redshift incompleteness occurs.
For comparison we also plot the directly computed median photometric redshift from the HUDF \citep[open circles,][]{cbs06}.
}
   \label{fi:zm_of_mag}
\end{figure}

A maximum likelihood analysis can only yield reasonable parameter constraints if the model is a good description of the data.
To test our assumption of a linear behaviour in (\ref{eq:redshift_zm_parametrisation}), we bin the matched galaxies in redshift magnitude bins and determine a single $z_\mathrm{m}$ for each bin using an additional likelihood fit with fixed \mbox{$(\alpha,\beta)=(0.563, 1.716)$}, see Fig.\thinspace\ref{fi:zm_of_mag}.
A linear $z_\mathrm{m}(m_{606})$ description is indeed in excellent agreement with the data in the magnitude range used for the joint fit.
Only at the bright end the large-scale structure peak at \mbox{$z\simeq 0.7$} induces an increased scatter.
However, the likelihood fit is much less affected by large-scale structure than the directly computed median redshift, which in contrast under-estimates the slope of the $z_\mathrm{m}(m_{606})$ relation for $z_\mathrm{m}\lesssim24.7$ (see Fig.\thinspace\ref{fi:zm_of_mag}).
This is the reason why \citetalias{hbb05}, who use the median redshift computed from spectroscopic data in the magnitude range \mbox{$21.8<m_{606}<24.4$}, derive a significantly flatter $z_\mathrm{m}(m_{606})$ relation 
\begin{equation}
\label{eq:zm_H05}
z_\mathrm{m}^\mathrm{H05}=-3.132+0.164\, m_{606} \quad (21.8<m_{606}<24.4)
\end{equation}
leading to an estimate of $z_\mathrm{m}=1.0\pm0.1$ for their shear catalogue.

In order to verify the applicability of (\ref{eq:redshift_zm_parametrisation}) for our fainter shear galaxies, we also plot $z_\mathrm{m}(m_{606})$ in Fig.\thinspace\ref{fi:zm_of_mag} computed from photometric redshifts for the HUDF \citep{cbs06}, finding a very good agreement.

Using the parameters $(\alpha,\beta,a,b)$ we construct the redshift distribution for all GEMS and GOODS galaxies in our shear catalogue from (\ref{eq:redshift_total}).
The resulting redshift distribution has a median redshift 
\mbox{$z_\mathrm{m}(\mathrm{GEMS+GOODS})=1.46 \pm 0.02(0.05)$},
 where the statistical errors stem from the uncertainty of $a$ and $b$.
Systematic uncertainties might arise from applying (\ref{eq:redshift_zm_parametrisation}) for galaxies up to 1.5 magnitudes fainter than the magnitude range used to determine the fit.
Additionally, the photometric redshift errors used in the maximum likelihood analysis do not take catastrophic outliers or systematic biases into account, but see \citet{gfs06} for a comparison to the spectroscopic subsample.
Furthermore the impact of the large-scale structure on the source redshift distribution will be slightly different for the whole GEMS field compared to the GOODS region.
We estimate the resulting systematic uncertainty as $\Delta z\simeq 0.1$, yielding 
\mbox{$z_\mathrm{m}(\mathrm{GEMS+GOODS})=1.46 \pm0.02(0.05)\pm0.10$}.
The constructed redshift distribution is well fit with a magnitude independent distribution (\ref{eq:redshift_distribution}) with \mbox{$(\alpha,\beta,z_0)=(0.537,1.454,1.832)$}.

Given that we derive the redshift parametrisation from the matched GOODS-MUSIC galaxies in the magnitude range \mbox{$21.75<m_{606}<26.25$}, while a low level of redshift incompleteness already occurs for \mbox{$m_{606}\gtrsim 25.75$} (see Fig.\thinspace\ref{fi:n_mag_music}), we repeat our analysis as a consistency check using only galaxies with \mbox{$21.75<m_{606}<25.75$} yielding a very similar redshift distribution with \mbox{$z_{m}=1.44$}.
We thus conclude that the low level of incompleteness does not significantly affect our analysis.

For the brighter galaxies in our shear catalogue with \mbox{$\mathrm{S}/\mathrm{N}>5, m_{606}<27.0$}, the constructed redshift distribution is expectedly shallower with \mbox{$z_\mathrm{m}=1.37 \pm0.02(0.05)\pm 0.08$} and can well be fit with a magnitude independent distribution (\ref{eq:redshift_distribution}) with \mbox{$(\alpha,\beta,z_0)=(0.529,1.470,1.717)$}.
Using our redshift parametrisation we also estimate the median redshift for the \citetalias{hbb05} shear catalogue yielding \mbox{$z_\mathrm{m}=1.25 \pm 0.02\pm 0.08$}.
Here we estimate slightly lower systematic errors due to the lesser extrapolation to fainter magnitudes.

In Sect.\thinspace\ref{se:cosmo_para_estimate} we will use our derived redshift distribution to constrain cosmological parameters marginalising over the statistical plus systematic error in $z_\mathrm{m}$.
Furthermore we will use this redshift distribution when we compare cosmic shear estimates for the GEMS and GOODS data with theoretical models.
The theoretical cosmic shear predictions shown in this paper are calculated for a flat $\Lambda$CDM cosmology according to the three year WMAP-only best-fitting values for 
\mbox{$(\Omega_\Lambda,\Omega_\mathrm{m},\Omega_\mathrm{b},h,n_\mathrm{s})=(0.76,0.24,0.042,0.73,0.95)$}
 \citep{sbd06} for different power spectrum normalisations $\sigma_8$ calculated using the non-linear correction to the power spectrum from \citet{spj03}.

At this stage we use the parallel data to test our pipeline and search for remaining systematics,
while presenting a cosmological parameter estimation in a future paper based on a larger data set.
Given the inhomogeneous depth and data quality
of the parallel data, this cosmological parameter estimation will require a thoroughly estimated, field dependent redshift distribution.
For the purpose of comparing the different estimators for shear and systematics to the expected shear signal in the current paper, we apply a simplified global redshift distribution estimated from the F775W magnitudes in GOODS-MUSIC catalogue.
Similarly to the F606W data we apply our likelihood analysis to all GOODS-MUSIC galaxies with \mbox{$22.0< m_{775}<26.0$} yielding best fitting parameters \mbox{$(\alpha,\beta,a,b)=(0.723, 1.402, 0.309, 0.395)$}, for which \mbox{$z_\mathrm{m}=0.9395 z_0$}.
The upper magnitude limit was chosen due to a similar turn-off point of $z_\mathrm{m}(m_{775})$ as in Fig.\thinspace\ref{fi:zm_of_mag} indicating redshift incompleteness.
To account for the different extinction in the parallel fields and the CDFS (\mbox{$A_\mathrm{775}^\mathrm{CDFS}=0.017 \,\mathrm{mag}$}), we apply an extinction correction based on the maps by \citet{sfd98}.

Using the extinction corrected magnitudes of all F775W galaxies in the parallel data shear catalogue, we construct a redshift distribution with 
\mbox{$z_\mathrm{m}=1.34$}, 
which can be fit with a magnitude independent distribution with 
\mbox{$(\alpha,\beta,z_0)=(0.746,1.163,1.191)$}.

   \begin{figure*}[htb]
   \sidecaption
   \centering
   \includegraphics[width=6.1cm]{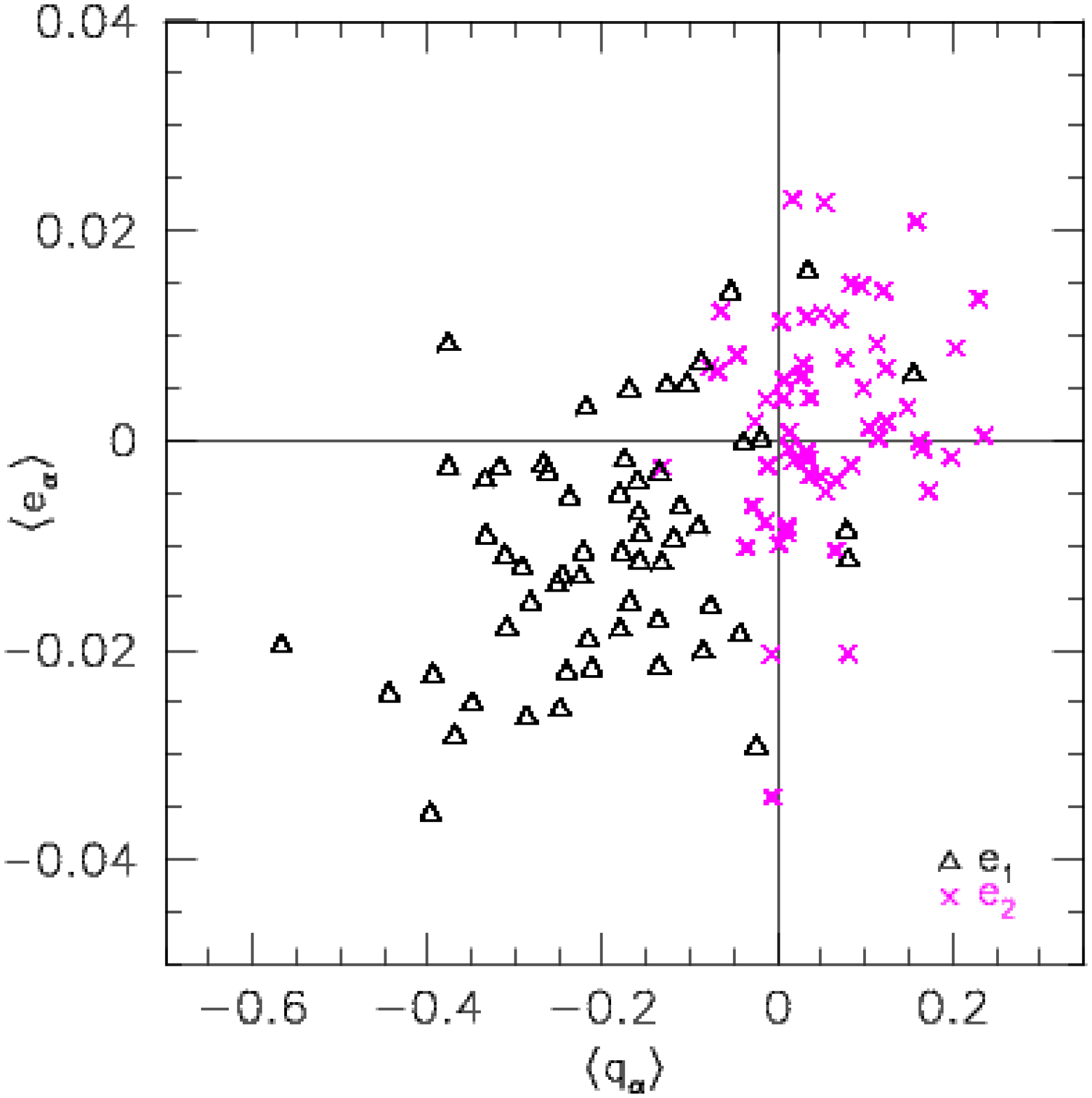}
   \includegraphics[width=6.1cm]{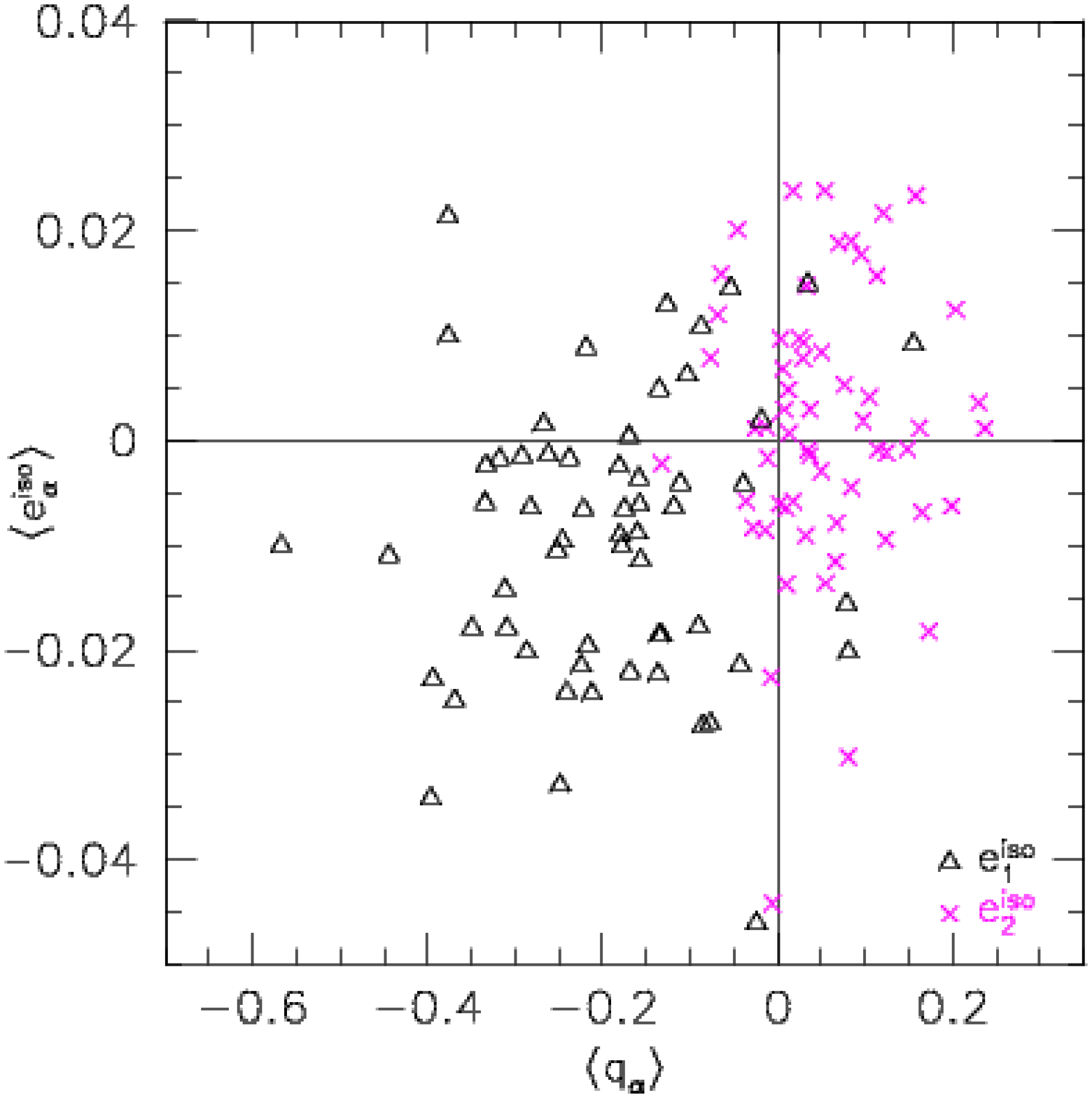}
   \caption{Mean galaxy ellipticity before ($\langle e_\alpha \rangle$, left) and after ($\langle e_\alpha^\mathrm{iso} \rangle$, right) PSF correction as a function of the mean PSF anisotropy kernel averaged over all galaxies in a field $\langle q_\alpha \rangle$, computed on a \textit{field-by-field} basis for the F775W parallel fields. The lack of a correlation after PSF correction (\mbox{$\mathrm{cor}=0.08$}) is a clear indication that PSF anisotropy residuals cannot be the origin for the negative average ellipticity $\langle e_1^\mathrm{iso} \rangle$.}
   \label{fi:gal:average_elli_average_psf}
    \end{figure*}

   \begin{figure*}[htb]
   \sidecaption
   \centering
   \includegraphics[width=6.1cm]{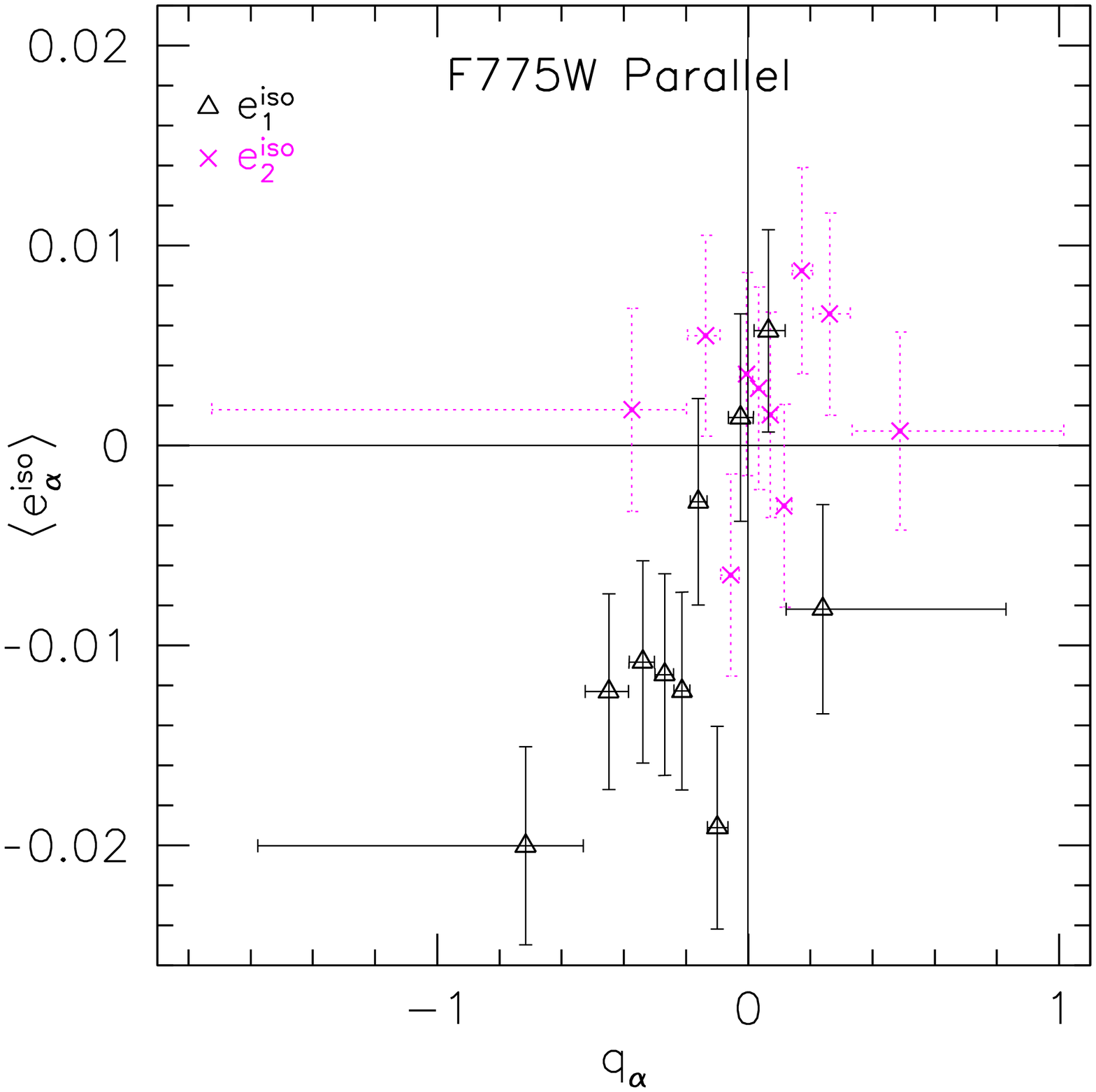}
   \includegraphics[width=6.1cm]{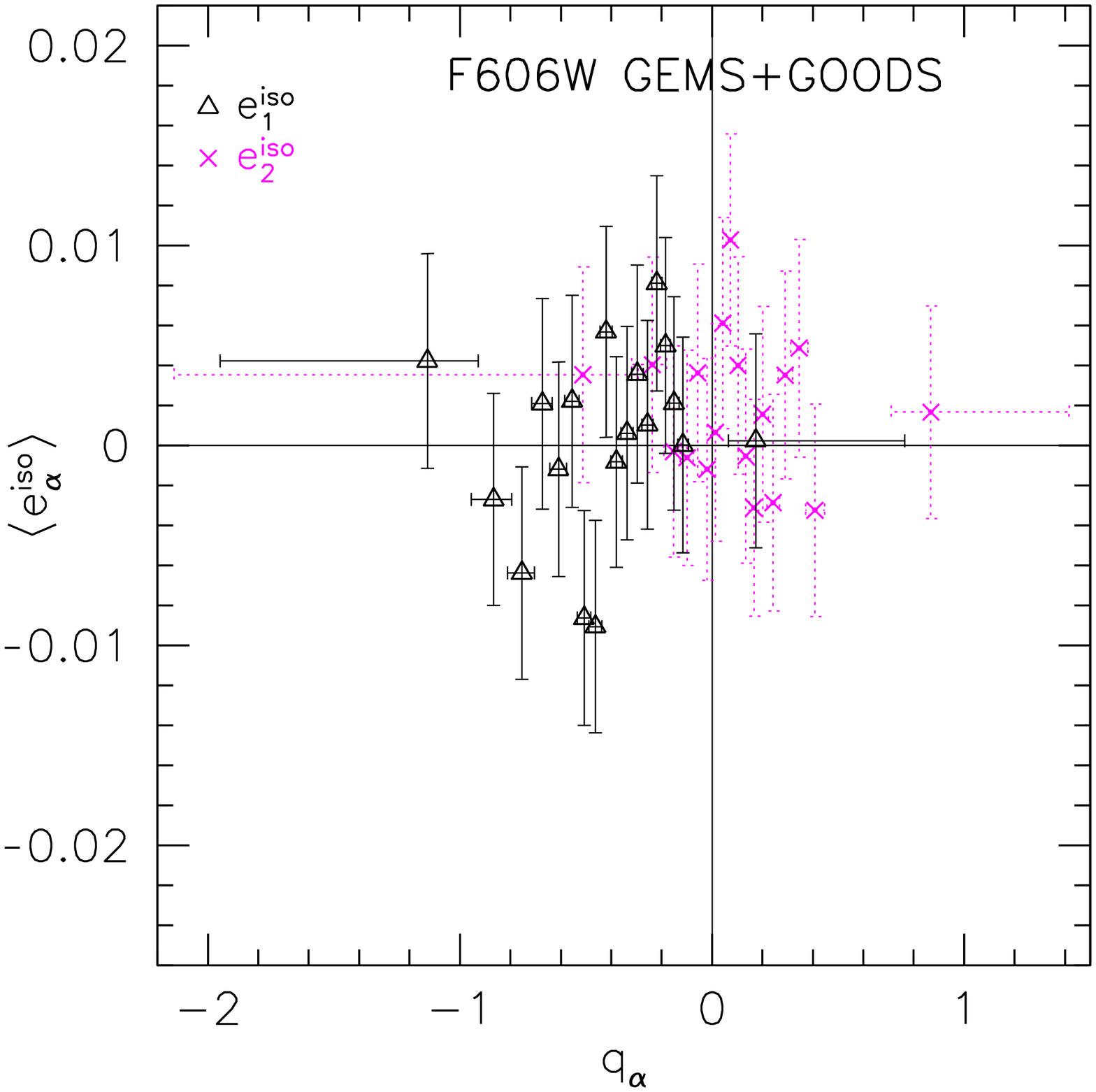}
   \caption{Mean PSF corrected galaxy ellipticity $\langle e_\alpha^\mathrm{iso} \rangle$ binned as a function of the PSF anisotropy kernel $q_\alpha$ for the parallel data (left) and the GEMS+GOODS data (right). The binning (indicated by the horizontal error-bars) was chosen such that all bins contain an equal number of galaxies. The lack of a correlation for the GEMS+GOODS data and $\langle e_2^\mathrm{iso} \rangle$ for the parallel data confirms the success of the PSF correction. The interpretation of the moderate correlation detected for $\langle e_1^\mathrm{iso} \rangle$ in the parallel data is ambiguous as it can also be caused by a position dependence of $\langle e_1^\mathrm{iso} \rangle$.}
   \label{fi:gal:average_elli_fct_q}
    \end{figure*}

\section{Cosmic shear estimates and tests for systematics}
\label{se:preliminary_cosmic_shear}
In this section we compute different cosmic shear statistics and perform a number of diagnostic tests to check for the presence of remaining systematics.
For the GEMS and GOODS data the plots in this section correspond to the larger galaxy set with $\mathrm{S}/\mathrm{N}>4$ including the faint galaxies which are stronger affected by the PSF.

\subsection{Average galaxy ellipticity}
\label{su:average_galaxy_elli}
For data uncontaminated by systematics the average galaxy ellipticity is expected to be consistent with zero. Any significant deviation from zero indicates an average alignment of the galaxies relative to the pixel grid.
We plot the average corrected but not rotated (see Sect.\thinspace\ref{se:galaxy_selection}) galaxy ellipticity $\langle e^\mathrm{iso}_\alpha \rangle$ for each field in Fig.\thinspace\ref{fi:gal:average_elli}.
Whereas the global average is essentially consistent with zero for the GEMS and GOODS data 
(\mbox{$\langle e^\mathrm{iso}_1\rangle=-0.0004\pm0.0011$}, \mbox{$\langle e^\mathrm{iso}_2\rangle=0.0012\pm0.0011$}),
the average $e^\mathrm{iso}_1$-component is significantly negative for the parallel data
(\mbox{$\langle e^\mathrm{iso}_1\rangle=-0.0084\pm0.0015$}, \mbox{$\langle e^\mathrm{iso}_2\rangle=0.0020\pm0.0015$}) 
corresponding to an average orientation in the direction of the $y$-axis.

\subsubsection{Could it be residual PSF contamination?}
There are different effects which could in principle cause such an average alignment:
For example one could speculate that our PSF fitting technique fails for the parallel data or that our implementation of the KSB+ formalism under-estimates the PSF anisotropy correction, e.g. due to neglected higher-order moments.
Yet, the average corrected galaxy ellipticity is consistent with zero for the GEMS and GOODS data, while the average uncorrected ellipticity is significantly non-zero for both datasets (parallel: \mbox{$\langle e_1\rangle=-0.0102\pm0.0012$},  \mbox{$\langle e_2\rangle=0.0028\pm0.0012$}; GEMS+GOODS: \mbox{$\langle e_1\rangle=-0.0090\pm0.0009$}, \mbox{$\langle e_2\rangle=0.0045\pm0.0009$)}.
Therefore this explanation becomes quite implausible, particularly as the average number of stars usable to derive the fit is higher for the parallel data (Fig.\thinspace\ref{fi:nstarsgalfields}).

To further test whether imperfect PSF correction could be the cause, we plot the mean galaxy ellipticity as a function of the mean PSF anisotropy kernel on a \textit{field-by-field} basis for parallel data in Fig.\thinspace\ref{fi:gal:average_elli_average_psf}. 
While there is a substantial correlation between $\langle q_\alpha \rangle$ and the mean uncorrected ellipticity $\langle e_\alpha \rangle$ (correlation \mbox{$\mathrm{cor}=\mathrm{cov}[\langle q_\alpha \rangle, \langle e_\alpha \rangle]/(\sigma_{\langle q_\alpha \rangle} \sigma_{\langle e_\alpha \rangle})=0.38$}), the mean PSF corrected ellipticity $\langle e_\alpha^\mathrm{iso} \rangle$ is basically uncorrelated with $\langle q_\alpha \rangle$ (\mbox{$\mathrm{cor}=0.08$}), clearly indicating that imperfect PSF correction is not the culprit here.

We also plot the mean corrected galaxy ellipticity $\langle e_\alpha^\mathrm{iso} \rangle$ computed in $q_\alpha$-bins in Fig.\thinspace\ref{fi:gal:average_elli_fct_q}.
The absence of a correlation both for the GEMS+GOODS data and additionally $\langle e_2^\mathrm{iso}\rangle$ in the parallel data again confirms the success of the PSF correction.
For the parallel data a moderate correlation is observed between $\langle e_1^\mathrm{iso} \rangle$ and $q_1$, which at first sight might be interpreted as an indication for imperfect PSF anisotropy correction.
However, it is important to keep in mind that $q_\alpha$ is position dependent.
Hence, if a different position dependent effect causes the non-zero $\langle e_1^\mathrm{iso}\rangle$ it will also mimic a dependence on $q_\alpha$.
From Fig.\thinspace\ref{fi:psf:short_term_variation} we find for example that highly negative values for $q_1$ appear mainly near medial $y$-positions close to the gap between the two chips.
Thus, the apparent correlation between $\langle e_1^\mathrm{iso} \rangle$ and $q_1$ shown in Fig.\thinspace\ref{fi:gal:average_elli_fct_q} could also be caused by a different effect which acts most strongly near the chip gap, such as CTE degradation (see Sect.\thinspace\ref{se:impact_cte}) or artefacts due to bad columns (see Sect.\thinspace\ref{su:dithering_impact}).
In this sense the \textit{field-by-field} comparison shown in Fig.\thinspace\ref{fi:gal:average_elli_average_psf} is a better test for imperfect PSF anisotropy correction, as it is independent of a possible position dependence.
Given the fact that this test does not show a significant indication for imperfect PSF anisotropy correction, we conclude that it is most likely not the explanation for the non-zero $\langle e_1^\mathrm{iso} \rangle$.
We investigate the position dependence further in Sect.\thinspace\ref{se:impact_cte} and compute the star-galaxy cross-correlation as an additional test for PSF anisotropy residuals in Sect.\thinspace\ref{se:star_gal_cross}.

   \begin{figure}
   \centering
   \includegraphics[width=6.5cm]{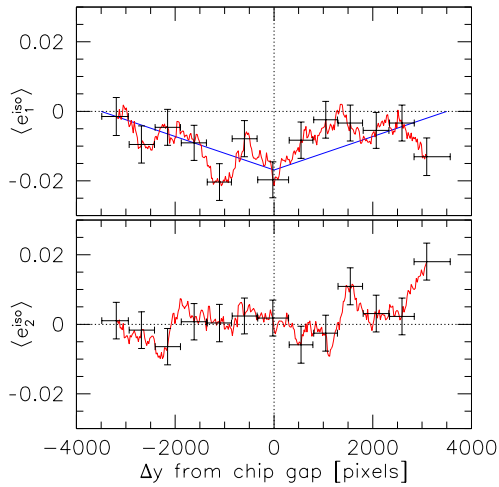}
 
   \caption{Average corrected galaxy ellipticity $\langle e^\mathrm{iso}_\alpha \rangle$ for the parallel F775W galaxy fields as a function of $\Delta y$, the $y$-position relative to the gap between the two camera chips.
The curve shows $\langle e^\mathrm{iso}_\alpha \rangle (\Delta y)$ box-averaged over 3000 galaxies. For certain $\Delta y$ the error-bars indicate the width of the averaging in $\Delta y$ and the error of the estimate. 
The straight lines indicate the expected dependence if the negative $\langle e^\mathrm{iso}_1 \rangle$ was purely caused by CTE degradation assuming a linear dependence of the mean ellipticity on the CTE charge loss. 
}
   \label{fi:average_e1iso_deltay}
    \end{figure}
   \begin{figure}
   \centering
   \includegraphics[width=6.5cm]{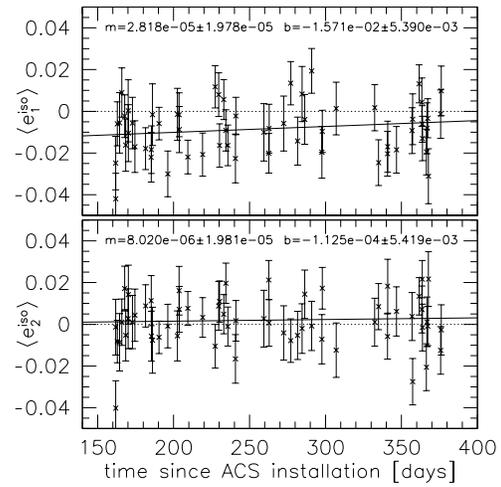}
 
   \caption{Average corrected galaxy ellipticity $\langle e^\mathrm{iso}_\alpha \rangle$ for the parallel F775W galaxy fields as a function of time since the installation of ACS on March 7, 2002.
The solid line shows a linear fit. If the negative $\langle e^\mathrm{iso}_1\rangle$ would be created by degradation of CTE an increase of the effect with time would be expected, which is not supported by the data.
}
   \label{fi:average_e1iso_time}
    \end{figure}

\subsubsection{Impact of CTE degradation}
\label{se:impact_cte}
Another possible explanation is a degradation of the charge-transfer efficiency (CTE) due to charge traps created by the continuous cosmic ray bombardment.
The ACS/WFC has two read-out amplifiers per chip, which are located in the four corners of the instrument.
Of major concern is the degradation of the parallel CTE causing charge trails behind objects in the readout direction, which also in the drizzled images is approximately parallel to the $y-$direction.
These charge trails lead to an average alignment of objects in the $y-$direction, corresponding to a negative average $e_1$ ellipticity component. 
As the depth of charge traps is limited, faint objects loose a larger fraction of their charges than bright ones, leading to a strong signal-to-noise dependence of the effect. Therefore the PSF correction estimated from high signal-to-noise stars does not provide a sufficient CTE correction for faint galaxies.

For a uniform distribution of charge traps the impact of CTE degradation depends linearly on the number of parallel transfers, so that
objects located near the gap between the two chips will be affected the most. 
\citet{mus05} find no significant difference in the parallel CTE for the two chips, indicating that also any impact on the weak lensing measurement should be symmetric between the two chips.
In Fig.\thinspace\ref{fi:average_e1iso_deltay} we plot $\langle e^\mathrm{iso}\rangle$ as a function of $\Delta y$, the $y$-position relative to the gap between the two camera chips. 
Although for the lower chip ($\Delta y<0$) $\langle e^\mathrm{iso}\rangle(\Delta y)$ roughly agrees with the linear trend expected for a CTE degradation, there are significant deviations for the upper chip ($\Delta y>0$).

   \begin{figure*}[htb]
   \sidecaption
   \centering
   \includegraphics[width=6.1cm]{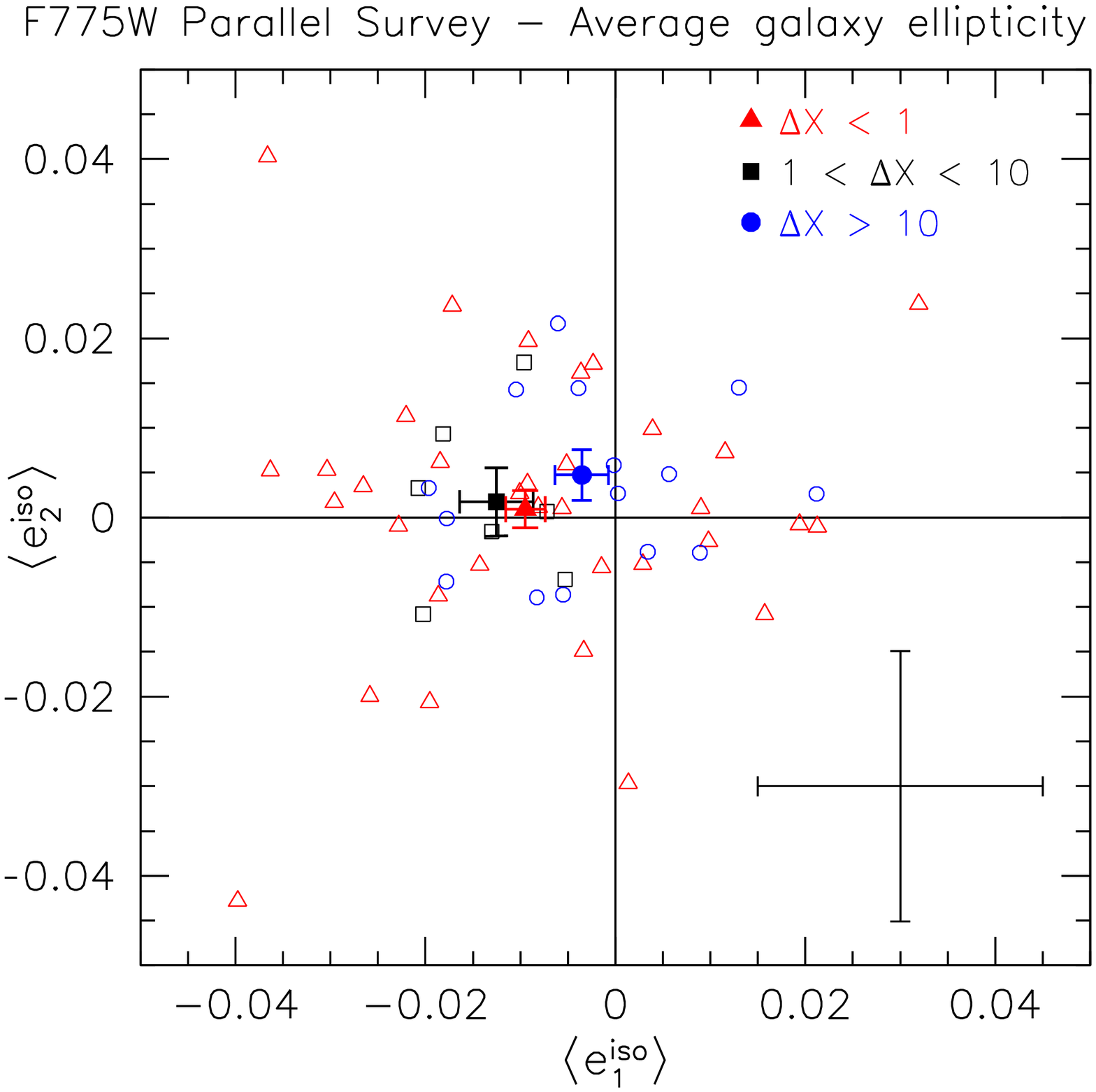}
   \includegraphics[width=6.1cm]{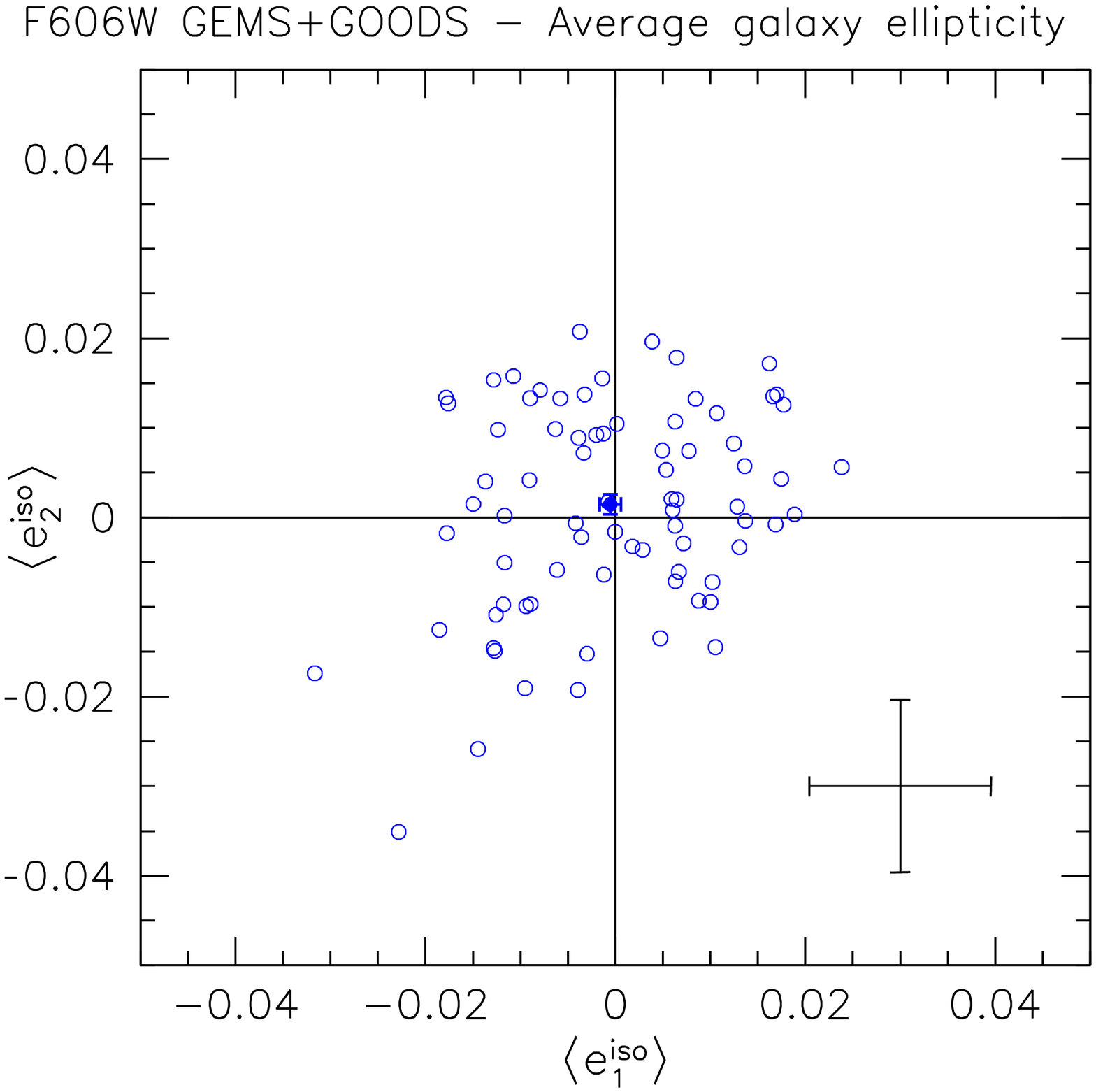}
   \caption{Average corrected galaxy ellipticity $e^\mathrm{iso}_\alpha$ for the parallel data (left) and the GEMS+GOODS data (right). The open symbols represent single field averages, whereas the bold symbols with error-bars ($1\sigma$) show global averages. The large error-bars in the lower right corner correspond to the average single field $1\sigma$ error, where the GEMS+GOODS error-bar is smaller compared to the parallel data error-bar due to the higher average galaxy number density (Fig.\thinspace\ref{fi:galaxy_N_exptime}). The parallel data was split according to the maximal dither between the exposures in the $x-$direction $\Delta X$ [pixels], as the $x-$dithering determines the possibilities to correct for bad columns.}
   \label{fi:gal:average_elli}
    \end{figure*}

Furthermore the ACS/WFC CTE decreases nearly linearly with time \citep{rim04,rie04,mus05} so that one would also expect a linear decrease of $\langle e^\mathrm{iso}_1\rangle$ with time, which is not in agreement with the data (Fig.\thinspace\ref{fi:average_e1iso_time}).
In addition, again, the discrepancy does not occur for the GEMS and GOODS data, which were taken nearly in the same time period as the parallel data.
We thus conclude that CTE degradation is not the dominant source for the observed negative $\langle e^\mathrm{iso}_1\rangle$.
Note that \citet{rma05,rma07} detect discrepancies between their focus-dependent \texttt{TINYTIM} model and stars in the COSMOS data, which they interpret to likely be caused by CTE degradation.
This is not in contradiction to our results, as the COSMOS data were taken at later epochs with significantly increased CTE degradation.

As a further test we also split the data shown in Figures\thinspace\ref{fi:average_e1iso_deltay} and \ref{fi:average_e1iso_time} into a low and a high signal-to-noise subset.
Here the observed dependencies are qualitatively unchanged, but at a lower significance, with a slightly larger absolute values of the negative $\langle e^\mathrm{iso}_1\rangle$ for the fainter sample: \mbox{$\langle e^\mathrm{iso}_1\rangle (\mathrm{S}/\mathrm{N}<7.5) = -0.0092\pm 0.0022$}, \mbox{$\langle e^\mathrm{iso}_1\rangle (\mathrm{S}/\mathrm{N}>7.5) = -0.0077\pm 0.0018$}.
If the effect was caused by CTE, one would probably expect a stronger dependence on the signal-to-noise ratio.

\subsubsection{Impact of dithering}
\label{su:dithering_impact}
In order to understand the origin of the negative $\langle e^\mathrm{iso}_1\rangle$ for the F775W parallel data it is helpful to consider the differences between the two surveys, as the problem does not occur for the F606W GEMS and GOODS images.
Besides the different filters and more homogeneous depth of the GEMS and GOODS tiles there are only two effects which can significantly affect the image quality:
Firstly the F775W fields are taken in parallel in contrast to the F606W data.
Although this could have some impact on the image PSF (Sect.\thinspace\ref{se:acs_parallel}), it is taken into account in our PSF correction scheme (Sect.\thinspace\ref{se:psf_correction_templates}).
Secondly the GEMS and GOODS data are well dithered, whereas most of the parallel fields were observed with no or only small dithering as defined by the primary observations.
To test the impact on the galaxy shape measurement we split the parallel fields in Fig.\thinspace\ref{fi:gal:average_elli} into three sets according to the maximal shift between the exposures in $x-$direction $\Delta X$.
Indeed $\langle e^\mathrm{iso}_1\rangle$ is almost consistent with zero for the well dithered fields with $\Delta X>10$ pixels (6 WFC pixels), whereas it is significantly negative for the less dithered fields.

Proper dithering is important to correct for bad or hot pixels, which otherwise create artifacts in the co-added frame.
Without dithering known bad pixels lead to output pixels receiving zero weight, which we set to zero pixel value, while unknown bad pixels such as spontaneously hot pixels or variable bias structures directly contribute with their bad pixel value. 

Bad pixels are not completely randomly distributed on the CCD chips, but sometimes occur as bad columns or clusters of bad pixels, which are preferentially aligned in the readout direction and therefore the $y-$direction. 
Thus, without proper dithering the shapes of faint objects containing bad columns or pixel clusters could possibly be influenced 
such that a slight average alignment in the $y$-direction is created and a negative $\langle e^\mathrm{iso}_1\rangle$ is measured.
We expect that faint galaxies are stronger, and due to their size more likely, affected than compact high signal-to-noise stars, which additionally might be rejected as noisy outliers during the PSF fitting, explaining why this effect is not taken into account by the PSF correction.

We try to minimise the impact of known bad pixels by rejecting galaxies containing low weight pixels within their \texttt{SExtractor} isophotal area (see Sect.\thinspace\ref{se:catalog_creation}).
However, also a bad column located near the edge of a galaxy image might bias the shape estimate without being rejected in this way.
Note that bad column segments appear with a higher density near the chip gap, which might qualitatively explain the $\Delta y$ dependence plotted in Fig.\thinspace\ref{fi:average_e1iso_deltay}.

Although the comparison shown in Fig.\thinspace\ref{fi:gal:average_elli} supports our interpretation that the negative $\langle e^\mathrm{iso}_1\rangle$ is caused by a lack of dithering, we will need to further investigate this effect on the basis of the complete ACS Parallel Survey for a final judgement, as it extends over a much larger time span allowing a clearer distinction from CTE effects.

So far we co-add parallel data observed within one visit to maximise the stability of the image conditions.
Due to the successful PSF correction for the two-epoch GOODS data (see also Sections \ref{se:star_gal_cross} and \ref{se:eb_measure}) we are confident that a combination of different visits will also be possible for parallel data with re-observations, which will reduce the number of fields with poor dithering. 
Additionally we are working on an improved search algorithm for galaxies which are affected by bad columns.

\subsection{Error estimates}
In the following subsections we compute several estimators for the cosmic shear signal and remaining systematics.
The statistical errors of these estimates are always computed in a similar way.
\subsubsection{Parallel data}
\paragraph{Bootstrapping on galaxy basis.}
To derive statistical weights for $\xi_\pm$ and $\langle M_\mathrm{ap}^2\rangle$, we generate for each field $i$ 200 bootstrap samples of the galaxy catalogue and compute $\xi_{\pm,ij}$ and $\langle M_\mathrm{ap}^2\rangle_{ij}$ for each angular bin $j$.
The weight $w_{ij}$ for this field and bin is then given as the inverse bootstrapping variance $w_{ij}=1/\sigma_{ij}^2$, yielding the combined estimates
\begin{equation}
\label{eq:combined_xi_map}
\xi_{\pm,j}=\frac{\sum_{i=1}^{N_{fields}} \xi_{\pm,ij} w_{ij}}{\sum_{i=1}^{N_{fields}} w_{ij}} \, , \quad
\langle M_\mathrm{ap}^2\rangle_{j}=\frac{\sum_{i=1}^{N_{fields}} \langle M_\mathrm{ap}^2\rangle_{ij} w_{ij}}{\sum_{i=1}^{N_{fields}} w_{ij}}\,.
\end{equation}
The estimate for the galaxy-star cross-correlation (see Sect.\thinspace\ref{se:star_gal_cross}) is calculated accordingly, with bootstrapping of the galaxy catalogue and a fixed stellar catalogue.

\paragraph{Bootstrapping on field basis.}
We determine the measurement error of the field combined estimates for $\xi_{\pm,j}$ and $\langle M_\mathrm{ap}^2\rangle_{j}$ from 300 bootstrap samples of our fields, combining the estimates for each realisation according to (\ref{eq:combined_xi_map}).
The error of the combined signal in each angular bin $j$ is then given by the bootstrap variance $\sigma_j^2$.
This error estimate accounts both for the shape noise and cosmic variance.

\subsubsection{GEMS and GOODS}
\paragraph{Bootstrapping on galaxy basis.}
For the combined GEMS and GOODS mosaic catalogue we analogously perform bootstrapping on galaxy basis to derive the shape noise error.
The errors plotted for the galaxy-star cross-correlation and the E-/B-mode decomposition within Sections\thinspace\ref{se:star_gal_cross} and \ref{se:eb_measure} correspond to this bootstrap variance.

For the cosmological parameter estimation in Sect.\thinspace\ref{se:cosmo_para_estimate}
 covariances are required, which additionally take sampling variance into account.
We compare covariances estimated directly from the data using a jackknife method with estimates from Gaussian realisations of the cosmic shear field.

\paragraph{Jackknife method.}
We use the modified jackknife method applied by \citetalias{hbb05}
to estimate the covariance matrix of the cosmic shear estimators.
In contrast to the bootstrapping on galaxy basis, the jackknife method applied includes an estimate for small-scale cosmic variance.
However, it must under-estimate cosmic variance on scales of the order of and larger than the field size.
We describe the algorithm in terms of the correlation functions $\xi_\pm$:
We first compute the correlation function $\xi_{\pm,j}$ in the angular bin $j$ from the complete galaxy catalogue. 
Next, we divide the whole survey into $N$ separate sub-regions on the sky, where for convenience we use the $N=78$ individual ACS tiles.
Then, the correlation function $\xi_{\pm,ij}$ is computed omitting the $i$-th subregion for $i=1,...,N$.
With
\begin{equation}
  \xi_{\pm,ij}^* = N \xi_{\pm,j} - (N-1) \xi_{\pm,ij} \, ,
\end{equation}
the jackknife estimate for $\xi_{\pm,j}$ is given by the average \mbox{$\hat{\xi}_{\pm,j}=\langle \xi_{\pm,ij}^* \rangle$}, 
and the jackknife estimate of the covariance between bins $j$ and $k$ can be computed as
\begin{equation}
\langle \Delta \xi_{\pm,j}  \Delta \xi_{\pm,k} \rangle = \frac{1}{N(N-1)} \sum_{i=1}^{i=N} 
\left( \xi_{\pm,ij}^* - \hat{\xi}_{\pm,j} \right)
\left( \xi_{\pm,ik}^* - \hat{\xi}_{\pm,k} \right) \, .
\end{equation}
Note that this jackknife method is expected to slightly underestimate the error even on scales much smaller than the field size due to the mixing of power between different scales in the non-linear regime.

\paragraph{Sampling variance from Gaussian random fields.}
Given that the GEMS and GOODS mosaic samples only one particular field in the sky, the large scale sampling variance errors cannot be determined from the data itself.
In order to derive a theoretical error estimate we have created 2000 $1^\circ\times 1^\circ$ Gaussian realisations of the shear field for a $\Lambda$CDM cosmology with $\sigma_8=0.7$ and the GEMS redshift distribution,
which we populate with $96$ galaxies $\mathrm{arcmin}^{-2}$ with ellipticities randomly drawn from our shear catalogue.
We then select a \mbox{$\sim 28\arcmin \times 28\arcmin$} subregion representing the actual masked geometry of the mosaic. 
From the sheared ellipticities we then compute the covariance matrix of the correlation functions from the different realisations \citep[see][]{sks04}.
This provides us with a robust estimate of the error covariance in the Gaussian limit also including the shape and shot noise contribution.
Note, however, that the Gaussian assumption strongly under-estimates the sampling variance for $\theta \lesssim 10^\prime$ 
\citep{kis05,swh07}, which we further discuss in Sect.\thinspace\ref{se:cosmo_para_estimate}.

\subsection{Star-galaxy cross-correlation}
\label{se:star_gal_cross}
An important diagnostic test for the effectiveness of the PSF anisotropy correction is given by the cross-correlation between uncorrected stellar ellipticities $e^*$ and corrected galaxy ellipticities $\gamma$,
which can be used as an estimate for residual PSF contamination.
Following \citet{bmr03} we compute
\begin{equation}
C^\mathrm{sys}(\theta)=\frac{\langle \gamma e^* \rangle (\theta)| \langle \gamma e^* \rangle (\theta)|}{ \langle e^* e^* \rangle(\theta)} \, .
\end{equation}
For the parallel data we substitute $e^*$ with the smearing corrected PSF model ellipticity
\begin{equation}
\label{eq:e_s_mod}
e^{*}_{\mathrm{mod},\alpha} \equiv  \frac{2 c_\mathrm{cal}}{\mathrm{Tr}P^g_{\mathrm{gal}}} P^{\mathrm{sm}}_{\alpha\beta,\mathrm{gal}} \, q_{\beta,\mathrm{total}}^{\mathrm{DRZ}}(x, y, r_\mathrm{g}) \, ,
\end{equation}
(see Eq.\thinspace\ref{eq:e_star_galaxy_elli}),
at all galaxy positions, which is necessary as $\langle e^* e^* \rangle$ is very noisy and undetermined in many bins due to the few stars present in most of the single parallel fields.
   \begin{figure*}[htb]
   \centering
   \includegraphics[width=7.5cm]{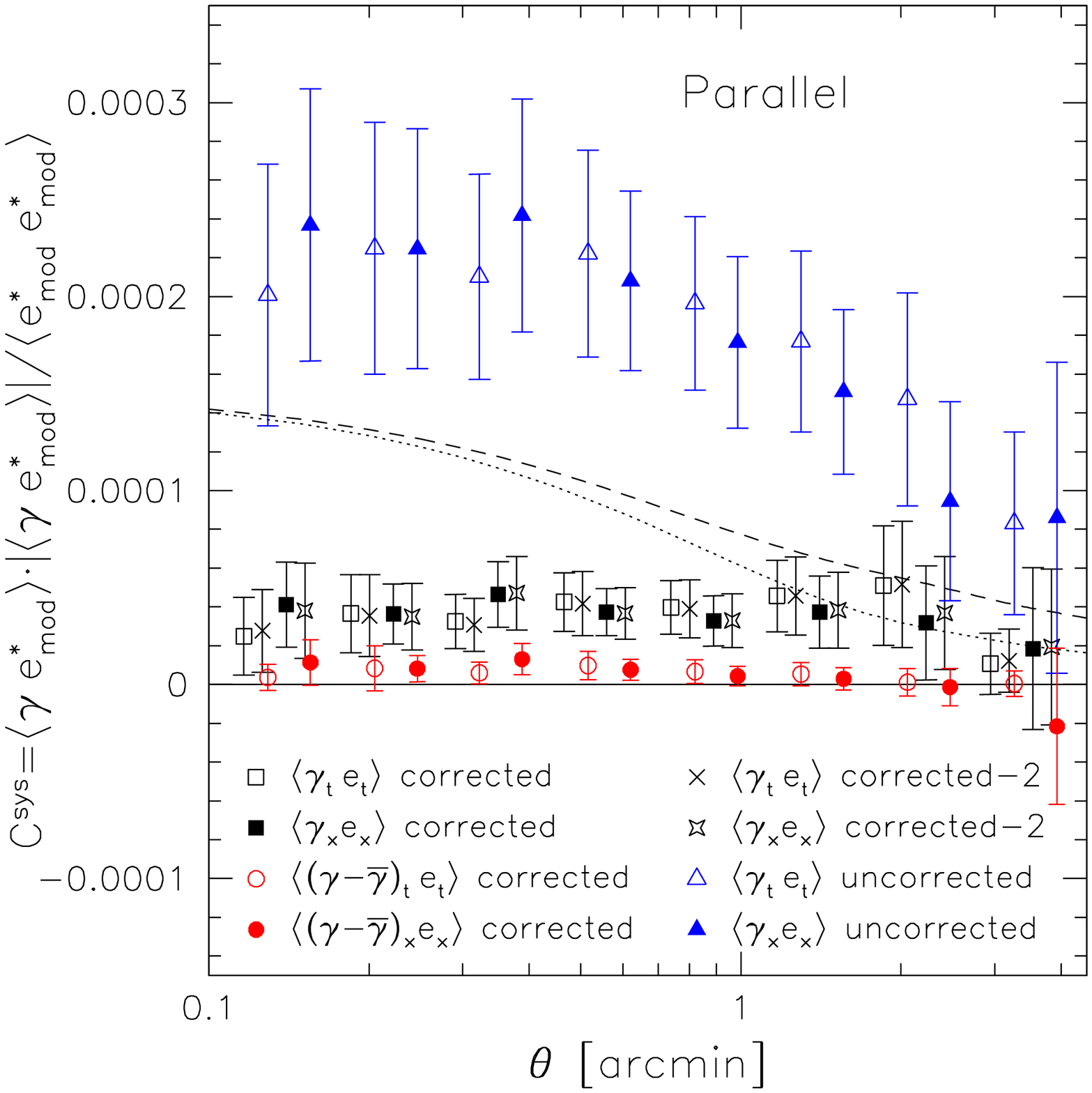}
   \includegraphics[width=7.5cm]{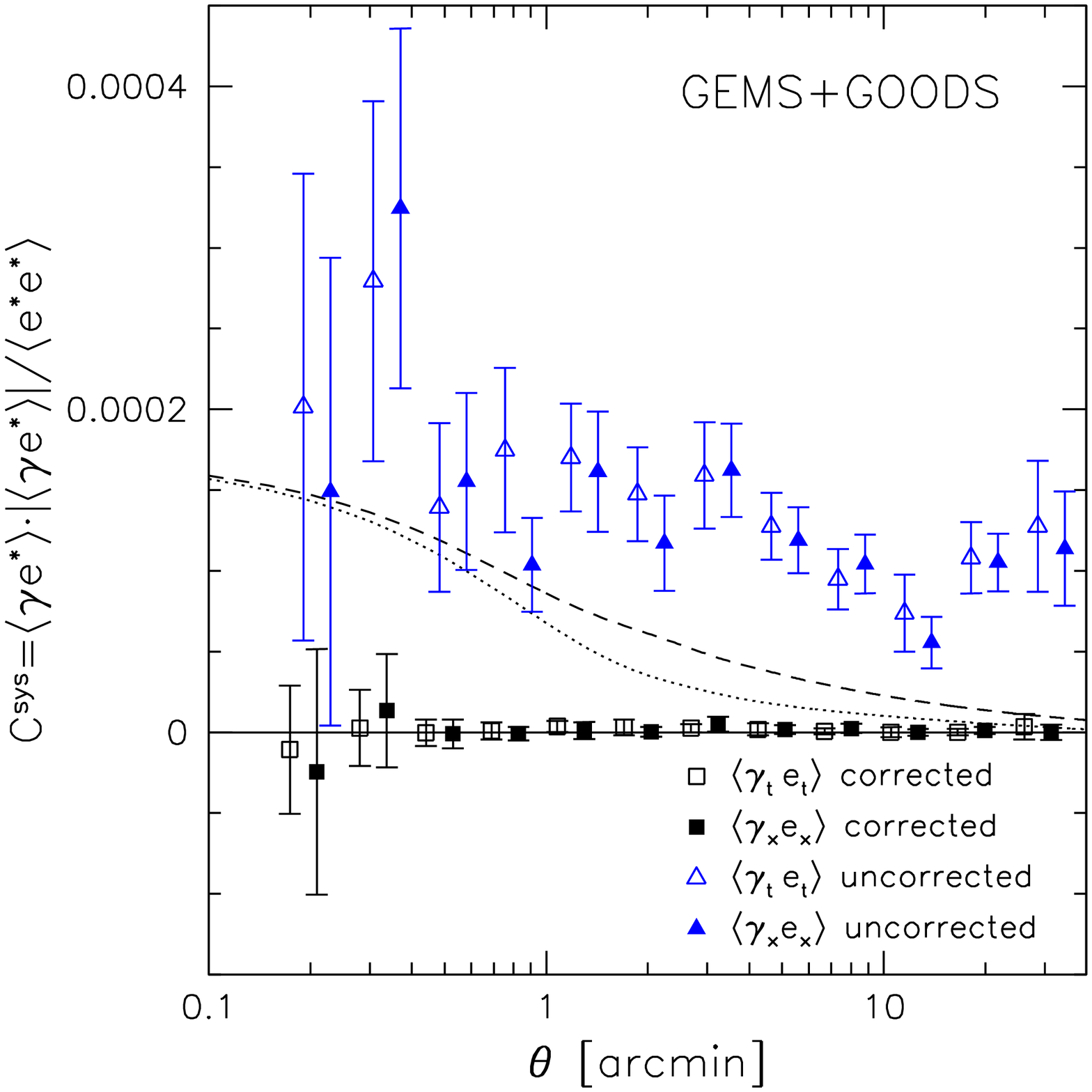}
   \caption{Star-galaxy cross-correlation $C^\mathrm{sys}$ for the parallel data (left) and the GEMS and GOODS data (right), where $C^\mathrm{sys}$ is calculated from the uncorrected stellar ellipticities $e^*$ for the GEMS and GOODS data and the PSF anisotropy model $e^*_\mathrm{mod}$ for the parallel data.
The squares show $C^\mathrm{sys}$ computed from the corrected galaxy ellipticities.
For the parallel data this can be compared to the crosses (stars), where the PSF correction was derived using the second-best fit PSF models. The negligible difference between the two indicates that the F775W stellar field exposures sample the PSF variations sufficiently well.
For comparison we also plot $C^\mathrm{sys}$ determined from the smearing but not anisotropy corrected galaxy ellipticities (triangles), and in case of the parallel data also computed from corrected galaxy ellipticities after subtraction of the mean corrected ellipticity (circles).
The different data sets are displayed with different $\theta$-offsets for clarity.
The dashed (dotted) line shows $\Lambda$CDM predictions for $\langle \gamma_t \gamma_t \rangle$ ($\langle \gamma_\times \gamma_\times \rangle$) for $\sigma_8=0.7$.}

   \label{fi:gal_star_crosscor}
    \end{figure*}

As can be seen from Fig.\thinspace\ref{fi:gal_star_crosscor}, $C^\mathrm{sys}$ is consistent with zero for the GEMS and GOODS data for all $\theta$ indicating that the PSF correction works very well for this dataset.
For comparison we also plot $C^\mathrm{sys}$ computed from the smearing but not anisotropy corrected galaxy ellipticities, which exceeds the theoretically expected cosmic shear signal, emphasising the need for proper PSF correction.

In contrast, $C^\mathrm{sys}$ is non-zero for the parallel data for most $\theta$.
Considering the results from Sect.\thinspace\ref{su:dithering_impact} we interpret this remaining systematic signal as cross-correlation between the (average) PSF pattern and the mean ellipticity component induced by the lack of dithering. 
This interpretation is supported by the fact that $C^\mathrm{sys}$ is almost consistent with zero when computed from the corrected galaxy ellipticities minus the mean ellipticity (Fig.\thinspace\ref{fi:gal_star_crosscor}), suggesting that the PSF correction also performs well for the parallel data.

The underlying assumption of our PSF correction algorithm is that the stellar fields sample the parameter space of PSF variations in the galaxy fields sufficiently well (see Sect.\thinspace\ref{se:discussion_psfalgorithm}).
To test this assumption we repeat the analysis always using the \textit{second}-best fit PSF model instead of the best fitting model.
If the sampling of the PSF variations was not sufficient, we would expect a significant impact on the PSF corrected ellipticities and particularly $C^\mathrm{sys}$ when switching to the  \textit{second}-best fit PSF model.
However, as the observed impact is negligible both for $C^\mathrm{sys}$ (left panel of Fig.\thinspace\ref{fi:gal_star_crosscor}) and the mean corrected galaxy ellipticity (\mbox{$\langle e^\mathrm{iso,mod2}_1\rangle=-0.0085\pm0.0014$}, \mbox{$\langle e^\mathrm{iso,mod2}_2\rangle=0.0018\pm0.0014$}, compare to Sect.\thinspace\ref{su:average_galaxy_elli}), the sampling of the PSF parameter space indeed seems to suffice.

\subsection{E-/B-mode decomposition}
\label{se:eb_measure}
   \begin{figure*}[htb]
   \centering
   \includegraphics[width=8cm]{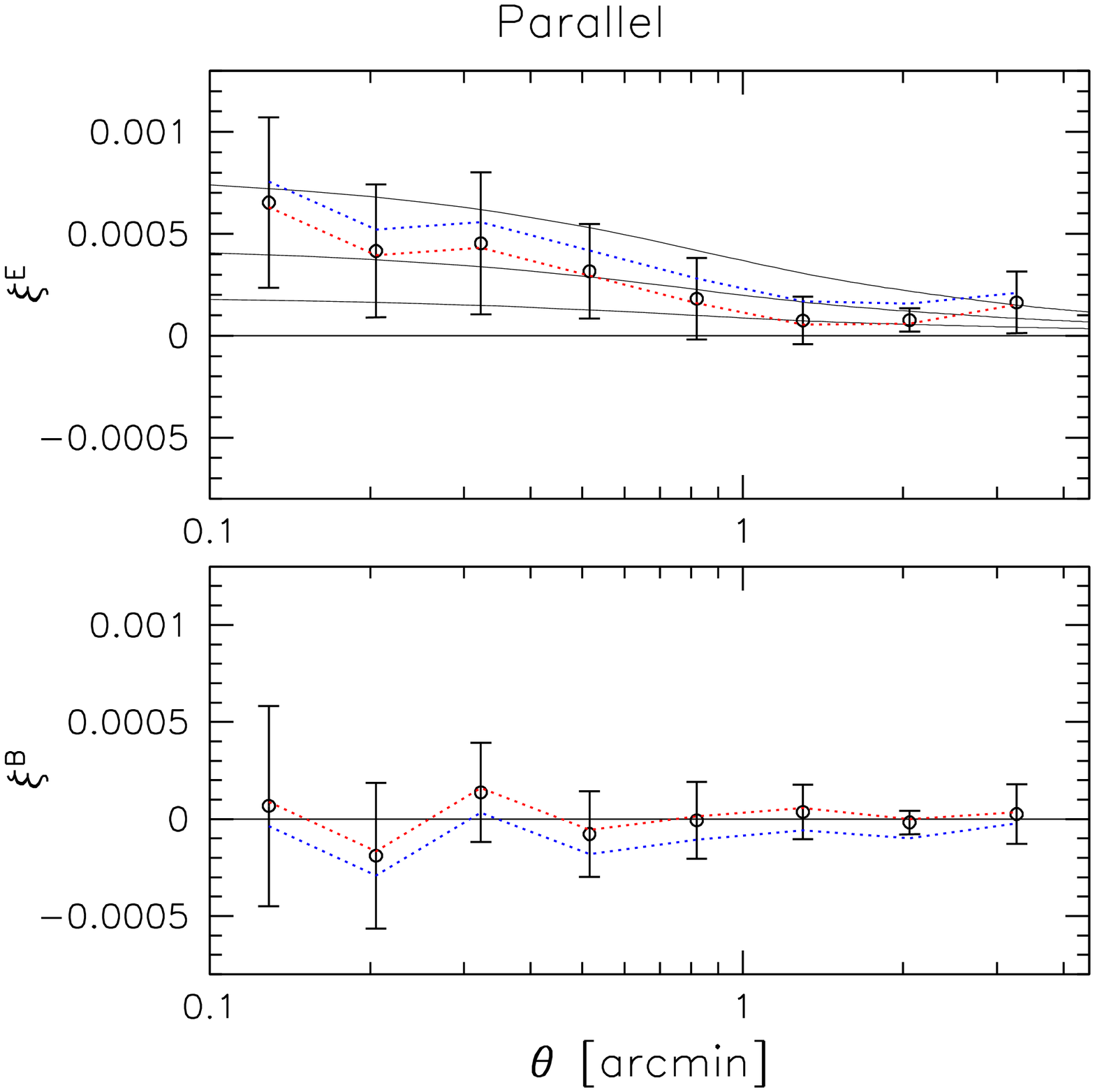}
   \includegraphics[width=8cm]{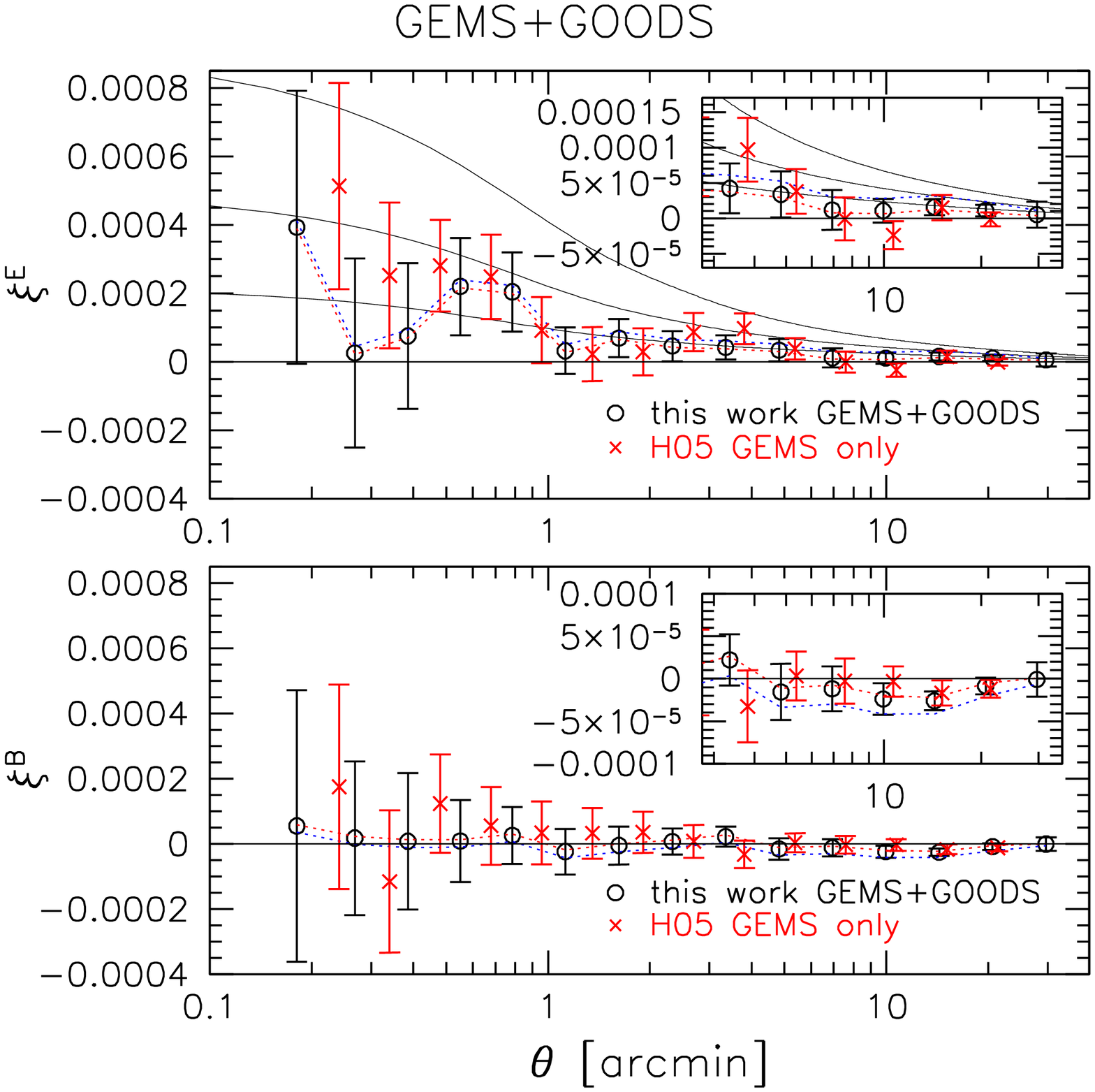}
   \caption{E-/B-mode decomposition of the correlation functions for the parallel data (left) and the combined GEMS and GOODS data (right). The open circles show $\xi^E$ and  $\xi^B$ computed using a fiducial $\Lambda$CDM model with \mbox{$\sigma_8=0.7$} for the extrapolation in Eq.\thinspace(\ref{eq:xi_prime}), whereas the dotted lines correspond to \mbox{$\sigma_8=1.0$} (upper line for $\xi^E$, lower line for $\xi^B$) and \mbox{$\sigma_8=0.6$} (lower line for $\xi^E$, upper line for $\xi^B$). 
The thin solid lines show $\Lambda$CDM predictions for \mbox{$\sigma_8=(0.6, 0.8, 1.0)$}.
In the right panels we also plot the \citetalias{hbb05} GEMS only estimate for $\xi^E$ and $\xi^B$ for \mbox{$\sigma_8=0.7$} (crosses). Note that the \citetalias{hbb05} catalogue is slightly shallower.
}
   \label{fi:xi_eb}
    \end{figure*}

  \begin{figure*}[htb]
   \sidecaption
   \centering
   \includegraphics[width=6.6cm]{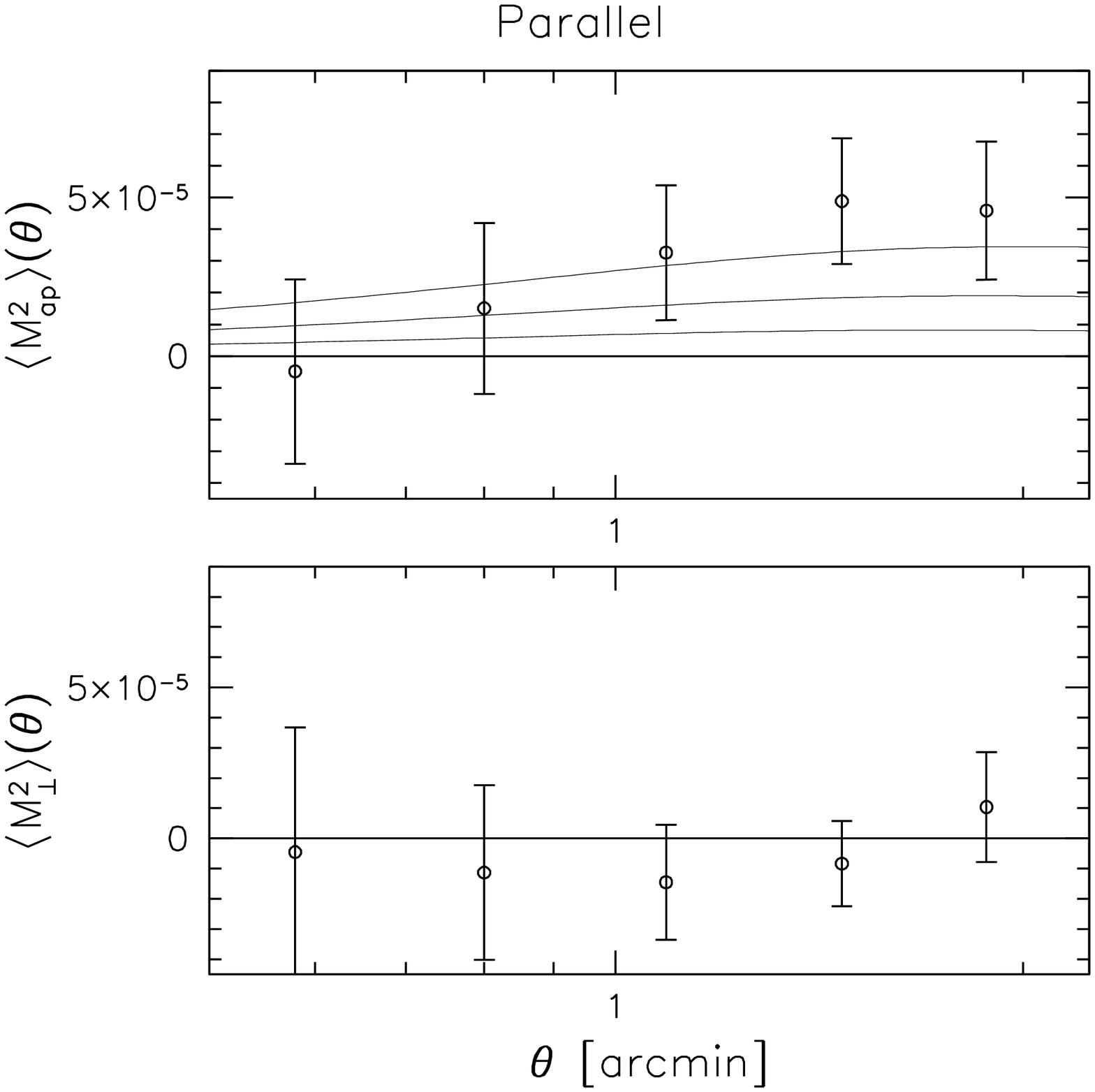}
   \includegraphics[width=6.6cm]{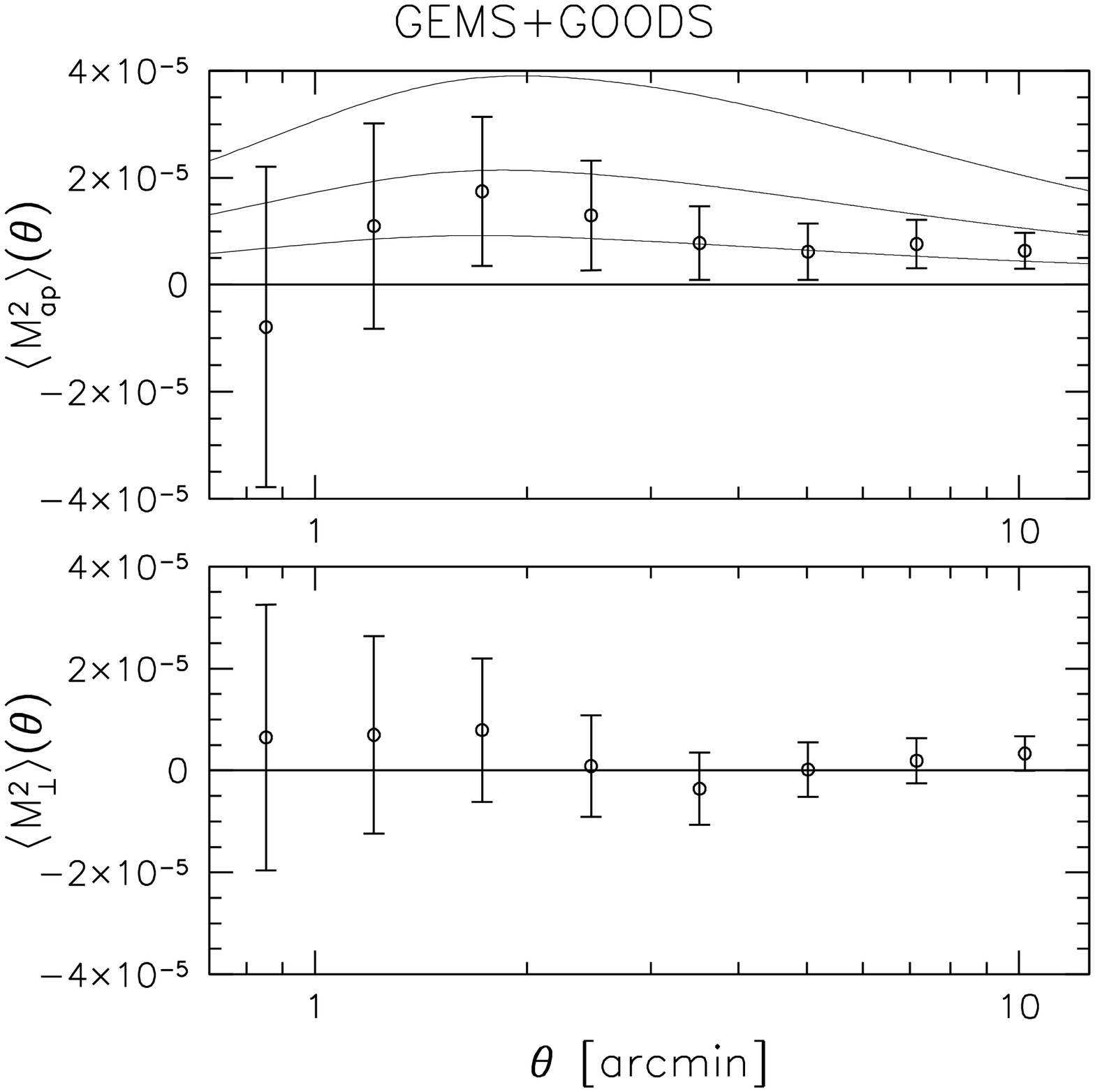}
   \caption{E-/B-mode decomposition of the aperture mass dispersion for the parallel data (left) and the combined GEMS and GOODS data (right). 
The thin solid lines show $\Lambda$CDM predictions for \mbox{$\sigma_8=(0.6, 0.8, 1.0)$}.
}
   \label{fi:map_eb}
    \end{figure*}

As a further test for contamination of the data with systematics we decompose the shear signal into E- and B-modes
using the shear correlation functions $\xi^E(\theta)$, $\xi^B(\theta)$ (Fig.\thinspace\ref{fi:xi_eb}) and
the aperture mass dispersion (Fig.\thinspace\ref{fi:map_eb}).
For this we first calculate $\xi_+(\theta)$ and $\xi_-(\theta)$ in 300 (1800) finite linear bins of width $\Delta \theta=0\farcs83$ ($1\farcs17$) from 1\arcsec to 4\farcm2 (35\arcmin) for the parallel (GEMS and GOODS) data.
$\xi^{E,B} (\theta)$ and $\langle M^2_{\mathrm{ap},\perp} \rangle (\theta)$ are then computed according to equations (\ref{eq:xi_eb},\ref{eq:map2ofxi},\ref{eq:mapbot2ofxi}) and logarithmically re-binned to reduce noise.

\subsubsection{$\xi^E$/$\xi^B$ decomposition}
As the computation of $\xi^{E,B} (\theta)$ requires knowledge of $\xi_-$ also for $\theta$ larger than the field size (see Sect.\thinspace\ref{se:theory:cosmic_shear_estimators}), we substitute the measured $\xi_-$ for $\theta>4\arcmin$ ($\theta>35\arcmin$) with theoretical predictions for a fiducial
$\Lambda$CDM cosmology with \mbox{$\sigma_8=0.7$}.
The impact of the fiducial cosmology on the E-/B-mode decomposition can be estimated by comparing $\xi^{E,B} (\theta)$ computed for \mbox{$\sigma_8=0.6$} and \mbox{$\sigma_8=1.0$} (dotted lines in Fig.\thinspace\ref{fi:xi_eb}).
While the difference is small for the GEMS and GOODS data (\mbox{$\sim 2\times 10^{-5}$}), the small size of the single ACS fields leads to a stronger cosmology dependence (\mbox{$\sim 1.5\times 10^{-4}$}) for the parallel data.
The B-mode component $\xi^B$ is consistent with zero for both datasets indicating that we are not subject to major contaminations with systematics.
The only exception is the slightly negative $\xi^B$ for the GEMS and GOODS data at large scales, which is an artefact of the discontinuity between the fiducial cosmological model and the low shear signal measured at large scales (see the E-mode signal and Sect.\thinspace\ref{se:su:correl_gems_goods}) in combination with the bootstrap errors, which do not take cosmic variance into account.

\subsubsection{$\langle M_{\mathrm{ap}}^2\rangle / \langle M_{\perp}^2\rangle$ decomposition}
Also the B-mode component of the aperture mass dispersion $\langle M^2_\perp \rangle (\theta)$ is consistent with zero for both datasets indicating the success of our PSF correction scheme (Fig.\thinspace\ref{fi:map_eb}).
Note that the E-/B-mode mixing due to incomplete knowledge of $\xi_\pm (\theta)$ for small $\theta$, which was recently discussed by \citet{kse06}, only leads to minor effects for the $\theta$ range considered here, since we truncate $\xi_\pm (\theta)$ only for \mbox{$\theta<\theta_\mathrm{min}=2\arcsec$}.
See \citet{sck07} for a E-/B-mode decomposition which can also be used for larger $\theta_\mathrm{min}$.
  \begin{figure*}[htb]
   \centering
   \includegraphics[width=8.0cm]{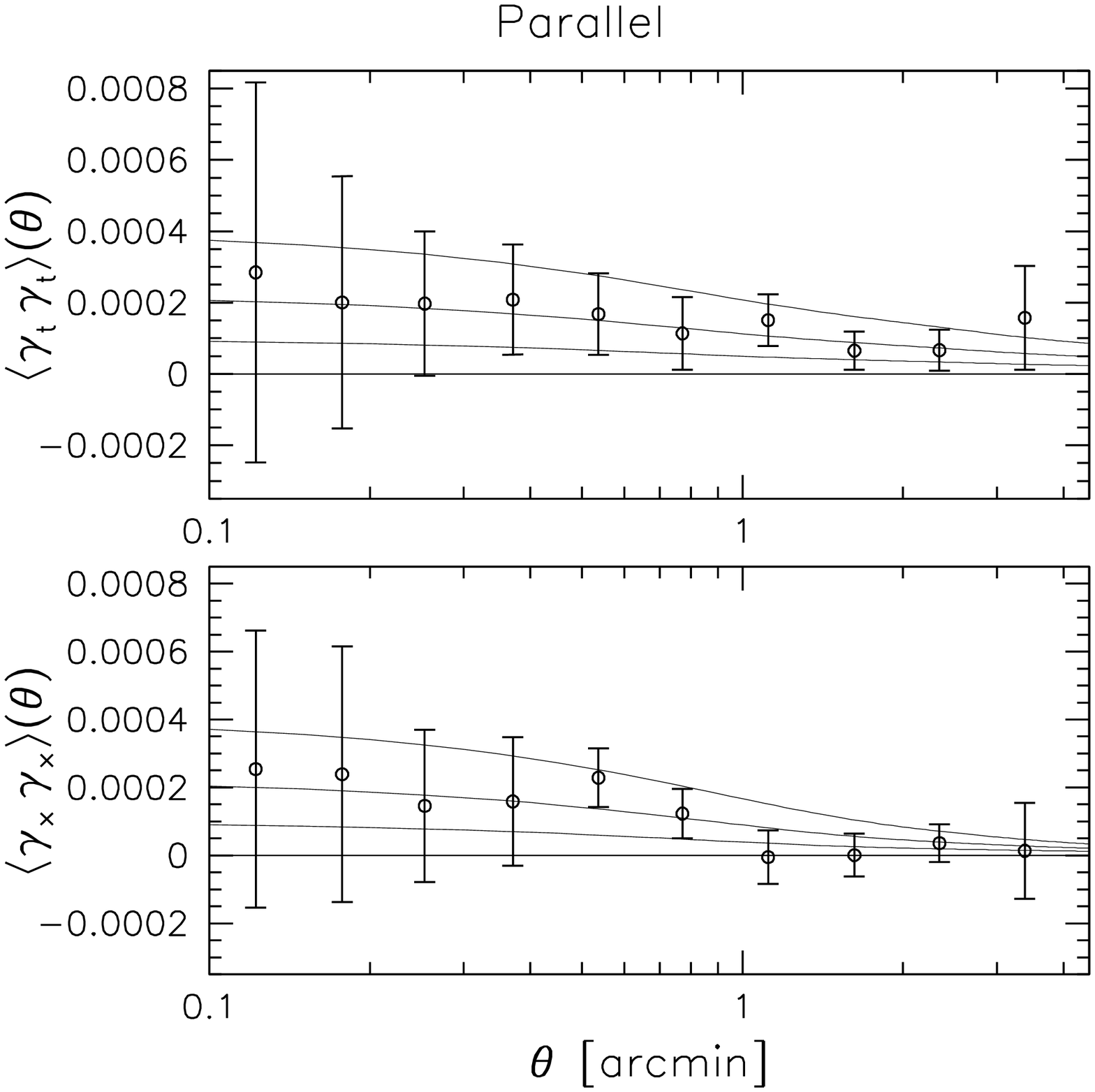}
   \includegraphics[width=8.0cm]{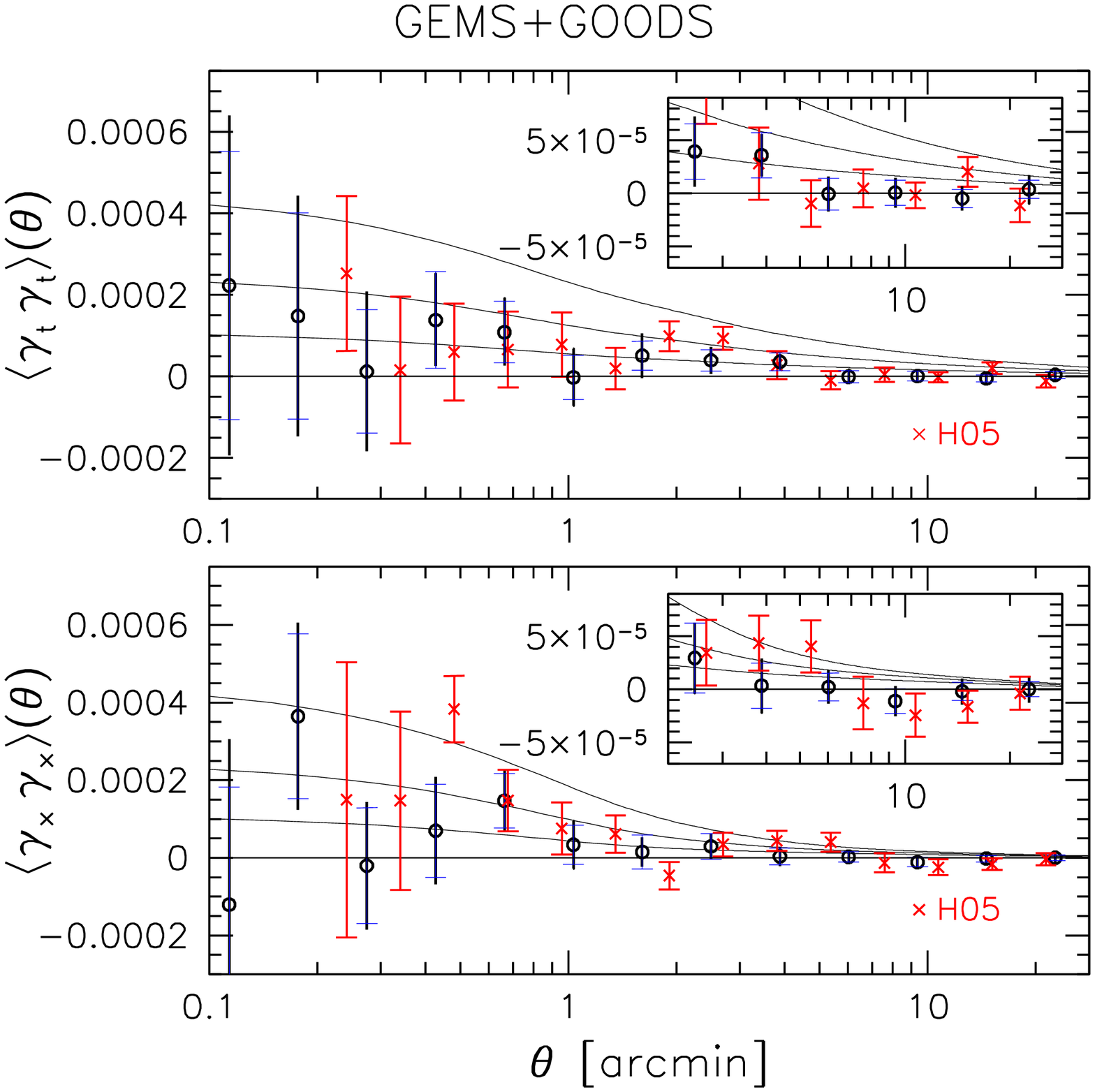}
   \caption{Two-point correlation functions $\langle \gamma_t \gamma_t \rangle$ and $\langle \gamma_\times \gamma_\times \rangle$ for the parallel data (left) and the combined GEMS and GOODS data (right). 
In the right panels we plot our estimate (open circles) both with the errors determined from Gaussian realisations (bold error-bars) and the Jackknife errors (thin caps),
and for comparison also
the \citetalias{hbb05} results (crosses).
The thin solid lines show $\Lambda$CDM predictions for \mbox{$\sigma_8=(0.6, 0.8, 1.0)$}.
Note the very low cosmic shear signal measured from the GEMS and GOODS data for large $\theta$.
}
   \label{fi:correl}
    \end{figure*}

\subsection{Shear correlation functions}
We plot our estimate for the logarithmically binned shear two-point correlation functions $\langle \gamma_t \gamma_t \rangle (\theta)$ and $\langle \gamma_\times \gamma_\times \rangle (\theta)$ in Fig.\thinspace\ref{fi:correl}.
Note that we use $\xi_\pm (\theta)$ for the cosmological parameter estimation in Sect.\thinspace\ref{se:cosmo_para_estimate},
but plot the equivalent data vectors  $\langle \gamma_t \gamma_t \rangle (\theta)$ and $\langle \gamma_\times \gamma_\times \rangle (\theta)$ in order to enable the comparison with \citetalias{hbb05}.

\subsubsection{GEMS and GOODS data}
\label{se:su:correl_gems_goods}
As we have shown in the previous sections, the GEMS and GOODS data are not contaminated with significant non-lensing signals.
We are therefore confident that the measured shear signal (right panel of Fig.\thinspace\ref{fi:correl}) is of cosmological origin.
While we detect significant shear correlations at small angular scales consistent with predictions for \mbox{$\sigma_8\sim0.6$}, both $\langle \gamma_t \gamma_t \rangle (\theta)$ and $\langle \gamma_\times \gamma_\times \rangle (\theta)$ are consistent with zero for $\theta \gtrsim 5\arcmin$,
which we interpret as caused by a large-scale under-density of the foreground structures in the CDFS.

There is good agreement between the error-bars determined from the jackknife method and from Gaussian realisations.
Only for scales of the order of the field size the jackknife method significantly under-estimates the modelled errors as it does not account for large-scale cosmic variance.
Note the good agreement of the data with the results from \citetalias{hbb05}.

\subsubsection{Parallel data}
While the measured shear correlation functions are roughly consistent with the plotted $\Lambda$CDM predictions for \mbox{$\sigma_8\sim 0.8$} (left panel of Fig.\thinspace\ref{fi:correl}),
one must be careful with its interpretation due to the
detected indications for remaining systematics (Sect.\thinspace\ref{su:average_galaxy_elli} and \thinspace\ref{se:star_gal_cross}), even if they do not show up as B-modes.
We thus postpone the cosmological interpretation of the parallel data shear signal to a future paper based on a larger data set with further corrections for the remaining systematics.

\section{Cosmological parameter estimation from the GEMS and GOODS data}
\label{se:cosmo_para_estimate}
Having shown that our GEMS and GOODS shear catalogues are not subject to significant non-lensing systematics, we use 
our estimate of the shear correlation functions, binned in 14 logarithmic bins for \mbox{$0\farcm058 <\theta<28\farcm1$},
in combination with the determined redshift distribution (Sect.\thinspace\ref{se:redshift_distribution})
for a cosmological parameter estimation
using a Monte Carlo Markov Chain (MCMC) technique \citep[see e.g. ][]{tdw05} as detailed in \citet{hss06}.
Here we utilise the covariance matrix derived from the Gaussian realisations.
This is motivated by the good agreement with the errors determined from the jackknife method at small scales indicating rather low impact of non-Gaussianity.
However, using ray-tracing simulations \citet{kis05} and \citet{swh07} found that Gaussian statistics strongly under-estimate the covariances also for GEMS like surveys, which we further discuss below.

For the parameter estimation we consider two simple $\Lambda$CDM cosmological models:
\begin{itemize}
  \item[\textbf{A}:] a $\Lambda$-universe with $\Omega_\mathrm{m},\Omega_\Lambda\in[0,1.5]$,
  \item[\textbf{B}:] a flat universe: $\Omega_\mathrm{m}+\Omega_\Lambda=1$ with $\Omega_\mathrm{m}>0$,
\end{itemize}
both with fixed \mbox{$(w,\Omega_\mathrm{b},n_\mathrm{s})=(-1,0.042,0.95)$}.
We assume a strong constraint $h=0.70\pm0.07$ for the Hubble parameter, as supported by the HST key project \citep{fmg01}
and compute the non-linear power spectrum using \texttt{halofit} \citep{spj03}, with the shape parameter calculated according to \citet{sug95}: $\Gamma=\Omega_\mathrm{m} h \exp{[-\Omega_\mathrm{b}(1+\sqrt{2 h}/\Omega_\mathrm{m})]}$, and the transfer function as given in \citet{ebw92}.
In the likelihood analysis we marginalise over the uncertainty in both $h$ and our redshift distribution.

  \begin{figure*}[htb]
   \centering
   \includegraphics[width=11.0cm,angle=270]{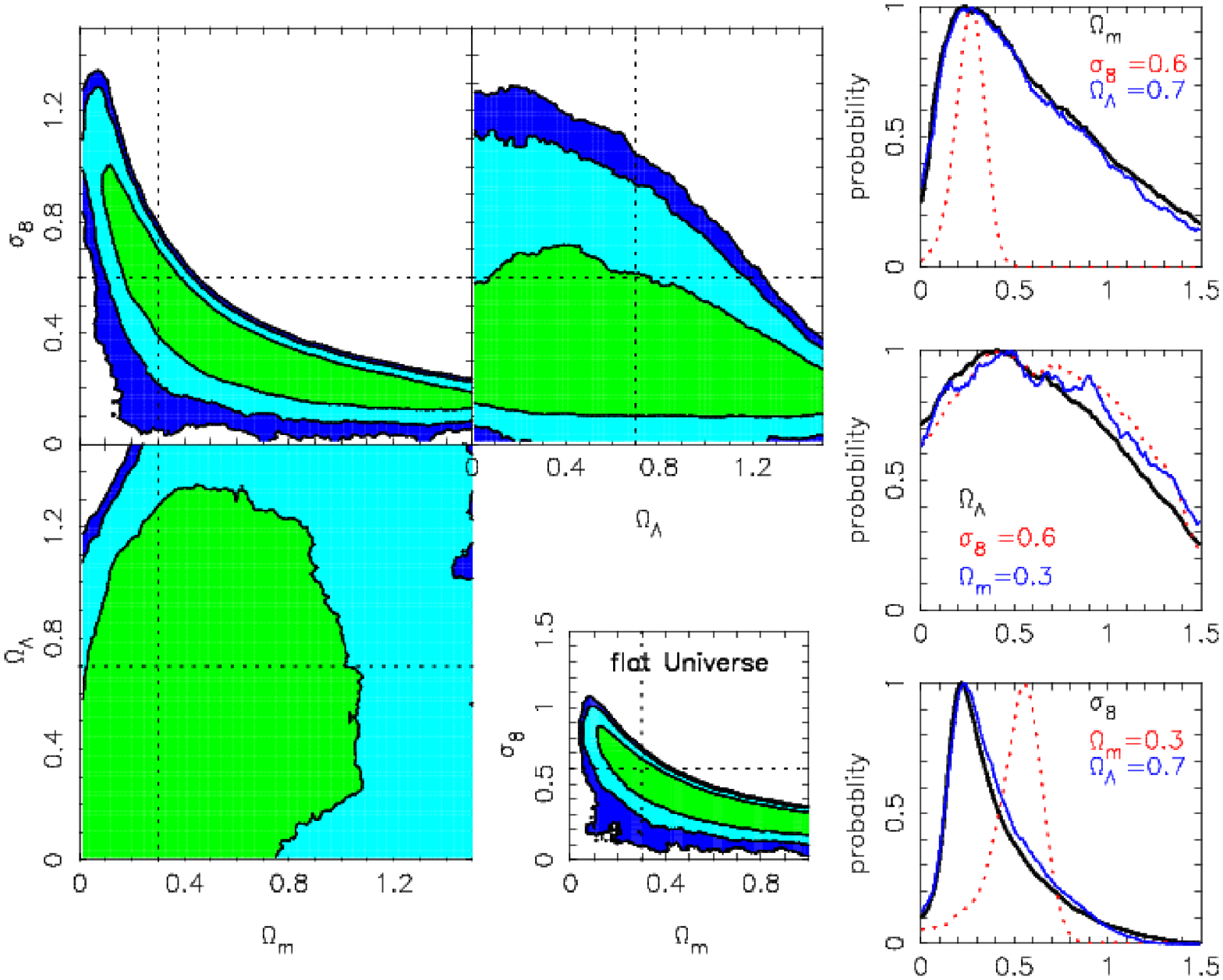}
   \caption{Constraints on $\sigma_8, \Omega_\mathrm{m},$ and $\Omega_\Lambda$ from the GEMS and GOODS data using all galaxies with \mbox{$\mathrm{S}/\mathrm{N}>4$}.
The three large contour plots show marginalised joint 2-dimensional $1,2,$ and $3\sigma$ likelihood contours for model \textbf{A}, whereas the small one was computed assuming flatness (model \textbf{B}).
For model \textbf{A} the marginalised probability is plotted on the right for $\Omega_\mathrm{m}$ (top), $\Omega_\Lambda$ (middle), and $\sigma_8$ (bottom), where the thick solid curves correspond to the total marginalised values, while the thin solid (dotted) lines correspond, from top to bottom, to fixed \mbox{$\Omega_\Lambda=0.7$} \mbox{$(\sigma_8=0.6)$}, \mbox{$\Omega_\mathrm{m}=0.3$} \mbox{$(\sigma_8=0.6)$}, \mbox{$\Omega_\Lambda=0.7$} \mbox{$(\Omega_\mathrm{m}=0.3)$}.
}
   \label{fi:mcmc_modelB}
    \end{figure*}

We plot the derived likelihood contours for $\sigma_8, \Omega_\mathrm{m},$ and $\Omega_\Lambda$ in Fig.\thinspace\ref{fi:mcmc_modelB}, where we use all galaxies with \mbox{$\mathrm{S}/\mathrm{N}>4$} (\mbox{$N_\mathrm{gal}=96 \,\mathrm{arcmin}^{-2}$}) corresponding to a median redshift \mbox{$z_\mathrm{m}=1.46\pm0.12$}.
For the more general model \textbf{A} the data only weakly constrain \mbox{$\Omega_\Lambda(\Omega_\mathrm{m}=0.3)=0.64^{+0.49}_{-0.41}$}, whereas
more stringent constraints are found for \mbox{$\sigma_8 (\Omega_\mathrm{m}=0.3)=0.52^{+0.11}_{-0.15}$}, or respectively, \mbox{$\Omega_\mathrm{m} (\sigma_8=0.6)=0.26^{+0.07}_{-0.09}$}, 
reflecting the marginalised 68\% confidence regions with strong priors on $\Omega_\mathrm{m}$ or $\sigma_8$ respectively.
Assuming flatness (model \textbf{B}) changes the estimates only marginally to \mbox{$\sigma_8 (\Omega_\mathrm{m}=0.3)=0.51^{+0.09}_{-0.13}$} and \mbox{$\Omega_\mathrm{m} (\sigma_8=0.6)=0.25^{+0.07}_{-0.08}$}, respectively.

Using the more conservative sample selection with \mbox{$\mathrm{S}/\mathrm{N}>5$}, \mbox{$m_{606}<27.0$}, \mbox{$N_\mathrm{gal}=72 \,\mathrm{arcmin}^{-2}$}, \mbox{$z_\mathrm{m}=1.37 \pm 0.10$} leads to a higher estimate of \mbox{$\sigma_8 (\Omega_\mathrm{m}=0.3)=0.59^{+0.11}_{-0.14}$} or \mbox{$\Omega_\mathrm{m} (\sigma_8=0.6)=0.30^{+0.08}_{-0.08}$} for model \textbf{A} without significantly affecting the error.
In principle, one would expect that the inclusion of the faint galaxies 
increases the signal-to-noise of the shear measurement as both the galaxy
number density and the lensing efficiency increase.
However, we can confirm the trend seen by \citetalias{hbb05} that the faintest galaxies appear to mainly add noise and  dilute the signal.
This is also consistent with the results from the STEP2 image simulations, 
where we find that the shear calibration of our KSB+ implementation is on average accurate to $\sim 3\%$, but shows a significant dependence on magnitude, with a slight over-estimation at the bright end and a $\sim 20\%$ under-estimation of the shear
for the faintest galaxies \citep{mhb07}.
Given the on average good calibration found for our analysis of the STEP2 simulations, which incorporate a cut \mbox{$\mathrm{S}/\mathrm{N}>4$}, we consider the estimate of $\sigma_8$ for the same cut to be more robust.
Yet, as the magnitude and size distribution, and additionally also the noise correlations are somewhat different for the STEP2 simulations and our data, we expect a slight remaining systematic error also for the average shear calibration.
Therefore, we use the difference of the two estimates for $\sigma_8$ as a conservative estimate of this uncertainty and take it into account as additional systematic error, yielding \mbox{$\sigma_8 (\Omega_\mathrm{m}=0.3)=0.52^{+0.11}_{-0.15}\mathrm{(stat)}\pm0.07\mathrm{(sys)}$}.
For a future shear tomography analysis this issue will need to be revisited, as it does not only require accurate shear calibration on average, but also over the whole magnitude range.

The constrained value for $\sigma_8$ is significantly lower than the estimates from other recent lensing surveys, e.g. \mbox{$\sigma_8=0.86 \pm 0.05$} \citep{smw06} and \mbox{$\sigma_8=0.85 \pm 0.06$} \citep{hmw06}, both from the CFHTLS for \mbox{$\Omega_\mathrm{m}=0.3$}, 
see 
\citet{hss06} for a compilation of recent estimates.
Our results are consistent with \mbox{$\sigma_8=0.8, \Omega_\mathrm{m}=0.3$} only at the $3\sigma$-level assuming Gaussian cosmic variance,
which we interpret as a substantial local under-density of the foreground structures in the CDFS.
In order to allow a clear comparison to the \citetalias{hbb05} results, who determine \mbox{$\sigma_8(\Omega_\mathrm{m}/0.3)^{0.65}=0.68\pm0.13$}, 
we recompute our redshift distribution using their $z_\mathrm{m}(\mathrm{mag})$ relation (\ref{eq:zm_H05}), yielding a median redshift $z_\mathrm{m}=1.12$ ($z_\mathrm{m}=1.07$) for the galaxies with \mbox{$\mathrm{S}/\mathrm{N}>4$} (\mbox{$\mathrm{S}/\mathrm{N}>5$}, \mbox{$m_{606}<27.0$}).
Then we repeat the cosmological parameter estimation assuming a redshift uncertainty $\Delta z_\mathrm{m}=0.1$ to be consistent with \citetalias{hbb05}.
For this redshift distribution we find \mbox{$\sigma_8 (\Omega_\mathrm{m}=0.3)=0.62^{+0.12}_{-0.16}$} (\mbox{$\sigma_8 (\Omega_\mathrm{m}=0.3)=0.66^{+0.11}_{-0.14}$}) in excellent agreement with the \citetalias{hbb05} results.
We thus conclude that our lower $\sigma_8$ estimate compared to \citetalias{hbb05} is mainly a result of our new redshift distribution based on the GOODS-MUSIC sample, and that the two independent shear pipelines yield consistent results (see also Sect.\thinspace\ref{su:shear_comparison}).

Our estimate of the statistical error includes the shape noise contribution, the estimated uncertainty of the redshift distribution, and a Gaussian estimate for cosmic variance.
Although there is good agreement of the errors from the jackknife method and Gaussian realisations at small scales, we expect to under-estimate cosmic variance due to non-linear evolution.
\citet{kis05} and \citet{swh07} found that the Gaussian approximation can lead to a substantial under-estimation for the correlation function covariance matrix in the non-linear regime.
Using a fitting formula found by \citet{swh07} we estimate that the diagonal elements of the $\xi_+$ covariance matrix will be under-predicted by a factor of $\sim 2.9$ for a single source redshift plane at \mbox{$z=1.4$}.
As this corresponds to the median redshift of our galaxies and since
our shear signal has the highest significance for $0\farcm6 \lesssim \theta \lesssim 5^\prime$ (see Fig.\thinspace\ref{fi:correl}), which (logarithmically averaged) roughly corresponds to a scale $\theta\sim 2^\prime$, we 
estimate very broadly that we on average under-estimate the cosmic variance contribution to the covariance matrix by a factor of $\sim 2.9$ leading to an error of $\sigma_8$ which is actually larger by $\sim \sqrt{2.9} \approx 1.7$.

Apart from the shear calibration uncertainty considered above,
further systematic errors might be introduced by intrinsic alignment of sources \citep{bth02,kis02,heh03,hbh04,hwh06,mhi06} or a correlation between the intrinsic ellipticities of galaxies and the density field responsible for gravitational lensing shear, detected by \citet{mhi06}.
Given the depth of the data analysed here, we however expect that the impact of these two effects will be small compared to the statistical uncertainties \citep[see ][]{hwh06}.
Further uncertainties arise from the limited accuracy of the predictions for the non-linear power spectrum.
Yet, given that the measured shear signal is particularly low for large $\theta$ (see Sect.\thinspace\ref{se:su:correl_gems_goods}), which are less affected by non-linear evolution, this cannot explain the low estimate of $\sigma_8$ for the GEMS and GOODS data. 

\section{Conclusions and outlook}
\label{se:conclusion}

We have presented a cosmic shear analysis of a first set of HST/ACS pure parallel observations and the combined GEMS and GOODS data of the CDFS.
We estimate that our new correction scheme for the temporally variable ACS PSF reduces the systematic contribution to the shear correlation functions due to PSF distortions to \mbox{$< 2 \times 10^{-6}$} for galaxy fields containing at least 10 stars.
This is currently the only technique taking the full time variation of the PSF between individual ACS exposures into account.
In the GEMS and GOODS data the success of the PSF correction is confirmed by a number of 
diagnostic tests indicating that the remaining level of systematics is consistent with zero.
For the parallel data we detect a low level of remaining systematics manifesting in a slight average alignment of the measured galaxy ellipticities in the $y-$direction,
which we interpret to be due to a lack of proper dithering.
We are currently further investigating this effect and exploring ways to correct for it, which will be necessary for the cosmic shear analysis of the complete set of ACS parallel observations.
Although the degradation of the ACS charge-transfer-efficiency has not been found to be a problem for the early data analysed in this work,
an in-depth analysis and correction will probably be required for the complete data set \citep[see also][]{rma05,rma07}.
Furthermore the parallel data are rather inhomogeneous regarding depth and extinction, raising the need for a well calibrated field-dependent redshift distribution.
It will also be necessary to carefully exclude any selection bias which might arise for certain classes of primary targets, particularly galaxy clusters.
Once these remaining obstacles are overcome, it will be possible to measure cosmic shear at small angular scales with unprecedented accuracy from the complete ACS Parallel Survey, with a strong reduction both of the shape noise and cosmic variance error due to many independent pointings.
The main limitation of the cosmological interpretation of the data might then arise from the current accuracy of theoretical predictions for the non-linear power spectrum at small scales.
An interesting comparison will be possible with the ACS COSMOS data \citep{mrl07}, from which cosmic shear can be measured on a wide range of angular scales.

Given the high demands concerning the control over systematics for cosmic shear measurements with ACS,
the derived technical expertise (see also \citetalias{hbb05}; \citealt{jwb05}; \citealt{rma05}; \citealt{rma07}; \citealt{lmk07}) will also be of benefit 
for other weak lensing studies with the instrument,
and possibly also other research fields requiring accurate PSF modelling.

Due to the weakness of the shear signal on the one hand, and the strong impact of poorly understood systematics on the other hand, an analysis of identical datasets with more than one independent pipeline is of great value to check the reliability of the algorithms employed.
In this work we have independently re-analysed the ACS observations of the GEMS and GOODS fields.
If we assume the same redshift parametrisation, our shear estimates are in excellent agreement with the earlier results found by \citetalias{hbb05}
 indicating the reliability of both lensing pipelines.
Such an independent comparison will also be highly desired both for the complete ACS Parallel Survey (Rhodes et al. in prep.) and the ACS COSMOS field \citep{mrl07}.
These comparisons, together with the results from the STEP project, will aid the preparations of
future space-based cosmic shear survey such as DUNE or SNAP, which will reach a very high statistical accuracy \citep{rmr04} requiring the continued advancement of improved algorithms such as shapelets (\citealt{bej02,reb03,mar05,kui06}; \citealt{nab07}).

Finally, we want to stress the possible impact of the field selection on a cosmic shear analysis:
The \textit{Chandra} Deep Field South was originally selected in a patch of the sky characterised by a low Galactic neutral hydrogen column density \mbox{($N_\mathrm{H}=8\times 10^{19} \mathrm{cm}^{-2}$)} and a lack of bright stars \citep{grt01}. 
Additionally, it neither contained known relevant extragalactic foreground sources nor X-ray sources from the ROSAT ALL-Sky Survey Catalogue\footnote{see \url{http://www.mpe.mpg.de/~mainieri/cdfs_pub/index.html}} excluding e.g. the presence of a low redshift galaxy cluster.
\citet{ami05} present a detailed analysis of compact structures in the CDFS showing the presence of a chain-like structure at $z=0.66$, a massive group at $z=0.735$ embedded into a galaxy wall extending beyond the $21^\prime \times 21^\prime$ field covered by the Vimos VLT Deep Survey \citep{fvp04}, and a further massive group at $z=1.098$ \citep[see also][]{gcd03,sbh04,vcd06}.
\citet{wmk04} identify a strong galaxy over-density at \mbox{$z\sim 0.15$}, which is too close to produce a significant lensing signal.
Given the lack of massive structures at lower redshifts $0.3\lesssim z \lesssim 0.6$ with high lensing efficiency, one would expect to measure a shear signal biased to lower values in this field as a result of strong sampling variance.
Therefore it is not surprising that our local single field estimate of \mbox{$\sigma_{8,\mathrm{CDFS}} (\Omega_\mathrm{m}=0.3)=0.52^{+0.11}_{-0.15}\mathrm{(stat)}\pm0.07\mathrm{(sys)}$} 
based on a source redshift distribution derived from the GOODS-MUSIC sample \citep{gfs06}, is 
incompatible at the \mbox{$\sim 3 \sigma$-}level assuming Gaussian cosmic variance with recent results of other weak lensing studies \citep[e.g.][]{hmw06,smw06}, which probe much larger regions on the sky.
\citet{kis05} and \citet{swh07} investigate the impact of non-Gaussianity on cosmic shear covariances.
From their results we broadly determine an under-estimation of the cosmic variance contribution to our error on $\sigma_8$ by a factor $\approx 1.7$, indicating that the CDFS is still an exceptionally under-dense field, but with a lower significance (\mbox{$\sim 2 \sigma$}) than under the Gaussian assumption.
Our $\sigma_8$ estimate is also significantly lower than the \citetalias{hbb05} results of \mbox{$\sigma_8(\Omega_\mathrm{m}/0.3)^{0.65}=0.68\pm0.13$} due to the deeper redshift distribution found in our analysis
with a median source redshift \mbox{$z_\mathrm{m}=1.46 \pm 0.12$}.
Recently \citet{pwp06} found a strong deficiency of faint red galaxies in the CDFS for the redshift range \mbox{$0.25 \lesssim z \lesssim 0.4$} indicating a substantial under-density, which is in excellent agreement with the low shear signal found in our analysis.

We believe that the CDFS represents a somewhat extreme case.
However, also other cosmic shear studies which observe a low number of small ``empty fields'' could be slightly biased just due to this prior selection.
Such a bias can of course be eliminated either with the observation of sufficiently large fields or truly random pointings,
which are realized in good approximation for a large fraction of the fields in the ACS Parallel Survey.

We plan to further investigate the peculiarity of the CDFS based on a shear tomography analysis with photometric redshifts derived for the full GEMS field, also using deep ground-based optical images from the MPG/ESO 2.2m telescope \citep{hed06} in combination with infrared images from the ESO 3.5m NTT \citep{omc06a,omc06b}.
If the low estimate for $\sigma_{8,\mathrm{CDFS}}$ indeed stems from an under-density of foreground structures we would expect an increased shear signal for a high-redshift sample of source galaxies due to the spectroscopically confirmed structures at $z=0.735$ and $z=1.098$.
Comparing the results with ray-tracing through N-body simulations we aim to further quantify the rarity of such an under-dense foreground field.

\begin{acknowledgements}

This work is based on observations made with the NASA/ESA Hubble Space Telescope, obtained from the data archives at the Space Telescope European Coordinating Facility and the Space Telescope Science Institute, which is operated by the Association of Universities for Research in Astronomy, Inc., under NASA contract NAS 5-26555.
We acknowledge extensive use of software packages developed by TERAPIX and the Leiden Data Analysis Center, and the photometric redshifts provided by the GOODS-MUSIC sample.
This research was partially conducted on the AIfA computer cluster.
We want thank the anonymous referee for his/her excellent comments and suggestions, which helped to improve this manuscript substantially.
We also thank Warren Hack and Anton Koekemoer for excellent support with MultiDrizzle, and 
Mike Jarvis, Richard Massey, 
Jason Rhodes, and Elisabetta Semboloni for fruitful discussions.
CH acknowledges support from a CITA national fellowship.
RAEF is affiliated to the Research and Science Support Department of the European Space Agency.
TS acknowledges support from the Studienstiftung des deutschen Volkes and the International Max Planck Research School for Radio and Infrared Astronomy.
This work was supported
by the German Ministry for Science and Education (BMBF) through DESY
under project 05AV5PDA/3,
and by the Deutsche Forschungsgemeinschaft under projects
SCHN 342/6--1, ER 327/2--1.

\end{acknowledgements}

\bibliographystyle{aa}
\bibliography{h4491,paper1c}
%\begin{thebibliography}{}
%\end{thebibliography}
\end{document}